\documentclass[aps,prd,nofootinbib]{revtex4}

\usepackage[english]{babel}
\usepackage{amsmath}
\usepackage{amssymb}
\usepackage{epsfig}
\usepackage{graphics,psfrag,rotating}
\usepackage{graphicx}
\usepackage{dcolumn}
\usepackage{bm}
\usepackage{overpic}
\usepackage{subfigure}

\begin{document}
\newcommand{\Od}{{\cal O}}
\newcommand{\lsim}   {\mathrel{\mathop{\kern 0pt \rlap
  {\raise.2ex\hbox{$<$}}}
  \lower.9ex\hbox{\kern-.190em $\sim$}}}
\newcommand{\gsim}   {\mathrel{\mathop{\kern 0pt \rlap
  {\raise.2ex\hbox{$>$}}}
  \lower.9ex\hbox{\kern-.190em $\sim$}}}


\author{J.\,A.\,R.\,Cembranos$^{(a,b)}$\footnote{E-mail: cembranos@physics.umn.edu},
A.\,de la Cruz-Dombriz$\,^{(b)}$\footnote{E-mail:
dombriz@fis.ucm.es}, A.\,Dobado$\,^{(b)}$\footnote{E-mail:
dobado@fis.ucm.es}, R.\,A.\,Lineros$\,^{(c)}$\footnote{lineros@to.infn.it}
and A.\,L.\,Maroto$\,^{(b)}$\footnote{E-mail:
maroto@fis.ucm.es}}

\affiliation{$^{(a)}$ William I. Fine Theoretical Physics Institute,
University of Minnesota, Minneapolis, MN 55455, USA and School of
Physics and Astronomy, University of Minnesota, Minneapolis, MN
55455, USA.}
\affiliation{$^{(b)}$Departamento de F\'{\i}sica
Te\'orica I, Universidad Complutense de Madrid, E-28040 Madrid,
Spain.}
\affiliation{$^{(c)}$ INFN sezione di Torino, I-10122
Torino, Italy and Dipartimento di Fisica Teorica, Universit\`{a} di
Torino, I-10122 Torino, Italy.}
\vspace{2.5cm}
\date{\today}

\title{
\vspace{12pt}
Photon spectra from WIMP annihilation} 

\begin{abstract}
If the present dark matter in the Universe annihilates into
Standard Model particles, it must contribute to the fluxes
of cosmic rays that are detected
on the Earth, and in particular, to the observed gamma ray fluxes.
The magnitude of such contribution
depends on the particular dark matter candidate, but certain
features of the produced
photon spectra
may be analyzed in a rather model-independent
fashion. In this work we provide the complete photon
spectra coming from WIMP
annihilation into Standard Model particle-antiparticle pairs obtained
by extensive Monte Carlo simulations. We present results
for each individual annihilation channel and provide
analytical fitting formulae for the different
spectra  for a wide range of WIMP masses.

\end{abstract}
\maketitle



\section{Introduction}
According to present observations of large scale structures, Cosmic Microwave
Background (CMB) anisotropies and light nuclei abundances,
the most important component of matter in the Universe cannot
be accommodated within the Standard Model (SM) of elementary particles.
Indeed, Dark Matter (DM) cannot be made of any of the known
particles, and this is one of the most appealing arguments
for the existence of new physics. Indeed DM  is a
required component not only on cosmological scales, but also
for a satisfactory description of rotational speeds of galaxies,
orbital velocities of galaxies in clusters, gravitational lensing
of background objects by galaxy clusters, such as the Bullet Cluster,
and the temperature distribution of hot gas in galaxies and
clusters of galaxies.
The experimental determination of the DM nature will require the
interplay of collider experiments \cite{colliders} and astrophysical
observations. These searches use to be classified in direct or
indirect searches (see \cite{Covi, Chen} for different
alternatives). Nevertheless,  non-gravitational evidence of its
existence and a concrete understanding of its nature still remain
elusive. Concerning direct searches, the elastic scattering of DM particles
from nuclei should lead directly to observable nuclear recoil signatures.
Although the number of DM particles which passes through
the Earth each second is quite large, the weak interactions between
 DM and the standard matter makes DM direct detection extremely difficult.

On the other hand, DM might be detected indirectly, by observing
their annihilation products  into SM particles. Thus, even if WIMPs
(Weakly Interacting Massive Particles) are stable, two of them may annihilate
into ordinary matter such as quarks, leptons and gauge bosons.
Their annihilation in different places (galactic halo, Sun, Earth, etc.)
produce cosmic rays to be discriminated through distinctive
signatures from the background. After WIMPs annihilation a
cascade process would occur. In the end the potentially observable
stable particles would be neutrinos, gamma rays, positrons
and antimatter (antiprotons, antihelium, antideuterions, etc.)
that may be observed through different devices. Neutrinos and gamma
rays have the advantage of maintaining their original
direction thanks to their null electric charges. On the contrary,
charged particles searches, such as those of positrons and
other antimatter particles, are hindered by propagation trajectories.

The detection of such indirect signals would not constitute
a conclusive evidence for DM since the uncertainties in
the specific DM interactions, DM densities and backgrounds
from other sources are not fully understood yet.
Nevertheless, this work precisely focuses on this kind of
detection as an indirect method to get information about the DM nature,
abundance and properties.

Photon fluxes in specific DM models are usually obtained
by software packages such as DarkSUSY and micrOMEGAs based on PHYTIA Monte Carlo
event generator. In general, the total photon spectrum obtained from
the addition of the contributions from different channels is obtained
for the particular SUSY model under consideration for a given
WIMP mass. In this sense,
it would be interesting to have a fitting function for the shape of
the spectra corresponding to each individual annihilation channel
and, in addition, determine the dependence of such spectra on the
 WIMP mass in  a
model independent way. This would allow to
apply the results to alternative candidates for which
software packages have not been developed,
and obtain photon fluxes for arbitrary WIMP candidates.
On the other hand, the information about channel contribution and
mass dependence can be very useful in order to identify gamma-ray
signals with specific WIMP candidates.

The paper is organized as follows: in  section II, we briefly review
the standard procedure for the calculation of gamma-ray fluxes from
WIMP pair annihilations. In section III, we comment on several
aspects of detectors and  backgrounds. Section IV is then devoted to
the details of specific simulations performed with PYTHIA. In
section V, we introduce the fitting formulae that will be used to
describe the spectra and in section VI the results for the
simulations, the fitted parameters and their dependence on the WIMP
mass are presented. Then, in section VII we provide some information
about the performed numerical codes obtained from our results and
available online. Section VIII is then devoted to the main
conclusions of the work. Finally, five appendices are provided in
section IX to illustrate the obtained results for some studied
annihilation channels.

\section{Gamma ray flux from DM annihilation}
Let us denote the DM mass by $M$ and its thermal averaged annihilation
cross-section  into  two SM particles (labelled by the subindex $i$)  by $\langle \sigma_{i} v \rangle$.Then the $\gamma$-ray flux from all possible annihilation channels is given by:
\begin{eqnarray}
\label{eq:integrand}
\frac{\text{d}\,\Phi_{\gamma}^{\text{DM}}}{\text{d}\,E_{\gamma}}
 &=& \frac{1}{4\pi M^2}
 \sum_i\langle\sigma_i v\rangle
\frac{\text{d}\,N_\gamma^i}{\text{d}\,E_{\gamma}}
 \;\;\;\, \times \;\;\;\,
 \frac{1}{\Delta\Omega} \int_{\Delta\Omega} {\rm d}\Omega
 \int_{\rm l.o.s.} \rho^2 [r(s)]\ {\rm d}s\;,
\\
&&
\underbrace{\;\;\;\;\;\;\;\;\;\;\;\;\;\;\;\;\;\;\;\;\;\;\;\;\;\;\;\;\;\;\;\;\;\;\;}
_{{\rm Particle~model~dependent}}\;\;\;\;\;\;\;\;\;\;
\underbrace{\;\;\;\;\;\;\;\;\;\;\;\;\;\;\;\;\;\;\;\;\;\;\;\;\;\;\;\;\;\;\;\;\;\;\;\;\;\;\;\;\;\;\;}_{{\rm Dark~matter~density~dependent}}\nonumber
\end{eqnarray}
where $\rho$ is the DM density as a function of distance from
its center $r$, which depends on the heliocentric distance
$s$. The integral is performed along the line of sight (l.o.s.) to the
target and averaged over the detector solid angle $\Delta\Omega$.

The first piece of the r.h.s. in \eqref{eq:integrand}
depends on the particular particle physics model for DM
annihilations. In particular, the self-annihilation cross sections is
mainly described
by the theory explaining the WIMP physics, whereas the number
of photons produced in each decaying channel per energy interval
involves decays and/or hadronization of unstable products,
for instance quarks and gauge bosons. Consequently,
the detailed study of these decay chains
and non-perturbative effects related to QCD is
a hard task to be accomplished by any analytical approach.
The second piece in  \eqref{eq:integrand} is a line-of-sight
integration through the DM density distribution. We
will discuss each of these pieces separately.

\subsection{Particle Physics model}
Although annihilation cross sections are not known, they are restricted
by collider constraints and direct detection. In addition,
the thermal relic density in the range $\Omega_{\rm CDM}
h^2 = 0.1123\pm 0.0035$ which is determined by fitting the standard
$\Lambda$CDM model to the WMAP7 data (Wilkinson Microwave Anisotropy
Probe results for 7 years of observations) \cite{Komatsu:2010fb},
the latest measurements from the BAO (Baryon Acoustic Oscillations)
in the distribution of galaxies \cite{Percival:2009} and the Hubble
constant ($H_0$) measurement \cite{Riess:2009pu}, do not allow an
arbitrary contribution from the DM gamma ray fluxes.

As already mentioned, the annihilation of WIMPs is closely related
to SM particle production. The time scale of an annihilation
process is shorter than typical astrophysical scales. This fact
implies that only stable or very long-lived particles survive
to the WIMP annihilations
and may therefore be observed by detectors.

For most of the DM candidates, the production of mono-energetic photons
is very suppressed.  The main reason for such a suppression comes
from the fact that DM is neutral.
Thus, it is usually assumed that the gamma-ray signal comes
fundamentally from secondary
photons originated in the cascade of decays of gauge bosons and jets produced from WIMP annihilations. These
annihilations would  produce in the end a broad energy distribution of photons,
which would be difficult to be distinguished from background.
However, the directional dependence of the gamma ray intensity coming from these
annihilations is mainly localized in point-like sources as will
be discussed in the following section. This fact could
therefore provide a distinctive signature.

In conclusion, for a particular DM candidate, an unique annihilation
channel may dominate, but in general, they all contribute. All those
channels contributions produce a broad energy gamma ray flux, whose
maximum constitutes a potential signature for its detection. Typically,
this peak is centered at an energy that is one order of magnitude lower
than the mass of the DM candidate.

On the other hand, a different strategy can be followed by taking
into account the fact that the cosmic ray background is suppressed
at high energies. Primary photons coming from the
Weicks\"{a}cker-Williams radiation dominate the spectrum at energies
close to the mass of the DM candidate and their signature is
potentially observable as a cut-off \cite{cutoff}. This approach
has the advantage of being less sensitive to electroweak corrections
which may be important if the mass of the DM candidate is larger
than the electroweak scale \cite{Ciafaloni:2010ti}.

\subsection{DM density directionality}

The line of sight integration can be obtained from:
\begin{equation}
\langle J \rangle_{\Delta \Omega} \doteq \frac{1}{\Delta \Omega}
\int_{\Delta \Omega} J(\psi) {\rm d}\Omega = \frac{2\pi}{\Delta\Omega}
\int_0^{\theta_{\rm max}} {\rm d}\theta\,\sin\theta
\int_{s_{\rm min}}^{s_{\rm max}}
{\rm d}s\,\rho^2 \left(\sqrt{s^2+s_0^2-2 s s_0 \cos\theta}\right)
\label{eq:jav}
\end{equation}
where
\begin{equation}
J(\psi) = \int_{\rm l.o.s.} {\rm d}s\,\rho^2(r).
\end{equation}
The angled brackets denote the averaging over the solid
angle $\Delta \Omega$, and $s_{\rm min}$ and $s_{\rm max}$ are the
lower and upper limits of the line-of-sight integration: $s_0
\cos\theta \pm \sqrt{r_t^2 - s_0^2 \sin^2\theta}$. In this formula
$s_0$ is the heliocentric distance and $r_t$ is the tidal radius.

Traditionally, the galactic center (GC) has attracted the attention
of this type of directional analysis since standard cusped
Navarro-Frenk-White (NFW) halos predict the existence of a very
important amount of DM in that direction \cite{BergstromUB,stoehr}.
However, this assumption is in contradiction with a substantial body
of astrophysical evidences \cite{evidences}, and a core profile is
not sensitive to standard DM candidates. On the contrary, cusped profiles
are not excluded for the Local Group dwarf spheroidals (dSphs) that constitute
interesting targets since they are much more dominated by DM.  In
this way, directional analysis towards Canis Major, Draco and
Sagittarius or Segue 1 \cite{dSphs} are more promising.

An alternative strategy takes advantage of the large field of view of FERMI,
that may be sensitive to the continuum photon flux coming from DM annihilation
at moderate latitudes ($|b| > 10^\circ$) \cite{stoehr}. Other proposed targets,
as the Large Magellanic Cloud (LMC) \cite{olinto}, are less interesting since
their central parts are dominated by baryonic matter.

\section{Detectors and backgrounds}

$\theta_{\rm max}$ in Eq. (\ref{eq:jav}) is the angle over which we average, and is
bounded from below by the experimental resolution of the particular
detector:
\begin{equation}
\Delta \Omega =2 \pi \int_0^{\theta_{\rm max}} {\rm d}\theta\, \sin\theta
= 2 \pi (1-\cos(\theta_{\rm max}) ).
\end{equation}
The quoted point spread function widths for the various experiments
are typically:  $0.4^\circ$ (EGRET), $0.1^\circ$ (CANGAROO-III, FERMI, HESS, MAGIC and VERITAS).
EGRET and FERMI are satellite
detectors with low energy thresholds ( about $100$ MeV), high energy
resolution ($\sim 15\%$) but only moderate angular precision. The
others are atmospheric Cerenkov telescopes (ACTs) with higher thresholds
($\approx$ 100 GeV) but better
angular resolution.  Typical reference sizes for the solid angle are
$\Delta \Omega=10^{-5}$ sr for ACTs and FERMI and $\Delta
\Omega=10^{-3}$ sr for EGRET.

There are different main sources of background for the signal under
consideration: hadronic, cosmic-ray electrons, localized astrophysical sources and the diffuse
$\gamma$-rays.  The latter is negligible
for ACTs, but only the last two are present for satellite experiments like
FERMI or EGRET.

For heavy WIMPs,  the produced high-energy gamma photons could be
in the range 30 GeV-10 TeV, detectable
by ACTs  such as HESS,
VERITAS or MAGIC.
On the contrary, for lighter WIMPs, the photon fluxes
would be in the range detectable by
space-based gamma ray observatories \cite{indirect}
such as EGRET, FERMI or AMS,
with better sensitivities around 30 MeV-300 GeV.

\section{Monte Carlo spectra generation: Technicalities}
In this section, we explicitly specify how gamma rays spectra have
been generated. We have used a widely known particle physics software,
PYTHIA (version 6.418) \cite{PYTHIA}, to obtain the results
we are about
to present.
In a first approximation, the WIMP annihilation is
described by two separated processes: The first one describes the
annihilation of WIMP particles and its output which are
particle-antiparticle SM pairs. The details are contained in the
theory describing the WIMP
physics. The second process considers the evolution (decays and/or
hadronization) of the SM unstable products, for instance, quarks and
gauge bosons. Unfortunately,
a first-principle description of this latter step
is too complex due to chain decays and non-perturbative QCD
effects.

As we mentioned above, in this work we have used PYTHIA
to generate the photon energy spectra starting from
pairs of SM particles, where each pair respects WIMP annihilation
quantum numbers  like neutral charge and color singlet. As will be described
below, we will allow for final state radiation from
charged particles to contribute to the photon spectra.
Due to the expected velocity dispersion of DM, we expect
most of the annihilations to happen quasi-statically. This fact offers
the center of mass (CM) frame as the most suitable frame to
produce the photon spectra. Hence, the process is
described by the total energy:
\begin{eqnarray}
E_{\text{CM}} \simeq 2\,M
\label{E_CM}
\end{eqnarray}
where $M$ is the mass of the \text{WIMP} particle. Therefore,
by considering different CM energies for the
SM particles pairs in each WIMP annihilation process we are
indeed studying different WIMP masses. The procedure to obtain
the photon spectra is thus straightforward, except for the particular
case of the $t$ quark. For any given  pair of SM particles which are
produced in the WIMP  annihilation, we count the number of photons
in each bin of energy and then normalize them to the total number
of simulated pair collisions. The bins which we have  considered
in the $x$ variable, $x\equiv E_{\gamma}/M$, are:  $[10^{-5},\,10^{-3}]$,
$[10^{-3},\,0.2]$, $[0.2,\,0.5]$, $[0.5,\,0.8]$ and  $[0.8,\,1.0]$.
Nevertheless, for some studied channels more precision was needed
in some particular energy intervals and additional bins were considered.

The number of simulated collisions in each bin  was fixed a priori
but it was changed if required in order to provide suitable statistics
in the number of produced photons. For instance, for the high energy
bins many collisions are required to get a significant number of
photons,  whereas for low-intermediate energy, many photons are
usually produced even for a small number of collisions. The total
number of photons corresponding to the different generated pairs
in terms of the WIMP mass are presented in Tables
\ref{table_events_W_Z_top}, \ref{table_events_tau}  and
\ref{table_events_quarks}. These results will be plotted in the
Appendix $\bf{E}$, Figure 9 at the
end of the paper.


The SM particle pairs decays generated are  $W$ and $Z$ gauge bosons,
$\tau$ and $\mu$ leptons and  $u$, $d$, $s$, $c$, $b$ and $t$ quarks. For each
annihilation channel we have studied the  gamma ray spectra produced
for different WIMP masses. The result of the simulations were  fitted
to analytical expressions as is described in the following section.
\begin{widetext}

\begin{table}
\centering \small{
\begin{tabular}{||c|c|c|c|c|c|c|c|c|c||}
%
\hline
Channel $\diagdown$ Mass (GeV) &  100 &   125 &  150  &   200   & 250    &  350     &  500  &  1000  \\
\hline
$W^+W^-$      & 5.21 &    -  &  1.91 &  6.85   &    -   &  7.83    & 2.91  & 2.85\\
$ZZ$          & 0.42 &  6.01 &  2.91 &  14.9   &    -   &  14.2    & 2.81  & 2.02\\
$t \bar t$    &   -  &    -  &    -  &  0.70   &  0.86  &  0.32    & 2.81  & 1.41\\
\hline
\end{tabular}
} \caption{\footnotesize{Total number of photons --
in $10^{7}$ units -- generated from  $W^+W^-$, $ZZ$ and $t \bar t$
channels for different WIMP masses.}}
\label{table_events_W_Z_top}
\end{table}
\begin{table}
\centering \small{
\begin{tabular}{||c|c|c|c|c|c|c|c|c|c||}
%
\hline
Channel $\diagdown$ Mass (GeV) &  25 &   50 &  100  &   200   & 500    &  1000     &  $10^{4}$  &  $5\cdot10^4$  \\
\hline
$\tau^+\tau^-$ & 2.25 &  2.25 & 2.23 & 1.07 & 2.81 & 2.33 & 8.41 & 7.80 \\
\hline
$\mu^+\mu^-$ & 2.25 &  2.25 & 2.23 & 1.07 & 2.81 & 2.33 & 8.41 & 7.80 \\
\hline
\end{tabular}
} \caption{\footnotesize{Total number of  photons -- in $10^{7}$
units -- generated from $\tau^+\tau^-$ and $\mu^+\mu^-$ channels
for different WIMP masses.}}
\label{table_events_tau}
\end{table}
\begin{table}
\centering \small{
\begin{tabular}{||c|c|c|c|c|c|c|c|c|c||}
%
\hline
Channel $\diagdown$ Mass (GeV)  &  50     &   100 &  200   &   500  &   1000 &  2000  &  5000    & 7000      &  8000\\
\hline
$u\bar u$   & 2.05    & 11.9  & 2.42  & 2.81 & 3.82  & 10.8 &  5.91   &  -        & 2.11\\
$d\bar d$   & 1.04    & 1.96  & 2.42  & 2.81 & 2.81  & 2.81  &  2.31   &   -       &  -  \\
$s\bar s$   & 15.3    & 2.00  & 1.97  & 2.81 & 9.82  & 2.71  &  2.71   & 11.0     &  - \\
$c\bar c$   & 2.41    & 1.99  & 16.8 & 2.81 & 2.81  & 3.81  &  12.0  &   -       & 3.00\\
$b\bar b$   & 11.7    & 1.91  & 2.62  & 2.61 & 8.81  & 2.20  &  3.81   &   -       & 1.70\\
%
\hline
\end{tabular}
} \caption{\footnotesize{Total number of  photons -- in $10^{7}$
units -- generated from  $u\bar u$, $d\bar d$, $s\bar s$, $c\bar c$
and $b\bar b$ channels for different WIMP masses.}}
\label{table_events_quarks}
\end{table}
\end{widetext}
%
%
\subsection{Final state radiation}
If the final state in the annihilation process contains charged
particles, there is a finite probability of emission of an
additional photon. This is discussed in detail in \cite{Bringmann}. In principle there are two types of contributions: that coming from photons directly radiated from the external legs, which is the final state radiation we have considered in the work, and that coming from virtual
particles exchanged in the WIMP annihilation process. The first kind of
contribution can be described for relativistic final states by means of
an universal Weizs\"{a}cker-Williams term fundamentally independent
from the particle physics model \cite{Bringmann}. On the other hand, radiation from virtual particles only takes place
in certain DM models and is only relevant in
particular cases, for instance, when the virtual particle mass is
almost degenerate with the WIMP mass. Even in these cases,
it has been shown \cite{Cannoni} that although this effect has to be
included for the complete evaluation of fluxes of high energy
photons from WIMP annihilation, its contribution is
relevant only in models and at energies where the lines contribution
is dominant over the secondary photons. For those reasons and
since the aim of the present work
is to provide model independent results for photon spectra,
 only final state radiation was included in our simulations.

\subsection{The case for $t$ quark decay}
The decay of top quark is not explicitly included in
\text{PYTHIA} package. We have approximated this process by its dominant
SM decay, i.e. each (anti) top
decays into $W^{+(-)}$ and (anti) bottom.
In order to maintain any non-perturbative effect, we work on an
initial four-particle state composed by $W^{+} b$ coming
from the top and  $W^{-} \bar{b}$ from antitop, which keeps all
kinematics and color properties from the original pair.
Starting from this configuration, we have forced decays and
hadronization processes to evolve as \text{PYTHIA} does and
therefore, the gamma rays spectra  corresponding to this channel
have also been included in our analysis.\\

\section{Analytical fits to PYTHIA simulation spectra}

In this section we present the  fitting functions used for the different
channels. According to the PYTHIA simulations
described in the previous section, three different parametrizations
were required in order to fit all available data from the studied
channels. The first one for quarks (except the top) and leptons.
Then,  a second one for gauge bosons $W$ and $Z$ and a third
one for the top.

\subsection{Quarks and leptons}
%
%
For quarks (except the top), $\tau$ and $\mu$ leptons,
the most general formula needed to reproduce the
 behaviour of the differential
number of photons per photon energy may
be written as:

\begin{eqnarray}
x^{1.5}\frac{\text{d}N_{\gamma}}{\text{d}x}\,=\, a_{1}\text{exp}\left(-b_{1} x^{n_1}-b_2 x^{n_2} -\frac{c_{1}}{x^{d_1}}+\frac{c_2}{x^{d_2}}\right) + q\,x^{1.5}\,\text{ln}\left[p(1-x)\right]\frac{x^2-2x+2}{x}
\label{general_formula}
\end{eqnarray}

In this formula, the logarithmic term
takes into account the final state radiation through the
Weizs\"{a}cker-Williams expression
\cite{Hooper2004,Bringmann}.
Nevertheless, initial radiation
is removed from our Monte Carlo
simulations in order to avoid wrongly counting their
possible contributions.

Strictly speaking, the $p$ parameter in the
Weizs\"{a}cker-Williams term in the previous
formula is $(M/m_{particle})^2$
where $m_{particle}$ is  the mass of the
charged particle that emits radiation. However in our case, it will be
a free parameter to be fitted since the radiation comes
from many possible charged particles, which are produced
along the decay and hadronization processes. Therefore we are
encapsulating all the bremsstrahlung effects in a
single Weizs\"{a}cker-Williams-like term.

Concerning the $\mu$ lepton, the expression above
\eqref{general_formula} becomes simpler
 since the exponential contribution is absent.
The $\mu^-$ decays in $e^-\bar\nu_{e}\nu_\mu$ with a
branching ratio of $\sim 1$ and therefore the only contribution
in addition to its own bremsstrahlung, is provided by the
radiation coming from the electron. The total gamma rays flux
is thus well fitted by:

\begin{eqnarray}
x^{1.5}\frac{\text{d}N_{\gamma}}{\text{d}x}\,=\, q\,x^{1.5}\,\text{ln}\left[p(1-x^{l})\right]\frac{x^2-2x+2}{x}
\label{general_formula_mu}
\end{eqnarray}
where the $l$ parameter in the logarithm is needed
in order to fit the simulations as will be seen in the corresponding
sections.

Let us mention at this stage that for the gamma rays obtained from
electron-positron pairs, the only
contribution is that coming from bremsstrahlung. Therefore,
the previous expression \eqref{general_formula_mu} is also valid
with $q=\alpha_{\text{QED}}/\pi$, $p=\left(M/m_{e^{-}}\right)^{2}$ and $l\equiv1$.
This choice of the parameters
corresponds  of course to the well-known
Weizs\"{a}cker-Williams formula.

\subsection{$W$ and $Z$ bosons}
For the $W$ and $Z$ gauge bosons, the parametrization used to
 fit the Monte Carlo simulation is:
\begin{eqnarray}
x^{1.5}\frac{\text{d}N_{\gamma}}{\text{d}x}\,=\, a_{1}\,\text{exp}\left(-b_{1}\, x^{n_1}-\frac{c_{1}}{x^{d_1}}\right)\left\{\frac{\text{ln}[p(j-x)]}{\text{ln}\,p}\right\}^{q}
\label{general_formula_W_Z}
\end{eqnarray}

This expression differs from the expression \eqref{general_formula} in
the absence of the  additive logarithmic contribution. Nonetheless,
this contribution acquires a multiplicative behaviour. The exponential
contribution is also quite simplified with only one positive and one
negative power laws. Moreover, $a_1$, $n_1$ and $q$ parameters
appear to be independent of the WIMP mass $M$
as will be seen in the corresponding section. The rest of parameters, i.e.,
$b_1$, $c_1$, $d_1$, $p$ and $j$, are WIMP mass dependent and will be
determined for each WIMP mass and for the $W$ and $Z$ separately.
In  both cases we have covered a WIMP mass range from
$100$ to $10^4$ $\text{GeV}$. Nonetheless, at masses higher
than 1000 $\text{GeV}$, we have observed no significant
change in the photon spectra for both particles.


\subsection{$t$ quark}
Finally, for the top, the required parametrization turned out to be:
%
%


\begin{eqnarray}
x^{1.5}\frac{\text{d}N_{\gamma}}{\text{d}x}\,=\, a_{1}\,\text{exp}\left(-b_{1}\, x^{n_1}-\frac{c_{1}}{x^{d_1}}-\frac{c_{2}}{x^{d_2}}\right)\left\{\frac{\text{ln}[p(1-x^{l})]}{\text{ln}\,p}\right\}^{q}
\label{general_formula_t}
\end{eqnarray}

Likewise the previous case for $W$ and $Z$ bosons,
gamma-ray spectra parametrization for the top is quite
different from that given by expression \eqref{general_formula}.
This time, the exponential contribution is more complicated than the
one in expression \eqref{general_formula_W_Z}, with one positive and
two negative power laws. Again, the  additive logarithmic contribution
is absent but it acquires a multiplicative behaviour. Notice the
exponent $l$
in the logarithmic argument, which is required to provide correct
fits for this particle.


The covered WIMP mass range for the top case was from $200$
to $10^5$ $\text{GeV}$. Nevertheless, at masses higher than
1000 $\text{GeV}$ we have observed again that there is no significant
change in the  gamma-ray spectra.



\section{Results from \text{PYTHIA} simulation}
In this section we present the results of our fit of the parameters given by
expressions  \eqref{general_formula}, \eqref{general_formula_W_Z}
and \eqref{general_formula_t} after having performed
the PYTHIA simulations described in section IV.
For each studied channel, we have considered the possibility of
parameters depending on the WIMP mass.

Once the parameters in expressions \eqref{general_formula},
\eqref{general_formula_W_Z} and \eqref{general_formula_t}
have been determined for each channel and different WIMP masses,
it is possible to study their
evolution with the WIMP mass $M$.
Some parameters in expressions are WIMP mass independent
and take values that depend on the studied channel. The rest are
WIMP mass dependent.

For some channels and in some range of WIMP masses, we observed
that this dependence was given by a simple power law. In fact, for a given
channel $(i)$  and a generic mass dependent $\mathcal{P}$ parameter, a simple
power-law scaling behavior would correspond to an expression like
%
\begin{eqnarray}
\mathcal{P}^{(i)}(M)\,=\,m_{\mathcal{P}^{(i)}} M^{n_{\mathcal{P}^{(i)}}}
\label{linear_fit}
\end{eqnarray}
with $m_{\mathcal{P}^{(i)}}$ and $n_{\mathcal{P}^{(i)}}$ constant values to be determined
for the different studied channels. Values of $m_{\mathcal{P}^{(i)}}$ and $n_{\mathcal{P}^{(i)}}$ and their range of validity
are  presented for each studied channel in the following.

\subsection{$W$ boson}
As commented above, the correct parametrization for the $W$ boson
simulations was given by expression \eqref{general_formula_W_Z}.
For this boson, there are five mass-dependent parameters:
$b_1$, $c_1$, $d_1$, $p$ and $j$ whose values are detailed in
Table \ref{table_W}. The mass independent parameters are
$a_1=25.8$, $n_1=0.51$ and $q=3.00$. The mass range considered for this boson
is 100 to $10^{5}$ $\text{GeV}$. In fact,
from $M=1000\,\text{GeV}$, the photon spectrum does not change.
The parameters obtained fit the enegy spectra from
$x=2\cdot10^{-4}$ till the end of the allowed interval.
It can be seen that for low masses the spectrum does not end
at $x=1$ but at smaller energies (e.g. $x\simeq0.78$ for
$M=100\,\text{GeV}$) and as masses get higher, the energy tail
approaches $x=1$.

Some of these results are presented in Figure 1 in Appendix $\bf{A}$ for four WIMP
masses: 100, 200, 350 and 1000 GeV. Besides,
mass dependent parameters $b_1$, $c_1$, $d_1$, $p$ and $j$ were presented
in the same Appendix in Figure 2.

Concerning the scaling behavior of these mass dependent parameters
given by expression \eqref{linear_fit}, we obtain that $b_1$, $c_1$ and $j$ parameters
scale with a simple power law of $M$ at high masses. In fact, $b_1$ and $c_1$ parameters
follow a two power-law behavior at low masses.
For $d_1$ parameter, we find that the sum of two power laws
covers this high masses interval,
whereas a simple power law at low masses is obeyed.
Parameter $p$ scales with two power laws in the whole studied mass interval.
These results are shown in Table \ref{table_W_fitting}.
\begin{table}
\centering \small{
\begin{tabular}{|c|c|c|c|c|c|}
\hline WIMP mass ($\text{GeV}$) & $b_1$ & $c_1$ & $d_1$ & $p$ & $j$ \\
\hline
100  &  9.48 & 0.651  & 0.292 & 973  &  0.790 \\
150  &  8.87 & 0.808  & 0.261 & 783  &  0.919 \\
200  &  8.64 & 0.882  & 0.250 & 684  &  0.955 \\
350  &  8.56 & 0.907  & 0.245 & 593  &  0.991 \\
500  &  8.51 & 0.917  & 0.244 & 560  &  0.996 \\
1000 &  8.45 & 0.931  & 0.242 & 535  &  1.000 \\
\hline
\hline
\end{tabular}\\
\begin{tabular}{|c|}
\hline
$a_1=25.8$ ; $n_1=0.510$ ; $q=3.00$\\
\hline
\end{tabular}
}
\caption{\footnotesize{\textbf{$W$ boson}:
$b_1$, $c_1$, $d_{1}$, $p$ and $j$ parameters corresponding to \eqref{general_formula_W_Z} in the $W^{+}W^{-}$ channel for
different WIMP masses. Mass independent parameters in \eqref{general_formula_W_Z} for this channel
are presented at the bottom of the table.
}}
\label{table_W}
\end{table}
\begin{widetext}
 \begin{table}
\centering \small{
\begin{tabular}{||c|c|c||}
\hline Parameter & WIMP mass interval (GeV) & Fitting power law(s)\\
\hline
$b_{1}$ & $100\leq M\leq200$ & $0.0433 M^{0.765} +  46.4 M^{-0.382}$ \\
               & $200 < M \leq 1000$ &   $9.29 M^{-0.0139}$\\
\hline
$c_{1}$ & $100\leq M\leq200$ & $-27 M^{-0.240} +  35.0 M^{-0.0643} -16.5$  \\
               & $200 < M \leq 1000$& $0.743 M^{0.0331}$  \\
\hline
$d_{1}$ & $100\leq M\leq 240$  &  $2.64\cdot 10^{-4} M^{1.03} +  2.28 M^{-0.470}$ \\
               & $240 < M \leq 1000$  &   $0.265 M^{-0.0137}$  \\
\hline
$p$ & $200\leq M\leq 1000$ & $10^5 M^{-1.13} +  285  M^{0.0794}$ \\
\hline
$j$ & $385\leq M \leq 1000$ & $0.943 M^{0.00852}$\\
\hline
\end{tabular}
} \caption{\footnotesize{Parameters corresponding to
\eqref{linear_fit} for $W$ boson. It can be seen that $p$
parameter follows two different power laws depending on the
WIMP mass interval. For the remaining  mass dependent
parameters there is a unique power law behavior in the WIMP mass
interval $350\leq M\leq1000$.}}
\label{table_W_fitting}
\end{table}
\end{widetext}
\subsection{$Z$ boson}

For the $Z$ boson the correct parametrization is again the
one given by expression \eqref{general_formula_W_Z}. For this boson
there are five mass-dependent parameters:
$b_1$, $c_1$, $d_1$, $p$ and $j$ which are detailed in
Table \ref{table_Z}. The mass independent parameters are
$a_1=25.8$, $n_1=0.5$ and $q=3.87$. The studied WIMP mass range for this boson
was from 100 to $10^{5}$ $\text{GeV}$. However, above
$M=1000\,\text{GeV}$ the energy spectrum does not change as can
be seen from our simulations.

The chosen parameters values fit the photon spectra
from $x=5\cdot10^{-4}$ till the end of the allowed interval. As for
the $W$ case, it can be seen that for low masses the
spectrum does not end
at $x=1$ but at smaller energies (e.g. $x\simeq0.7$ for
$M=100\,\text{GeV}$) and as masses get higher, the high-energy tail
approaches $x=1$.

Concerning the power-law scaling of the parameters with $M$, we
obtained that  parameters $b_1$, $c_1$, $d_1$ ad $j$ follow a simple power-law behavior
for high WIMP masses. Parameter $p$ follow a two sum power-law behavior for masses higher than
170 GeV. Concerning $d_1$ parameter, the whole accessible WIMP mass interval is covered by different
either one or two power laws. These results can be seen in Table \ref{table_Z_fitting}.
%
\begin{table}
\centering \small{
\begin{tabular}{||c|c|c|c|c|c||}
\hline WIMP mass (\text{GeV}) & $b_1$ & $c_1$ & $d_1$ & $p$ & $j$\\
\hline
100  & 10.3  & 0.498  & 0.323 & 7010  &   0.702 \\
125  & 9.74  & 0.612  & 0.294 & 4220  &   0.836 \\
150  & 9.49  & 0.675  & 0.280 & 3850  &   0.894 \\
200  & 9.28  & 0.734  & 0.268 & 3630  &   0.943 \\
350  & 9.02  & 0.800  & 0.257 & 3380  &   0.978 \\
500  & 8.95  & 0.813  & 0.255 & 3260  &   0.988 \\
1000 & 8.91  & 0.819  & 0.254 & 3140  &   0.997 \\
\hline
\end{tabular}
\\
\begin{tabular}{|c|}
\hline
$a_1=25.8$ ; $n_1=0.5$ ; $q=3.87$\\
\hline
\end{tabular}
} \caption{\footnotesize{\textbf{$Z$ boson}:
$b_1$, $c_1$, $d_1$, $p$
and $j$ parameters corresponding to  \eqref{general_formula_W_Z} in the $ZZ$ channel for
different WIMP masses.  Mass independent parameters in \eqref{general_formula_W_Z} for this channel
are presented at the bottom of the table.}}
\label{table_Z}
\end{table}
\begin{widetext}
 \begin{table}
\centering \small{
\begin{tabular}{||c|c|c||}
\hline Parameter & WIMP mass interval (GeV) & Fitting power law(s)\\
\hline
$b_{1}$ & $500\leq M\leq1000$ & $9.36 M^{ -0.00710}$ \\
\hline
$c_{1}$ & $465\leq M\leq1000$ & $0.765 M^{ 0.00980}$ \\
\hline
$d_{1}$ & $100\leq M\leq 191$        & $0.00999 M^{0.530}+ 21.5 M^{-1.01}$  \\
               & $191 < M\leq 360$            & $2.02\cdot 10^{-9} M^{2.56}+ 0.491 M^{-0.115}$ \\
               & $360 < \leq M\leq 1000$  & $0.272 M^{-0.00990}$ \\
\hline
$p$ & $170\leq M\leq1000$ &  $8550 M^{-0.166} + 0.476 M^{0.984}$\\
\hline
$j$ & $350\leq M \leq 1000$ & $0.884 M^{0.0175}$  \\
\hline
\end{tabular}
} \caption{\footnotesize{Parameters corresponding to
\eqref{linear_fit} for the $Z$ boson. It can be seen that
all mass dependent parameters for the  $Z$ boson follow a simple
power-law scaling at intermediate and high masses.}}
\label{table_Z_fitting}
\end{table}
\end{widetext}

\subsection{$t$ quark}

For the top, there are six mass dependent parameters:
$b_1$, $n_1$, $c_2$, $p$ $q$ and $l$ which are detailed in
Table \ref{table_top}.  The mass independent parameters are
$a_1=290$, $c_1=1.61$, $d_1=0.19$ and $d_2=0.845$. The mass range for this quark is
from 200 to $10^5$ $\text{GeV}$. Nevertheless, from
1000 GeV onwards, the photon spectra do not change as was proven
by considering several higher masses. The chosen parameters fit the
 spectra from $x=10^{-4}$ till the end of the allowed interval. Again for low masses, the
spectra do not end at $x=1$ but at smaller energies (e.g. $x\simeq 0.7$ for
$m=200\,\text{GeV}$ ) and, as masses get higher, the spectral tail
approaches $x=1$.

Some of these results are presented graphically in Figure 3, Appendix $\bf{B}$
for four WIMP masses: 200, 250, 500 and 1000 GeV . Also in this Appendix,
mass dependent parameters $b_1$, $n_1$, $c_2$, $p$, $q$ and $l$ are plotted in
Figure 4.

Concerning the scaling behavior of the $c_2$, $p$, $q$ and $l$ parameters, they obey
a simple power law in the whole accessible WIMP mass range.
Nevertheless, for $b_1$ and $c_1$ parameters the simple power law
behavior starts from masses bigger than 350 GeV. These results can
be seen in Table \ref{table_top_fitting}.
\begin{table}
\centering \small{
\begin{tabular}{||c|c|c|c|c|c|c||}
\hline WIMP mass (\text{GeV}) & $b_1$ & $n_1$ & $c_2$ & $p$ & $q$ & $l$ \\
\hline
200  & 14.4  & 0.477 & $3.34\cdot10^{-4}$ & 1.34  & 1.76 & 4.42\\
250  & 13.5  & 0.457 & $1.54\cdot10^{-4}$ & 1.95  & 1.96 & 4.14\\
350  & 13.0  & 0.448 & $5.99\cdot10^{-5}$ & 3.78  & 2.32 & 3.74\\
500  & 12.8  & 0.442 & $1.69\cdot10^{-5}$ & 7.40  & 2.75 & 3.36\\
1000 & 12.4  & 0.436 & $1.80\cdot10^{-6}$ & 30.0  & 3.85 & 2.72\\
\hline
\end{tabular}
\\
\begin{tabular}{|c|}
\hline
$a_1=290$ ; $c_1=1.61$ ; $d_1=0.19$ ; $d_2=0.845$\\
\hline
\end{tabular}
}
\caption{\footnotesize{\textbf{$t$ quark}: $b_1$, $n_1$, $c_2$, $p$, $q$ and $l$ parameters corresponding to
\eqref{general_formula_t} in the $t\bar{t}$ channel for different WIMP masses.
Mass independent parameters in \eqref{general_formula_t} for this channel
are presented at the bottom of the table.
}}
\label{table_top}
\end{table}
\begin{widetext}
 \begin{table}
\centering \small{
\begin{tabular}{||c|c|c||}
\hline Parameter & WIMP mass interval (GeV) & Fitting power law(s) \\
\hline
$b_{1}$ & $200\leq M \leq 350$ & $9.32 M^{0.0507} + 11.0\cdot 10^{6} M^{-2.91}$  \\
               & $350< M \leq 1000$   &  $16.4 M^{-0.0400}$ \\
\hline
$n_{1}$ & $200 \leq M < 300$ &  $21.4 M^{-0.818}+0.00867 M^{0.589}$ \\
               & $300\leq M \leq 1000$ & $0.559 M^{-0.0379}$ \\
\hline
$c_{2}$ & $200\leq M \leq 1000$ & $8910 M^{ -3.23}$ \\
\hline
$p$ & $200\leq M <1000$ & $5.78\cdot 10^{-5} M^{1.89}$ \\
\hline
$q$ & $200\leq M \leq 1000$ & $0.133 M^{0.488}$  \\
\hline
$l$ & $200\leq M \leq 1000$ & $21.9 M^{-0.302}$ \\
\hline
\end{tabular}
} \caption{\footnotesize{Parameters corresponding to
\eqref{linear_fit} for the $t$ quark. It can be seen that all mass
dependent parameters for the $t$ quark follow a simple power-law scaling
behavior at intermediate and high WIMP masses.
Parameters $b_1$ and $n_1$ follow the simple power law
for $M>300$ GeV whereas the rest of parameters
presented do so in the interval $200\leq M \leq 1000$.}}
\label{table_top_fitting}
\end{table}
\end{widetext}
\subsection{Leptons and quarks}
For the rest of the quarks and leptons,
the parametrization given in \eqref{general_formula} is
completely valid. Now we present results for $\tau$ and $\mu$ leptons
and all quarks except for the top.


\subsubsection{$\tau$ lepton}
%
%
%
For the $\tau$ lepton, there are only two mass dependent
parameters in the spectra fitting function
\eqref{general_formula}: $n_1$ and $p$.
The remaining parameters are mass independent for this particle
and their values are $a_1=14.7$, $b_1=5.40$, $b_2=5.31$, $n_2=1.40$,
$c_1=2.54$, $d_1=0.295$, $c_2=0.373$, $d_2=0.470$ and $q=0.00260$.  These results
are presented in Table \ref{table_tau}.

In this case, the WIMP mass interval considered ranges from $25$ to
$5\cdot10^{4}$ GeV. For masses higher than $5\cdot10^{4}$ GeV,
the spectra do not seem to change, within the
statistical uncertainties, with respect to that corresponding to
$5\cdot10^{4}$ GeV.

The $n_1$ parameter scales with the WIMP mass
as a simple power law  for $M<5\cdot10^{4}$ GeV.
For the other mass dependent $p$ parameter, the
power-law behavior is valid in two separated intervals
with an inflection point in the behavior
at $M=1000$ GeV. These results can be seen in Table
\ref{table_tau_fitting}.

Some of these results are presented graphically in Figure 5, Appendix
$\bf{C}$ for four WIMP masses: 25, 100, 1000 and $5\cdot10^4$ GeV.
Also in this Appendix, mass dependent parameters $n_1$ and $p$ are presented in Figure 6.

For this particle, it is worth mentioning the
increasing contribution of the logarithmic term in
\eqref{general_formula}  as the WIMP mass increases. This fact can be seen in
the presented plots from $x=0.5$ onwards. As a consequence,
the values of $p$ parameter increase as the WIMP masses increase.

\begin{widetext}
\begin{table}
\centering \small{
\begin{tabular}{||c|c|c||}
%
\hline WIMP mass (\text{GeV}) & $n_{1}$ & $p$  \\
\hline
25              & 10.1   & 221      \\
50              & 10.0   & 767      \\
100             & 9.91   & 2520     \\
200             & 9.80   & 8660     \\
500             & 9.67   & $4.01\cdot10^4$   \\
1000            & 9.57   & $1.35\cdot10^5$   \\
$10^{4}$        & 9.25   & $4.80\cdot10^6$   \\
$5\cdot10^{4}$  & 9.14   & $5.44\cdot10^{7}$ \\
\hline
\end{tabular}
\\
\begin{tabular}{|c|}
\hline
$a_1=14.7$ ; $b_1=5.40$ ; $b_2=5.31$ ; $n_2=1.40$ ; \\
$c_1=2.54$ ; $d_1=0.295$ ; $c_2=0.373$ ; $d_2=0.470$ ; $q=0.00260$\\
\hline
\end{tabular}
} \caption{\footnotesize{\textbf{$\tau$ lepton}:
$n_1$ and $p$ parameters corresponding to  \eqref{general_formula}
in the $\tau^{+}\tau^{-}$ channel for different WIMP masses. Mass independent
parameters in \eqref{general_formula} for this channel
are presented at the bottom of the table.
}}
\label{table_tau}
\end{table}
 \begin{table}
\centering \small{
\begin{tabular}{||c|c|c|c||}
\hline Parameter & WIMP mass interval (GeV) & Fitting power law(s) \\
\hline
$n_{1}$ & $25\leq M<10^{4}$                        & $10.6 M^{-0.0148}$  \\
               & $10^{4}\leq M\leq 5\cdot 10^{4}$  &  $-7.00 M^{-1.99}+179 M^{-0.763} + 9.09$  \\

\hline
$p$ & $25\leq M<1000$                          &  $0.773 M^{1.75}$ \\
        & $1000\leq M \leq 5\cdot 10^{4}$ &   $3.07 M^{1.55}$   \\
\hline
\end{tabular}
} \caption{\footnotesize{Parameters for expression
\eqref{linear_fit} for $\tau$ lepton. It can be seen that $n_1$ and $p$
parameters follow power-law behaviors.}}
\label{table_tau_fitting}
\end{table}
\end{widetext}
\subsubsection{$\mu$  lepton}
%
%
For the $\mu$ particle and according to
expression \eqref{general_formula_mu}, there are only three mass
dependent parameters: $q$, $p$ and $l$. These values are presented
in Table \ref{table_mu}. In this case, the considered range for
WIMP masses is from $25$ to $5\cdot10^{4}$ GeV.

The scaling of the $p$ parameter with the WIMP mass
shows two well differentiated regimes, with different asymptotic power laws:
one from $M=25$ GeV  to $M=100$ GeV, and another from
$M=750$ GeV to $M=5\cdot10^4$ GeV.
On the other hand, $q$ and $l$ parameters
present a sum of two power laws evolution
in the whole studied WIMP mass range.

As for the $\tau$ lepton, the flux of photons increases
as the WIMP mass increases. In this case, the $q$ parameter
increases as the WIMP masses do so, instead of the $p$ parameter
as was the case for the $\tau$.

%
\begin{widetext}
\begin{table}
\centering \small{
\begin{tabular}{||c|c|c|c||}
%
\hline WIMP mass (\text{GeV}) & $p$ & $q$ & $l$  \\
\hline
25                & 9510                               & $3.37\cdot10^{-3}$ & 0.787 \\
50                & 23600                             & $3.40\cdot10^{-3}$ & 0.642 \\
100     &   54600                                    & $3.45\cdot10^{-3}$ & 0.579 \\
200     &   $1.12\cdot10^5$                 & $3.50\cdot10^{-3}$ & 0.548 \\
500     &   $2.54\cdot10^5$                 & $3.61\cdot10^{-3}$ & 0.523 \\
1000    &   $4.13\cdot10^5$                & $3.70\cdot10^{-3}$ & 0.511 \\
5000    &   $1.18\cdot10^6$                & $3.91\cdot10^{-3}$ & 0.484 \\
$10^4$   &  $1.84\cdot10^6$              & $4.00\cdot10^{-3}$ & 0.474 \\
$5\cdot10^4$ & $5.29\cdot10^6$      & $4.23\cdot10^{-3}$ & 0.454 \\
\hline
\end{tabular}
} \caption{\footnotesize{\textbf{$\mu$ lepton}:
Parameters corresponding to  \eqref{general_formula_mu}
for different WIMP masses. All parameters in expression \eqref{general_formula_mu} are WIMP mass dependent.}}
\label{table_mu}
\end{table}
\end{widetext}
\begin{widetext}
\begin{table}
\centering \small{
\begin{tabular}{||c|c|c||}
\hline Parameter & WIMP mass interval (GeV) & Fitting power law(s)\\
\hline
$p$ & $25\leq M\leq 100$ & $176 M^{1.25}$  \\
    & $750\leq M\leq 5\cdot10^4$ & $4530 M^{0.653}$ \\

\hline
$q$ & $25\leq M<5\cdot10^{4}$ & $0.00230 M^{-0.911} + 0.00291 M^{0.0348}$  \\
\hline
$l$ & $25\leq M<5\cdot10^{4}$ & $0.626  M^{-0.0300} + 16.4  M^{-1.34}$  \\
\hline
\end{tabular}
} \caption{\footnotesize{Parameters for expression
\eqref{linear_fit} for $\mu$ lepton. It can be seen that $p$
parameter follows simple power laws in two different WIMP
mass regimes. Nevertheless, for $q$ and $l$ parameters a sum of two power laws accounts
for the WIMP mass dependence
in the whole studied WIMP mass interval.}}
\label{table_mu_fitting}
\end{table}
\end{widetext}
\subsubsection{$u$ quark}
%

The mass independent parameters
are $a_1=5.58$, $b_2=5.50$, $c_1=0.315$, $c_2=0.0$
(therefore $d_2$ is irrelevant)
and $q=9.30\cdot10^{-4}$. The mass dependent parameters
are  $b_1$, $n_1$, $n_2$, $d_1$ and $p$. These results are presented
in Table \ref{table_quark_u}. The analyzed mass range for this
quark is from 50 to 8000 $\text{GeV}$.

The spectra of the two highest studied masses (5000 and 8000 GeV)
clearly differ in the low energy interval. Therefore no
conclusion can be made about the existence of an asymptotic high masses
limit in the spectral shapes.
Concerning the mass evolution of the parameters for this quark,
we observe simple power-law behaviors
for both $b_1$ and $p$ parameters
in the whole studied WIMP mass range interval.
On the other hand, $n_1$, $n_2$ and $d_1$ parameters
are fitted by a sum of two power laws in the  studied range.
These resuls can be seen in Table \ref{table_quark_u_fitting}.
The chosen values for the parameters
turn out to fit the  spectra
from $x=5\cdot10^{-4}$ till the end of the
allowed energy interval. Nevertheless,
for some masses, the fit also applies  for
lower energies, i.e. lower $x$ values, up to $10^{-4}$.

\begin{table}
\centering \small{
\begin{tabular}{||c|c|c|c|c|c|c||}
%
\hline WIMP mass (\text{GeV}) & $b_{1}$ & $n_{1}$ & $n_{2}$ & $d_1$ & $p$ \\
\hline
%
50   & 3.60  & 2.77  & 0.585    & 0.383 &  129  \\
100  & 3.75  & 2.64  & 0.551   & 0.355 &  225  \\
200  & 3.88  & 2.54  & 0.521   & 0.332 &  409  \\
500  & 4.04  & 2.44  & 0.490   & 0.308 &  856  \\
1000 & 4.18  & 2.40  & 0.472  & 0.293 & 1540 \\
2000 & 4.34  & 2.39  & 0.463  & 0.281 & 2800 \\
5000 & 4.55  & 2.38  & 0.450  & 0.266 & 6000 \\
8000  & 4.67 & 2.34 &  0.448  & 0.259 & 8900\\
\hline
\end{tabular}
\\
\begin{tabular}{|c|}
\hline
$a_1=5.58$ ; $b_2=5.50$ ; $c_1=0.315$ ; $c_2=0.0$ ; $q=9.30\cdot10^{-4}$.\\
\hline
\end{tabular}
} \caption{\footnotesize{\textbf{$u$ quark}:
$b_{1}$, $n_{1}$, $n_{2}$, $d_1$ and $p$ parameters corresponding to
expression \eqref{general_formula} when applied to $u\bar{u}$ channel
for different WIMP masses. Mass independent parameters in
\eqref{general_formula} for this channel
are presented at the bottom of the table.
}}
\label{table_quark_u}
\end{table}
\begin{table}
\centering \small{
\begin{tabular}{||c|c|c||}
\hline Parameter & WIMP mass interval (GeV) & Fitting power law(s)\\
\hline
$b_{1}$ & 50 $\leq M \leq 8000$ & $2.96 M^{0.0506}$ \\
\hline
$n_{1}$ & $50\leq M\leq 8000$ & $2.91 M^{-0.351} + 1.90 M^{0.0172}$\\
\hline
$n_{2}$ & $50\leq M\leq 8000$ & $0.0587 M^{0.146} + 0.848 M^{-0.145}$\\
\hline
$d_{1}$ & $50 \leq M\leq 8000$ & $0.317 M^{-0.0300} + 0.403 M^{-0.351}$\\
\hline
$p$ & $ 50 \leq M \leq 8000$ & $4.74 M^{0.839}$ \\
\hline
\end{tabular}
} \caption{\footnotesize{Parameters corresponding to \eqref{linear_fit} for $u$ quark.
$b_1$ and $p$ parameters follow a simple power-law behavior in the whole studied WIMP mass interval. $n_1$, $n_2$ and $d_1$ parameters follow a sum of two power laws in the whole mass interval.
}}
\label{table_quark_u_fitting}
\end{table}
\subsubsection{$d$ quark}
For this channel there are five mass-dependent
parameters: $b_1$, $n_1$ $n_2$, $c_1$ and $p$. The
 mass independent parameters are $a_1=5.20$, $b_2=5.10$,
$d_1=0.410$, $c_2=0.0260$, $d_2=0.570$ and $q=1.40\cdot10^{-4}$. All
parameters in this case are presented in Table \ref{table_quark_d}.
The mass range studied for this quark was from 50 to 5000 $\text{GeV}$.

In this channel, no conclusion can be drawn
about the existence of an asymptotic high mass limit in
the spectral shape.
The chosen parameters provide good fits
 from $x=2\cdot10^{-3}$ for $M=50\,\text{GeV}$
whereas for the rest of masses the fits work
very well till $x=5\cdot10^{-4}$.
%

%


 The scaling of  $b_1$ with $M$ is given by a simple
power-law in the whole mass interval,
whereas $n_1$, $n_2$ and $c_1$ follow a sum of two power-law behavior
in the whole studied mass. Finally, the $p$ parameter presents a power-law
behavior for $M> 50$ GeV. These results can be seen
in Table \ref{table_quark_d_fitting}.

\begin{table}
\centering \small{
\begin{tabular}{||c|c|c|c|c|c|c||}
%
\hline WIMP mass (\text{GeV}) & $b_{1}$ & $n_{1}$ & $n_2$ & $c_1$  & $p$\\
\hline
50    & 4.09  & 2.69  & 0.561  & 0.327  & 17.7 \\
100   & 4.24  & 2.47  & 0.522  & 0.293  & 66.2 \\
200   & 4.39  & 2.34  & 0.480  & 0.258  & 166  \\
500   & 4.56  & 2.28  & 0.448  & 0.228  & 483  \\
1000  & 4.75  & 2.25  & 0.426  & 0.212  & 1270 \\
2000  & 4.91  & 2.24  & 0.409  & 0.200  & 3130 \\
5000  & 5.11  & 2.23  & 0.394  & 0.187  & 10200\\
\hline
\end{tabular}
\\
\begin{tabular}{|c|}
\hline
$a_1=5.20$ ; $b_2=5.10$ ; $d_1=0.410$ ; $c_2=0.0260$ ; $d_2=0.570$ ; $q=1.40\cdot10^{-4}$\\
\hline
\end{tabular}
} \caption{\footnotesize{\textbf{$d$ quark}:
$b_{1}$, $n_{1}$, $n_2$, $c_1$ and $p$ parameters corresponding to
expression \eqref{general_formula} when applied to $d\bar{d}$ channel
for different WIMP masses. Mass independent parameters in \eqref{general_formula} for this channel
are presented at the bottom of the table.
}}
\label{table_quark_d}
\end{table}
\begin{widetext}
 \begin{table}
\centering \small{
\begin{tabular}{||c|c|c||}
\hline Parameter & WIMP mass interval (GeV) & Fitting power law(s)\\
\hline
$b_{1}$ & $50\leq M \leq5000$ & $3.39 M^{0.0485}$ \\
\hline
$n_{1}$ & $50 \leq M\leq5000$ & $21.8 M^{-0.993} + 2.25 M^{-0.00113}$  \\
\hline
$n_{2}$ & $50\leq M\leq 5000$ & $0.848 M^{-0219} + 0.161 M^{0.0573}$ \\
\hline
$c_{1}$ & $50\leq M \leq 5000$ & $0.722 M^{-0.270} + 0.0544 M^{0.0874}$ \\
\hline
$p$ & $50<M\leq5000$ & $0.168 M^{1.29}$ \\
\hline
\end{tabular}
} \caption{\footnotesize{Parameters corresponding to
\eqref{linear_fit} for $d$ quark.  As can be seen, $b_1$ parameter follows a simple power law in the whole accessible mass interval.
$n_1$, $n_2$ and $c_1$ parameters follow a sum of two power laws in the whole accessible mass interval. Finally, $p$ parameter follows a power law for $M>50$ GeV.
}}
\label{table_quark_d_fitting}
\end{table}
\end{widetext}
\subsubsection{$s$ quark}
%

For the $s$ quark,  there are just four mass dependent
parameters $b_1$, $n_2$, $d_1$ and $p$. The mass  independent parameters for this particle
in \eqref{general_formula} are $a_1=4.83$, $n_1=2.03$,
$b_2=6.50$, $c_1=0.335$, $c_2=0.0$ ($d_2$ is irrelevant as
for the $u$ quark) and  $q=2.40\cdot10^{-4}$.
All these parameters are detailed in Table \ref{table_quark_s}.
The studied mass range for this quark is between 50 and 7000 $\text{GeV}$.

As in the $d$ quark case, no conclusion can be drawn about the existence of an
asymptotic high mass limit in the spectral shape.
The scaling with $M$ of the parameters for this quark
is a simple power law for $b_1$ parameter for masses higher than 1000 GeV, the sum of two power laws for $n_2$ and $d_1$ parameters  in the whole studied WIMP mass, and two power laws for $p$ parameter: one for masses smaller than 1000 GeV and another for masses higher than 1000 GeV.
These results are shown in Table \ref{table_quark_s_fitting}.
\begin{widetext}
\begin{table}
\centering \small{
\begin{tabular}{|c|c|c|c|c|c|c|}
%
\hline WIMP mass (\text{GeV}) & $b_{1}$ & $n_{2}$ & $d_1$ & $p$ \\
\hline
50     & 4.78  & 0.719   & 0.367   & 186 \\
100    & 5.31  & 0.669   & 0.332   & 409 \\
200    & 5.43  & 0.648   & 0.315   & 605 \\
500    & 5.60  & 0.612   & 0.290   & 1180\\
1000   & 5.73  & 0.592   & 0.276   & 1980\\
2000   & 5.87  & 0.575   & 0.263   & 3320\\
5000   & 6.07  & 0.557   & 0.249   & 6500\\
7000   & 6.12  & 0.548   & 0.244   & 7570\\
%
\hline
\end{tabular}
\\
\begin{tabular}{|c|}
\hline
$a_1=4.83$ ; $n_1=2.03$ ; $b_2=6.50$ ; $c_1=0.335$ ; $c_2=0.0$ ; $q=2.40\cdot10^{-4}$\\
\hline
\end{tabular}
} \caption{\footnotesize{\textbf{$s$ quark}:
$b_{1}$, $n_{2}$, $d_1$ and $p$ parameters corresponding to
expression \eqref{general_formula} when applied to $s\bar{s}$ channel
for different WIMP masses. Mass independent parameters in
\eqref{general_formula} for this channel
are presented at the bottom of the table.
}}
\label{table_quark_s}
\end{table}
\end{widetext}
\begin{widetext}
 \begin{table}
\centering \small{
\begin{tabular}{||c|c|c||}
\hline Parameter & WIMP mass interval (GeV) & Fitting power law(s)\\
\hline
$b_{1}$ & $M>1000$  & $4.54M^{0.0339}$ \\
\hline
$n_{2}$ & $50\leq M \leq 7000$ & $3.68 M^{-1.01} + 0.744 M^{-0.0352}$ \\
\hline
$d_{1}$ & $50 \leq M \leq 7000$ & $0.621 M^{-0.674} +0.414 M^{-0.0588}$  \\
\hline
$p$ & $50\leq M \leq 100$ & $4.01 M^{0.981}$\\
         & $100<M\leq 7000$ &  $12.8 M^{0.732}$\\
\hline
\end{tabular}
} \caption{\footnotesize{Parameters corresponding
to  \eqref{linear_fit} for $s$ quark. As can be seen, $b_1$ parameter
follows a simple power-law behavior for masses higher than 1000 GeV. $n_2$ and $d_1$ parameters follow the sum
of two power laws for the whole studied WIMP mass interval. Finally, $p$ parameter presents two power laws: one for
masses smaller than 1000 GeV and another for masses higher than 1000 GeV.
}}
\label{table_quark_s_fitting}
\end{table}
\end{widetext}
\subsubsection{$c$ quark}
%

As for the $d$ quark, there are five
mass dependent parameters. In this case
$b_1$, $n_1$, $c_1$, $d_1$ and $p$ which
are presented in Table \ref{table_quark_c}.
The mass independent parameters are
$a_1=5.58$, $b_2=7.9$, $n_2=0.686$, $c_2=0.0$
(therefore $d_2$ is irrelevant) and $q=9.00\cdot 10^{-4}$.

Likewise the $u$ quark, the studied mass range was
from 50 to 8000 GeV and  again no conclusion can
be made about the existence of an asymptotic high mass
limit for the spectral shape.
Higher masses simulations would be thus required also in this case.

The scaling  of $b_1$ and $n_1$ with $M$ shows a simple
power-law  behavior
in the considered range. For $c_1$ and $p$, the single
power-law evolution is only valid for masses above 200 GeV. Finally,
 $d_1$ parameter follows a sum of two power laws in the studied mass range.
These results are shown in Table \ref{table_quark_c_fitting}.
%
\begin{widetext}
\begin{table}
\centering \small{
\begin{tabular}{||c|c|c|c|c|c|c||}
%
\hline WIMP mass (\text{GeV}) & $b_{1}$ & $n_{1}$ & $c_1$ & $d_1$ & $p$ \\
\hline
50   & 5.93 & 2.35 &  0.239 & 0.428 & 210   \\
100  & 5.48 & 2.08 &  0.283 & 0.374 & 379   \\
200  & 4.98 & 1.86 &  0.330 & 0.330 & 673   \\
500  & 4.50 & 1.65 &  0.378 & 0.288 & 1230  \\
1000 & 4.00 & 1.50 &  0.406 & 0.264 & 2110  \\
2000 & 3.70 & 1.35 &  0.432 & 0.245 & 4050  \\
5000 & 3.27 & 1.17 &  0.470 & 0.221 & 8080  \\
8000 & 3.08 & 1.11 &  0.494 & 0.208 & 12000 \\
\hline
\end{tabular}
\\
\begin{tabular}{|c|}
\hline
$a_1=5.58$ ; $b_2=7.90$ ; $n_2=0.686$ ; $c_2=0.0$ ;  $q=9.00\cdot10^{-4}$\\
\hline
\end{tabular}
} \caption{\footnotesize{\textbf{$c$ quark}:
$b_{1}$, $n_{1}$, $c_1$, $d_1$ and $p$ parameters corresponding to
expression \eqref{general_formula} in the $c\bar{c}$ channel
for different WIMP masses.  Mass independent parameters in \eqref{general_formula}
for this channel
are presented at the bottom of the table.
}}
\label{table_quark_c}
\end{table}
\end{widetext}
\begin{widetext}
 \begin{table}
\centering \small{
\begin{tabular}{||c|c|c||}
\hline Parameter & WIMP mass interval (GeV) & Fitting power law(s)\\
\hline
$b_{1}$ & $50 \leq M \leq 8000$ & $9.90 M^{-0.130}$ \\
\hline
$n_{1}$ & $50\leq M \leq 8000$ & $4.14 M^{-0.148}$  \\
\hline
$c_{1}$  & $500\leq M \leq 8000$ & $0.210 M^{0.0951}$ \\
\hline
$d_{1}$ & $50 \leq M\leq 8000$ & $1.50 M^{-0.632} + 0.479 M^{-0.0942}$ \\
\hline
$p$ & $200 <M\leq 8000$ & $8.11 M^{0.812}$ \\
\hline
\end{tabular}
} \caption{\footnotesize{Parameters corresponding to
\eqref{linear_fit} for $c$ quark. It can be seen that the
mass dependent parameters follow
a power-law  behavior for intermediate and high WIMP masses.
In particular the $d_1$ parameter follows a sum of two power-law behavior in he whole accessible WIMP mass range.
}}
\label{table_quark_c_fitting}
\end{table}
\end{widetext}
\subsubsection{$b$ quark}

%
%

For the $b$ quark, the required gamma rays spectra
parametrization is the one given by expression \eqref{general_formula}.
For this particle, the mass independent parameters
are $a_1=10.0$, $b_2=11.0$, $c_2=0.0151$, $d_2=0.550$,
$q=2.60\cdot10^{-4}$. The mass dependent parameters are
$b_1$ $n_1$, $n_2$, $c_1$, $d_1$ and $p$.
Their values are presented in Table \ref{table_quark_b}.

The studied mass range is from 50 to 8000 GeV.
Unlike previous particles for
which the spectra did not change remarkably for very high masses,
in the present case
no conclusion can be drawn about the existence of an
asymptotic high mass limit.

Concerning the scaling behavior of the parameters
for this quark, we observe that the behavior depends
both on the WIMP mass and on the considered parameter.
Thus  $b_1$ and $n_1$ no longer scale with a single power-law  for
$M$ higher than 100 GeV. For $n_2$, two simple power laws
can be seen, one from 50 to 1000 GeV (not included)
and a second one from 1000 (included) to 8000 GeV.
$c_1$ shows also a power-law behavior but only  up to 50 GeV.
Finally, both $d_1$ and $p$,
scale with simple power laws from 500  GeV up.
These results are summarized in Table \ref{table_quark_b_fitting}.

Some of these results are presented graphically in Figure 7, Appendix $\bf{D}$ for four
WIMP masses: 50, 200, 1000 and 5000 GeV . Also
in this Appendix, mass dependent parameters $b_1$, $n_1$, $n_2$, $c_1$, $d_1$ and $p$ are plotted in Figure 8.
%
\begin{widetext}
\begin{table}
\centering \small{
\begin{tabular}{||c|c|c|c|c|c|c||}
%
\hline WIMP mass (\text{GeV}) & $b_1$ & $n_1$ & $n_2$ & $c_1$ & $d_1$ & $p$ \\
\hline
50   &  19.5   & 6.48  &  0.710 & 0.365  & 0.393    &   57.8 \\
100  &  17.1   & 5.80  &  0.695 & 0.403  & 0.360    &   138  \\
200  &  13.1   & 5.01  &  0.680 & 0.415  & 0.340    &   281  \\
500  &  8.76   & 4.04  &  0.660 & 0.431  & 0.319    &   623  \\
1000 &  6.00   & 3.36  &  0.647 & 0.447  & 0.305    &   1030 \\
2000 &  4.60   & 2.85  &  0.640 & 0.460  & 0.294    &   1620 \\
5000 &  3.00   & 2.26  &  0.634 & 0.479  & 0.280    &   2670 \\
8000 &  2.35   & 2.00  &  0.629 & 0.490  & 0.274    &   3790 \\
\hline
\end{tabular}
\\
\begin{tabular}{|c|}
\hline
$a_1=10.0$ ; $b_2=11.0$ ; $c_2=0.0151$ ; $d_2=0.550$ ; $q=2.60\cdot10^{-4}$\\
\hline
\end{tabular}
}
\caption{\footnotesize{\textbf{$b$ quark}:
$b_1$, $n_1$, $n_2$, $c_1$, $d_1$ and $p$ parameters corresponding
to expression \eqref{general_formula} in the $b\bar{b}$ channel
for different WIMP masses. Mass independent parameters
in \eqref{general_formula} for this channel
are presented at the bottom of the table.
}}
\label{table_quark_b}
\end{table}
\end{widetext}
\begin{widetext}
 \begin{table}
\centering \small{
\begin{tabular}{||c|c|c||}
\hline Parameter & WIMP mass interval (GeV) & Fitting power law(s)\\
\hline
$b_{1}$ & $100 < M \leq8000$ & $152 M^{-0.462}$ \\
\hline
$n_{1}$ & $100 < M \leq8000$ & $18.7 M^{-0.248}$ \\
\hline
$n_{2}$ & $50 \leq M < 1000$      & $0.805 M^{-0.0319}$  \\
        & $1000\leq M  \leq8000$      & $0.707 M^{-0.0129}$ \\
\hline
$c_{1}$ & $50<M\leq8000$ & $0.328 M^{ 0.0447}$ \\
\hline

$d_{1}$ & $50 \leq M < 600$        & $0.474 M^{-0.0639} + 37.1 M^{-1.87}$ \\
               & $600 \leq M \leq8000$ & $0.449 M^{-0.0552}$  \\
\hline
$p$ & $200 < M \leq8000$ & $11.8 M^{0.641}$ \\
\hline
\end{tabular}
} \caption{\footnotesize{Parameters corresponding to
\eqref{linear_fit} for the $b$ quark. All mass dependent
parameters for the $b$ quark
follow simple power-law scalings  at intermediate and high energies.  Only
$n_2$ and $d_1$ parameters follow power-law behaviors at low WIMP masses.
}}
\label{table_quark_b_fitting}
\end{table}
\end{widetext}
\newpage
\section{Numerical codes}

All the calculations performed in this investigation are available at the website\\
$\text{http://teorica.fis.ucm.es/}$$\sim$$\text{PaginaWeb/downloads.html}$

At this site, we provide the $\tt Mathematica$~\cite{mathematica}
files that contain the fitting expressions \eqref{general_formula},
\eqref{general_formula_W_Z} and \eqref{general_formula_t} for
$x^{1.5}\text{d}N_{\gamma}/\text{d}x$ presented in this paper when
applied for each studied channel. Let us remind that these
parametrizations are valid in the corresponding WIMP masses
intervals mentioned in the corresponding sections. Also in these
files, the fitting formulae for mass dependent parameters in each
channel are presented.

%
%

\section{Conclusions}

In this work, we have extensively studied the photon spectra coming from
WIMP pair annihilation into SM particle particle-antiparticle
pairs for all the
phenomenologically relevant channels. The covered WIMP mass range
has been optimized for each particular
channel taking into account mass thresholds, statistics, and
saturation of the Monte Carlo simulation. For instance, for light
quarks it was from 50 GeV to 8000 GeV, for leptons it was from 25
GeV to 50 TeV, for gauge bosons from 100 GeV to 1000 GeV and for $t$
quark from 200 to 1000 GeV. All simulated spectra covered the whole
accessible energy interval, from extremely low energetic photons till
photons with one half of the available total center of mass energy.

Once the spectra were obtained, analytical expressions were proposed
to fit the simulation data. Three different fitting functions appeared to
be valid depending on the studied channel: one for light quarks and
leptons, another for gauge bosons and finally one for $t$ quark
very similar to the latter. Those
expressions depended on  either WIMP mass dependent or independent
parameters. For WIMP mass independent parameters, their
values did nevertheless depend on the considered annihilation
channel whereas for
WIMP mass dependent ones, their evolutions with WIMP mass were
parametrized from the obtained values by continuous and smooth
curves.

In addition to a better understanding of the different channels for
photon production from DM annihilation, the use of these
fitting functions  found in these analyses can save an important
amount of computing time and resources: Monte Carlo simulations
do not need to be repeated each time that
a particular photon spectrum needs to be known for a given
channel and center of mass energy. This fact is
particularly important for high energy photons, whose production rate is very suppressed
and would require large computation times and to store big amounts of data. Our research was thus
able to present very good statistics for those energies.


By having used extensive PYTHIA Monte Carlo simulation we have been able to obtain
relatively simple parametrizations of these spectra and fit the corresponding parameters.
As our analysis is model independent, it
could be useful, both for theoreticians and experimentalists,
interested in the indirect DM detection through gamma rays.
Given some theoretical model, and the corresponding
velocity averaged annihilation cross sections for the different channels,
our formulae make it possible to obtain the expected
photon spectrum for each particular theoretical model in a
relatively simple way. In this sense, further work is in progress
to extend our analysis to other stable particles like positron or
neutrinos but these results will be presented elsewhere.

%
%
\begin{acknowledgments}
This work has been supported in part by MICINN (Spain) project
numbers FIS 2008-01323 and FPA 2008-00592, CAM/UCM 910309, MEC grant
BES-2006-12059, DOE grant DE-FG02-94ER40823 and MICINN
Consolider-Ingenio MULTIDARK CSD2009-00064. We are particularly
grateful to Prof. Jonathan Feng for his continuous and encouraging
help during 2007 summer at UC Irvine. We would also like to thank
Dr. Mario Bondioli for his preliminary help with simulation software
and Dr. Mirco Cannoni, Prof. Mario E. Gomez and Prof. Mikhail
Voloshin for interesting discussions about different aspects of the
physical meaning of the spectra behavior. Dr. Abelardo Moralejo drew
our attention of some particular aspects of bremsstrahlung. Finally,
fruitful discussions were held with Daniel Nieto about detectors
technicalities and improvement of the numerical codes.

RL acknowledges financial support given by Ministero dell'Istruzione,
dell'Universit\`{a} e della Ricerca (MIUR), by the University of Torino (UniTO),
by the Istituto Nazionale di Fisica Nucleare (INFN) within the
Astroparticle Physics Project, and by the Italian Space Agency (ASI)
under contract Nro: I/088/06/0.

AdlCD wants to thank David Fern\'andez (UCM) for his technical
support with the performed simulations and Dr. Javier Almeida for his help to manage simulation results.
Scientific discussions about other available simulation softwares were held with Beatriz Ca\~{n}adas.
\end{acknowledgments}


\section{APPENDICES}
In this section we present simulations for some of the studied channels:
$W^{+}W^{-}$, $t\bar{t}$, $\tau^{+}\tau^{-}$ and $b\bar{b}$. For these
channels four simulated spectra are presented together with the proposed
fit formulae.
For each channel, evolution with WIMP mass of mass dependent parameters
have been plotted. The final appendix $\bf{E}$ shows the running with
the WIMP mass of the total number of photons per WIMP pair annihilation.
\vspace{2cm}
\begin{center}
\bf{A. Plots for $W$ gauge boson}
\end{center}

\begin{widetext}
\begin{figure}[!hbp]
en\subfigure[ \hspace{1ex} Photon spectrum for $M=100\,\text{GeV}$
for $W^+W^-$ channel. ]{
\begin{overpic}[width=8.520cm]{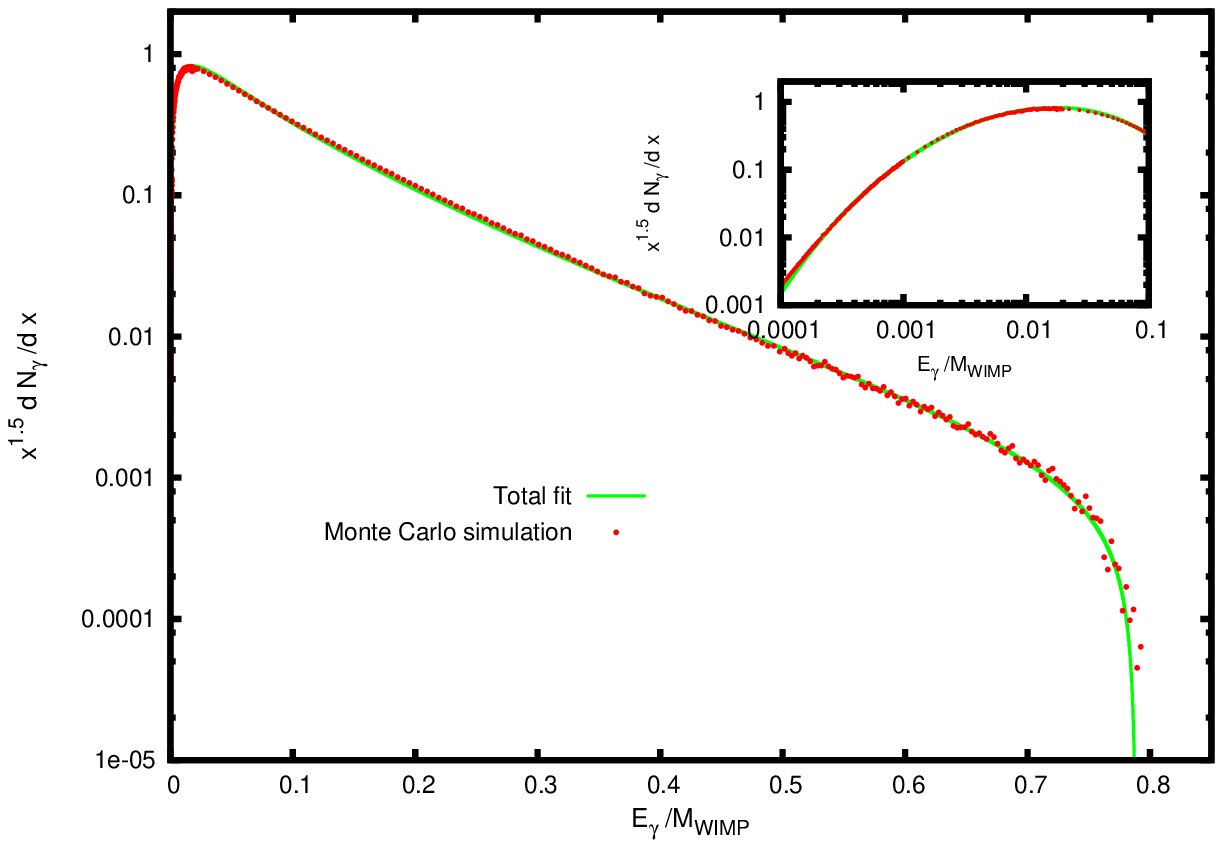}
\label{W100}
\end{overpic}
}
\subfigure[ \hspace{1ex} Photon spectrum for $M=200\,\text{GeV}$ for $W^+W^-$ channel.]{
\begin{overpic}[width=8.520cm]{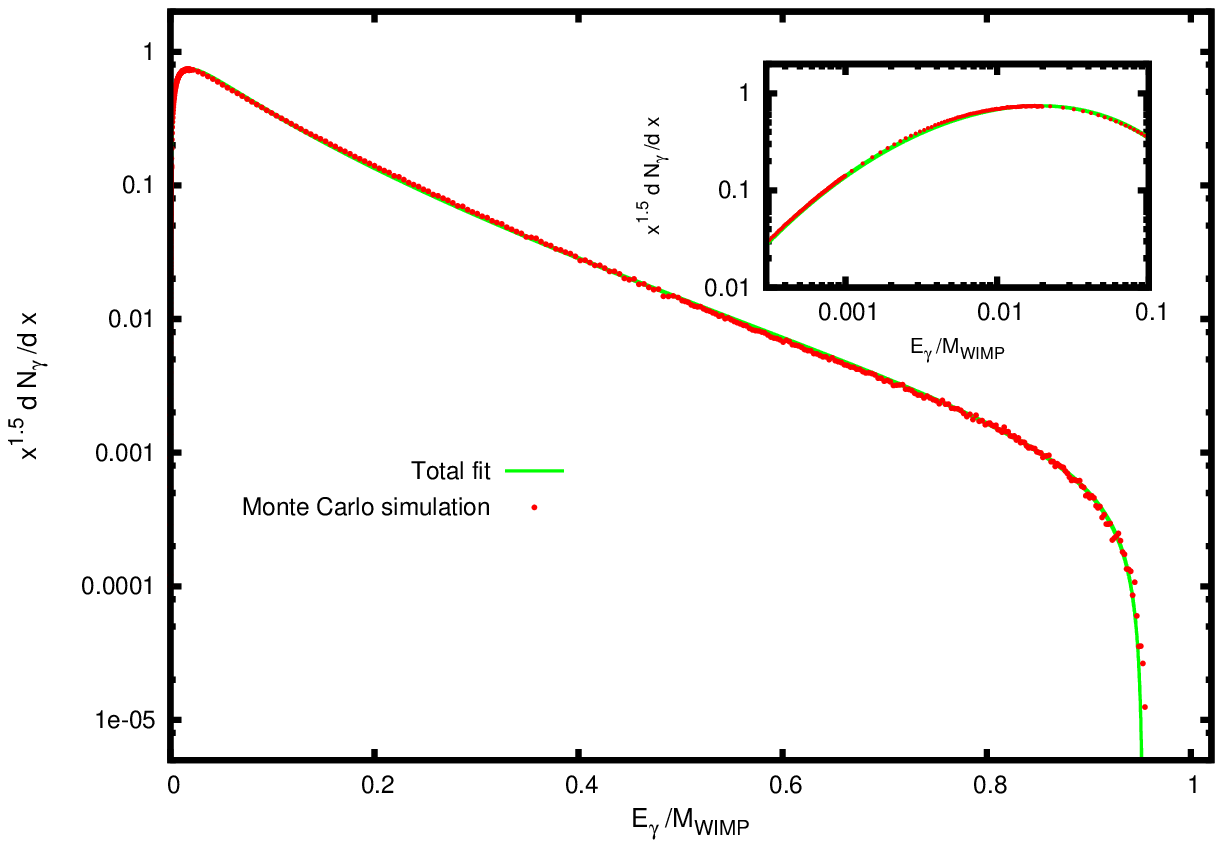}
\end{overpic}
}
\end{figure}
\vspace{-0.4cm}
\begin{figure}[!hbp]
\subfigure[ \hspace{1ex} Photon spectrum for $M=350\,\text{GeV}$ for $W^+W^-$ channel.]{
\begin{overpic}[width=8.520cm]{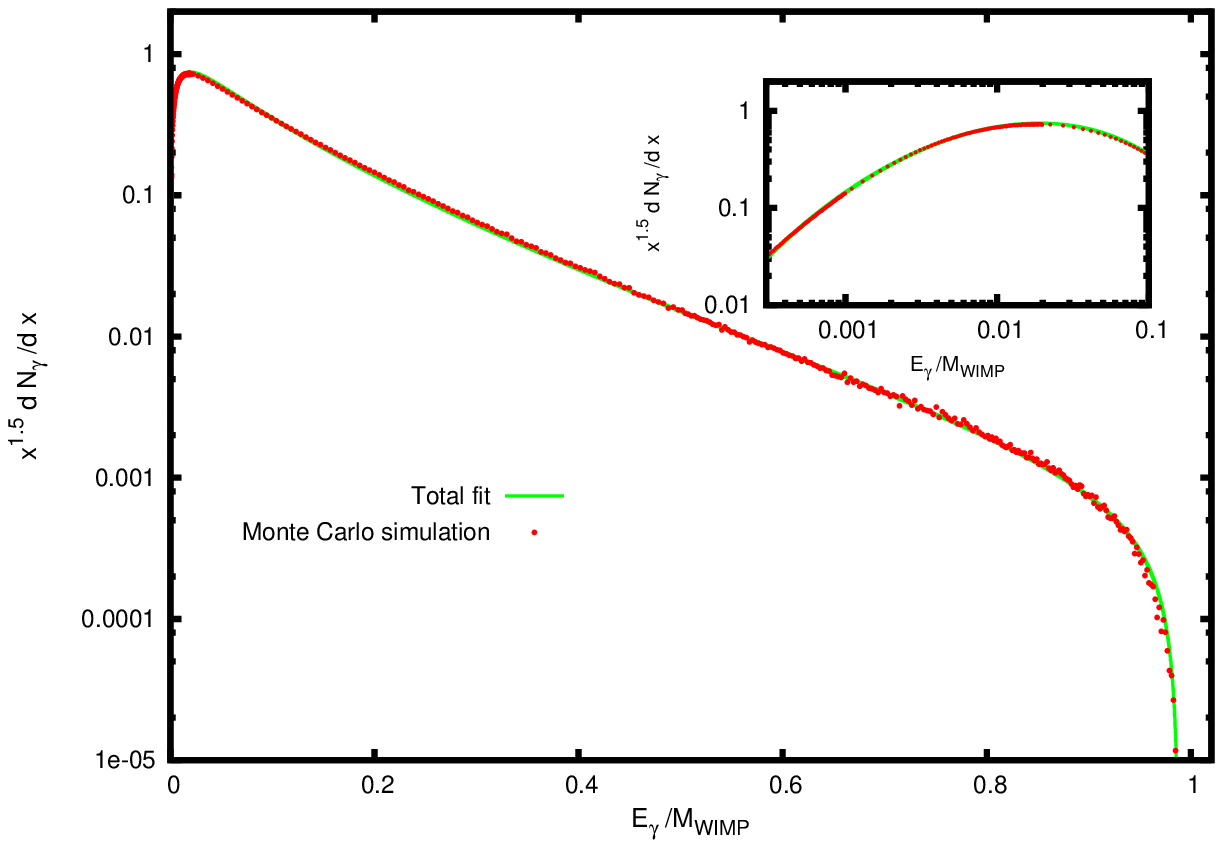}
\end{overpic}
}
\subfigure[ \hspace{1ex} Photon spectrum for $M=1000\,\text{GeV}$ for $W^+W^-$ channel.]{
\begin{overpic}[width=8.520cm]{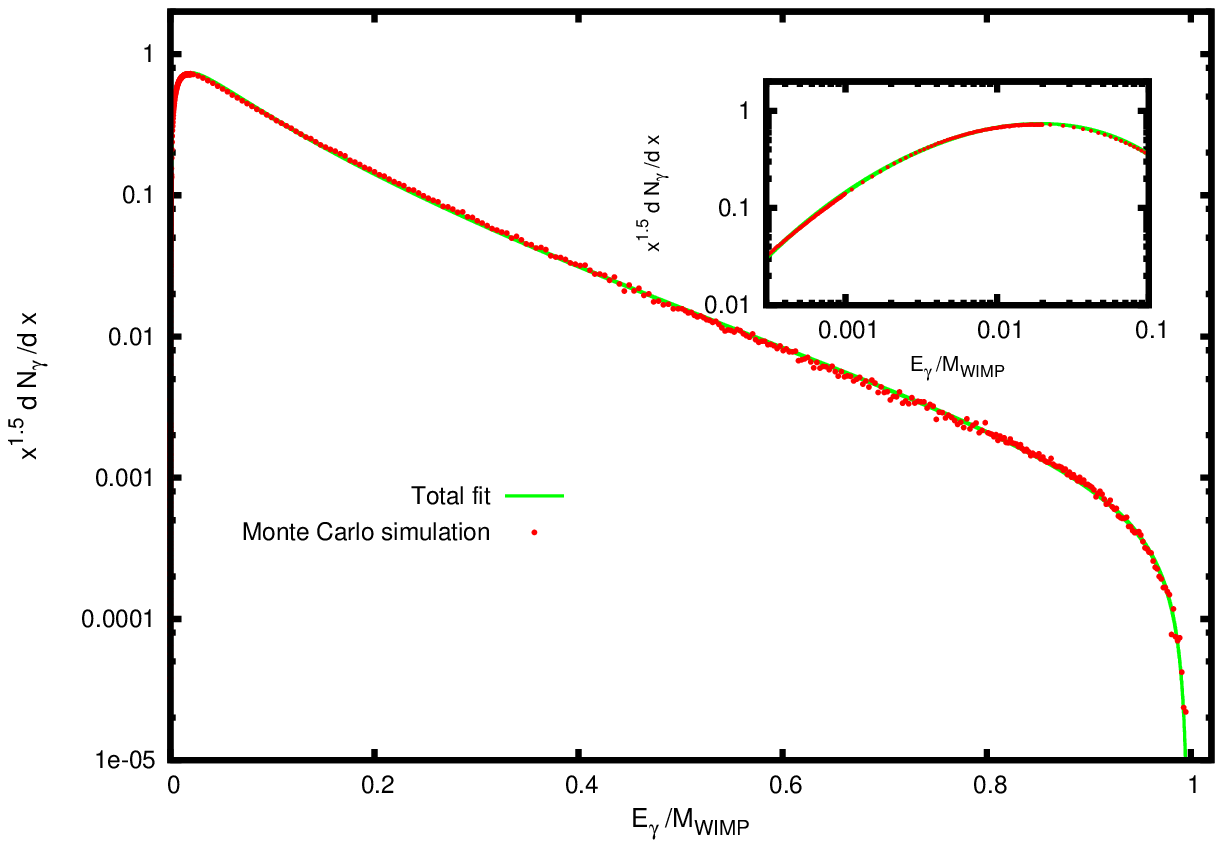}
\end{overpic}
}
\label{W_figures_1}
\caption{Photon spectra for four different WIMP masses (50, 200,
1000 and 5000 GeV) in the $W^+W^-$  channel. Red dotted points are PHYTIA simulations and solid lines correspond to the proposed fitting functions.}
\end{figure}
\newpage
\vspace{-0.4cm}
\begin{figure}[!hbp]
\subfigure[ \hspace{1ex} $b_1$ parameter of expression \eqref{general_formula} for $W^+W^-$ channel.]{
\begin{overpic}[width=8.520cm]{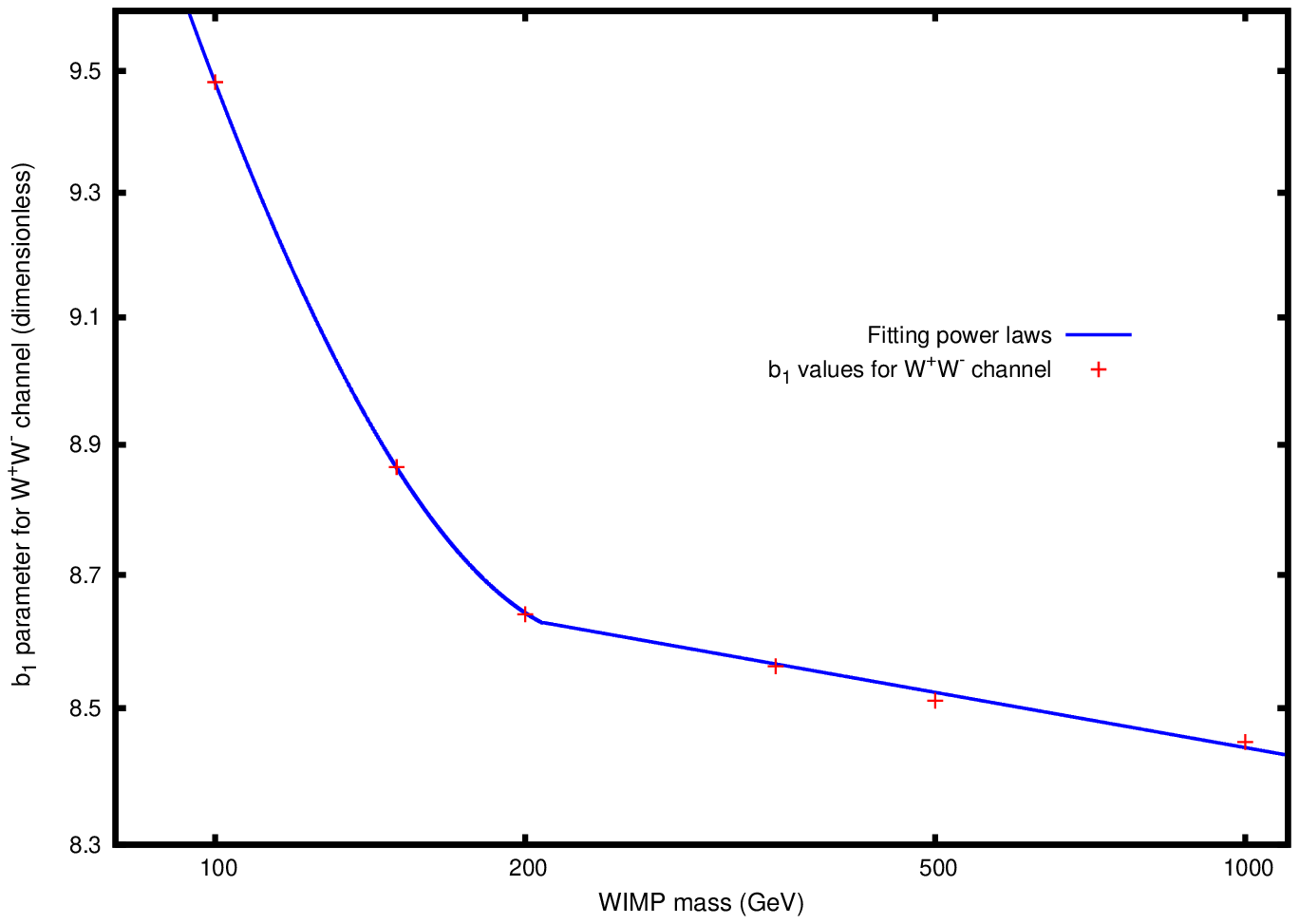}
\end{overpic}
}
\subfigure[ \hspace{1ex} $c_1$ parameter of expression \eqref{general_formula} for $W^+W^-$ channel.]{
\begin{overpic}[width=8.520cm]{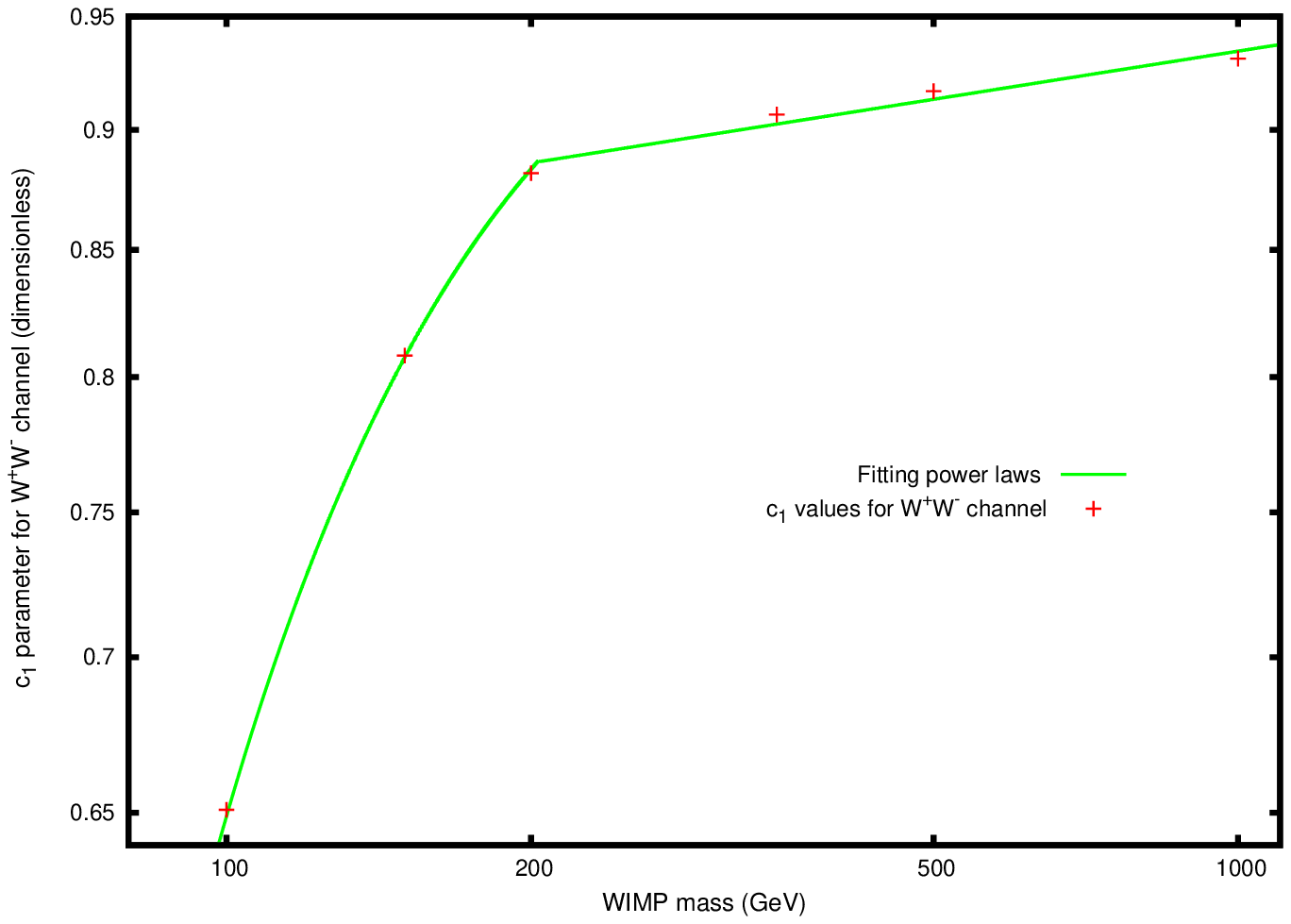}
\end{overpic}
}
\end{figure}
\vspace{-0.4cm}
\begin{figure}[!hbp]
\subfigure[ \hspace{1ex} $d_1$ parameter of expression \eqref{general_formula} for $W^+W^-$ channel.]{
\begin{overpic}[width=8.520cm]{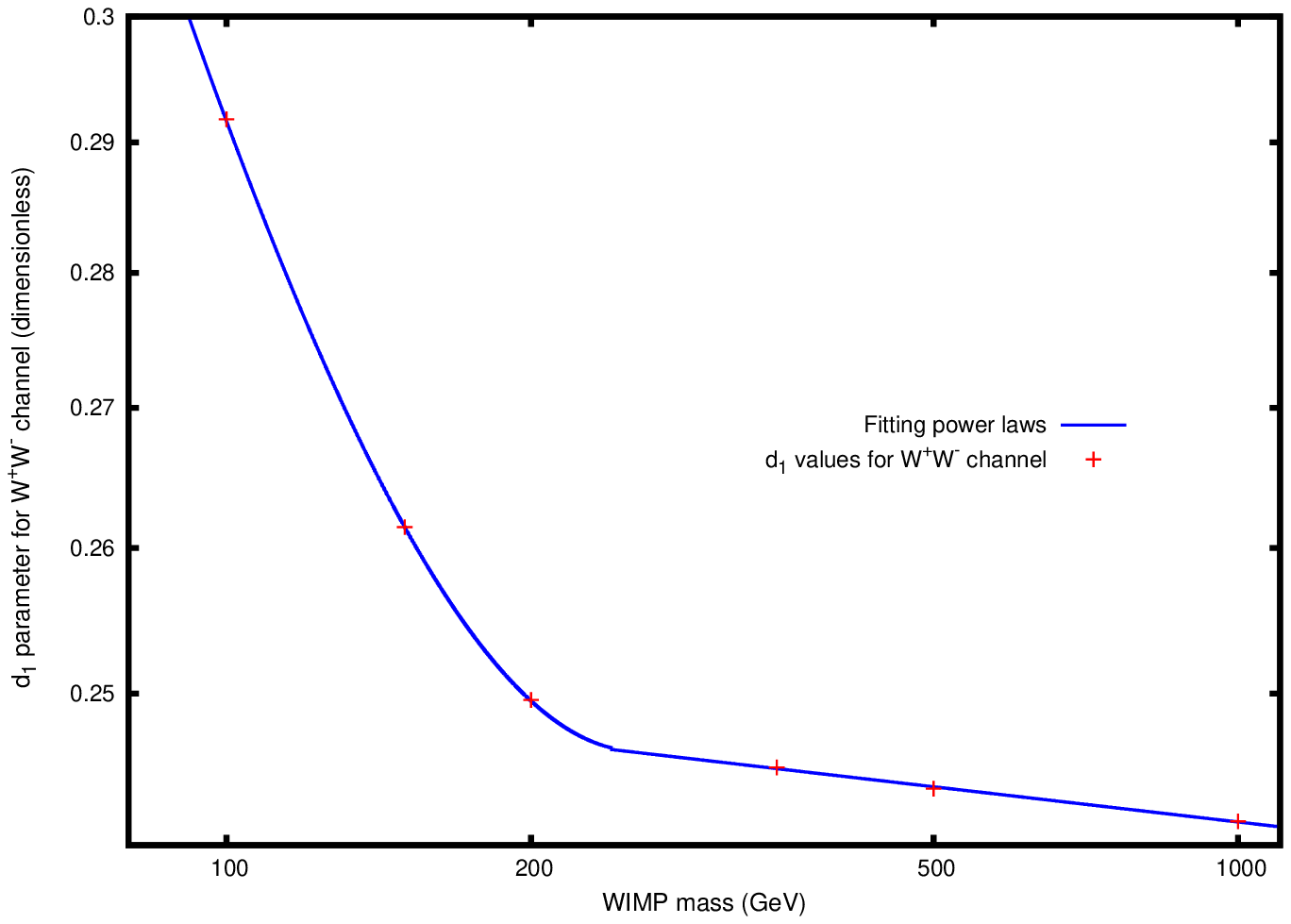}
\end{overpic}
}
\subfigure[ \hspace{1ex} $p$ parameter of expression \eqref{general_formula} for $W^+W^-$ channel.]{
\begin{overpic}[width=8.520cm]{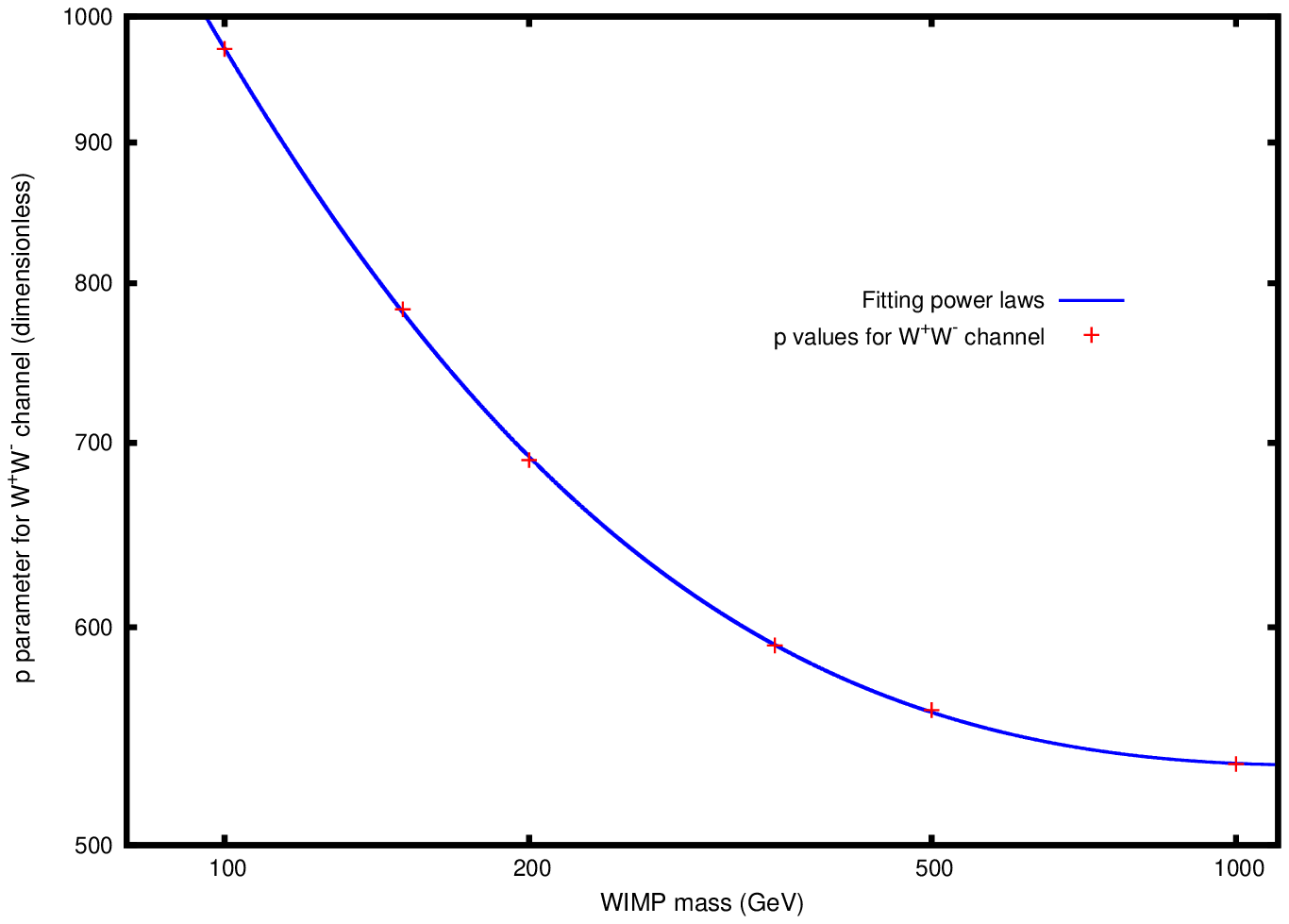}
\end{overpic}
}
\end{figure}
\vspace{-0.4cm}
\begin{figure}[!hbp]
\subfigure[ \hspace{1ex} $j$ parameter of expression \eqref{general_formula} for $W^+W^-$ channel.]{
\begin{overpic}[width=8.520cm]{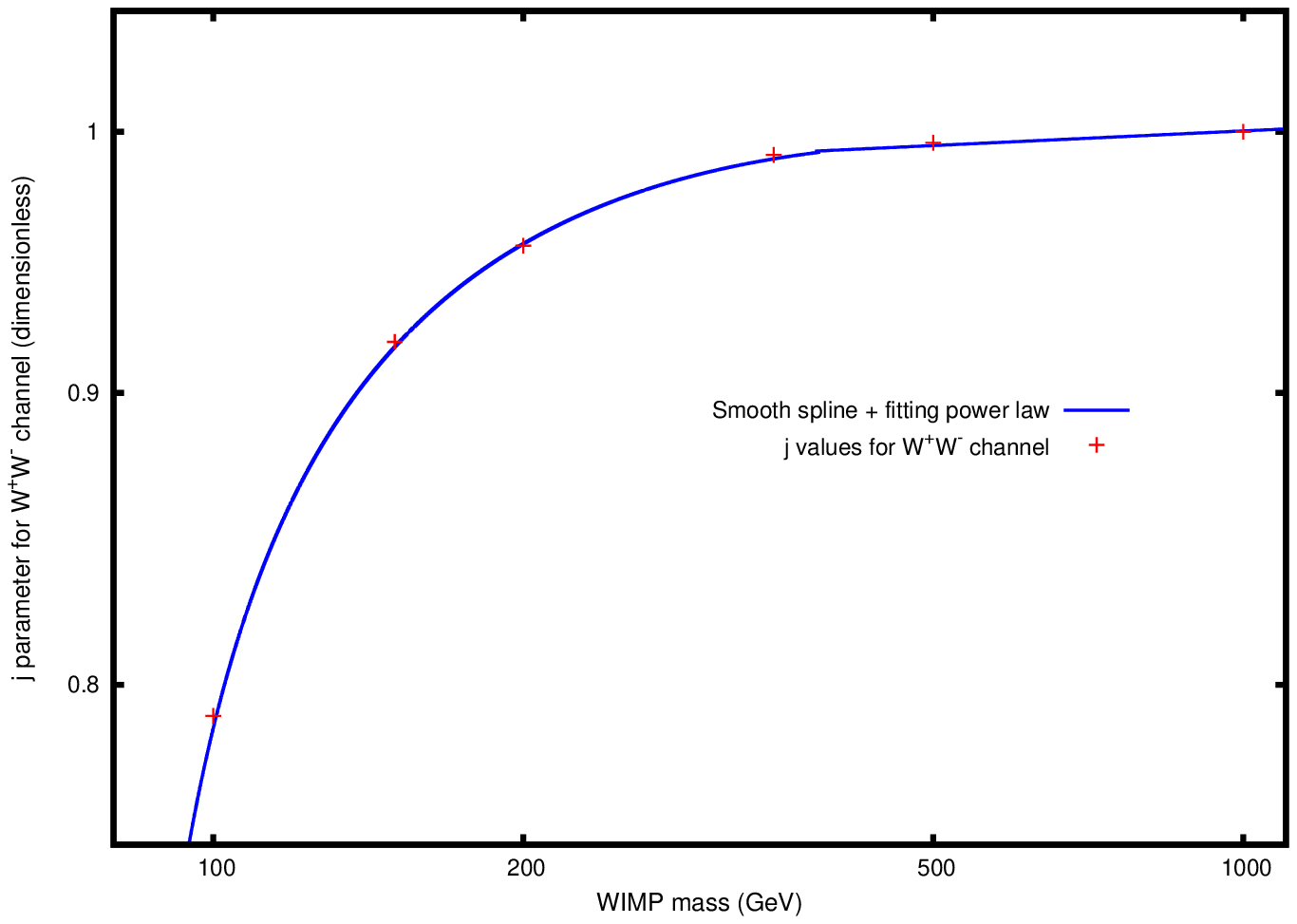}
\end{overpic}
}
\label{W_figures_2}
\caption{Mass dependence of $b_1$, $c_1$, $d_1$, $p$ and $j$ parameters for
$W^+W^-$  channel. Crossed points are parameters values found after the fitting process for each WIMP mass and solid lines correspond to the proposed fitting functions.}

\end{figure}

\end{widetext}
\newpage
\begin{center}
\bf{B. Plots for $t$ quark}
\end{center}

\begin{figure}[!hbp]
\subfigure[ \hspace{1ex} Photon spectrum for $M=200\,\text{GeV}$ for $t \bar t$ channel.
]{
\begin{overpic}[width=8.520cm]{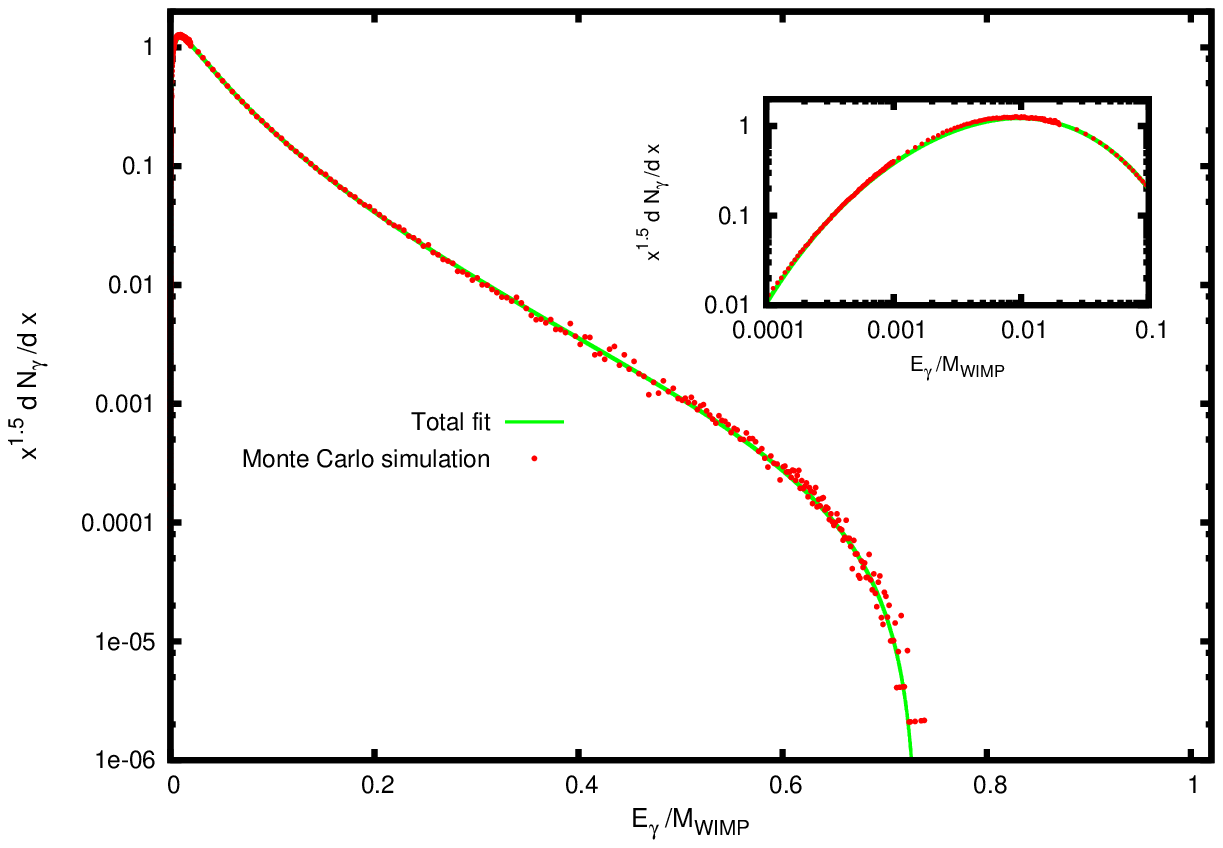}
\end{overpic}
}
\subfigure[ \hspace{1ex} Photon spectrum for $M=250\,\text{GeV}$ for $t \bar t$ channel.]{
\begin{overpic}[width=8.520cm]{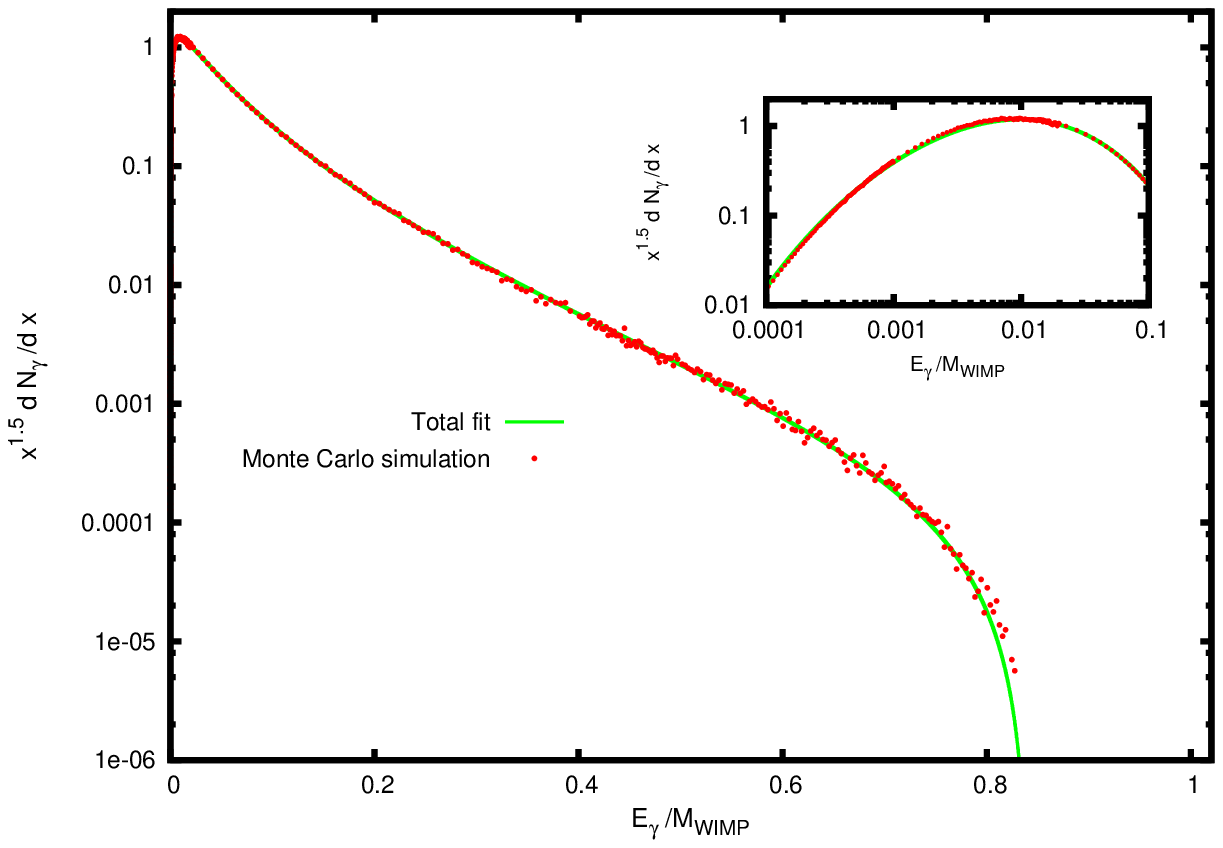}
\end{overpic}
}
\subfigure[ \hspace{1ex} Photon spectrum for $M=500\,\text{GeV}$ for $t \bar t$ channel.]{
\begin{overpic}[width=8.520cm]{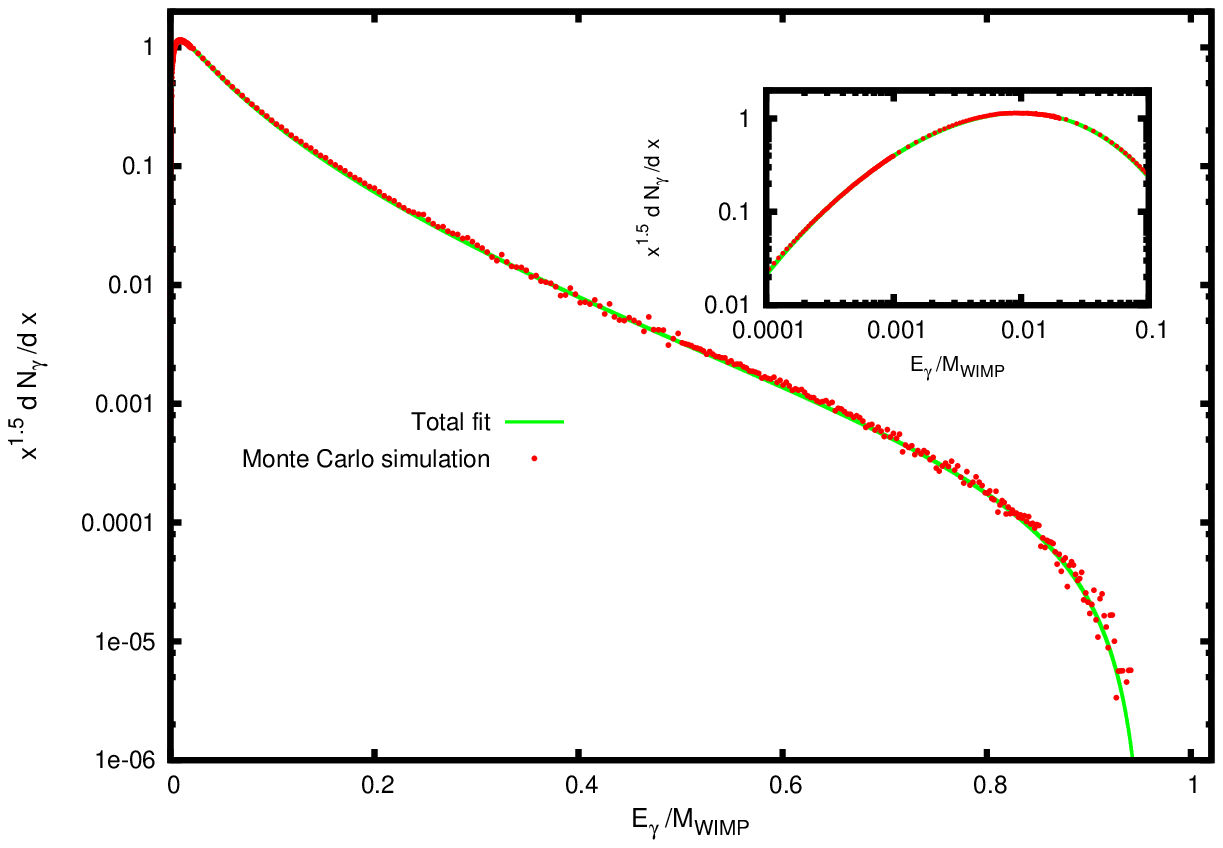}
\end{overpic}
}
\subfigure[ \hspace{1ex} Photon spectrum for $M=1000\,\text{GeV}$ for $t \bar t$ channel.]{
\begin{overpic}[width=8.520cm]{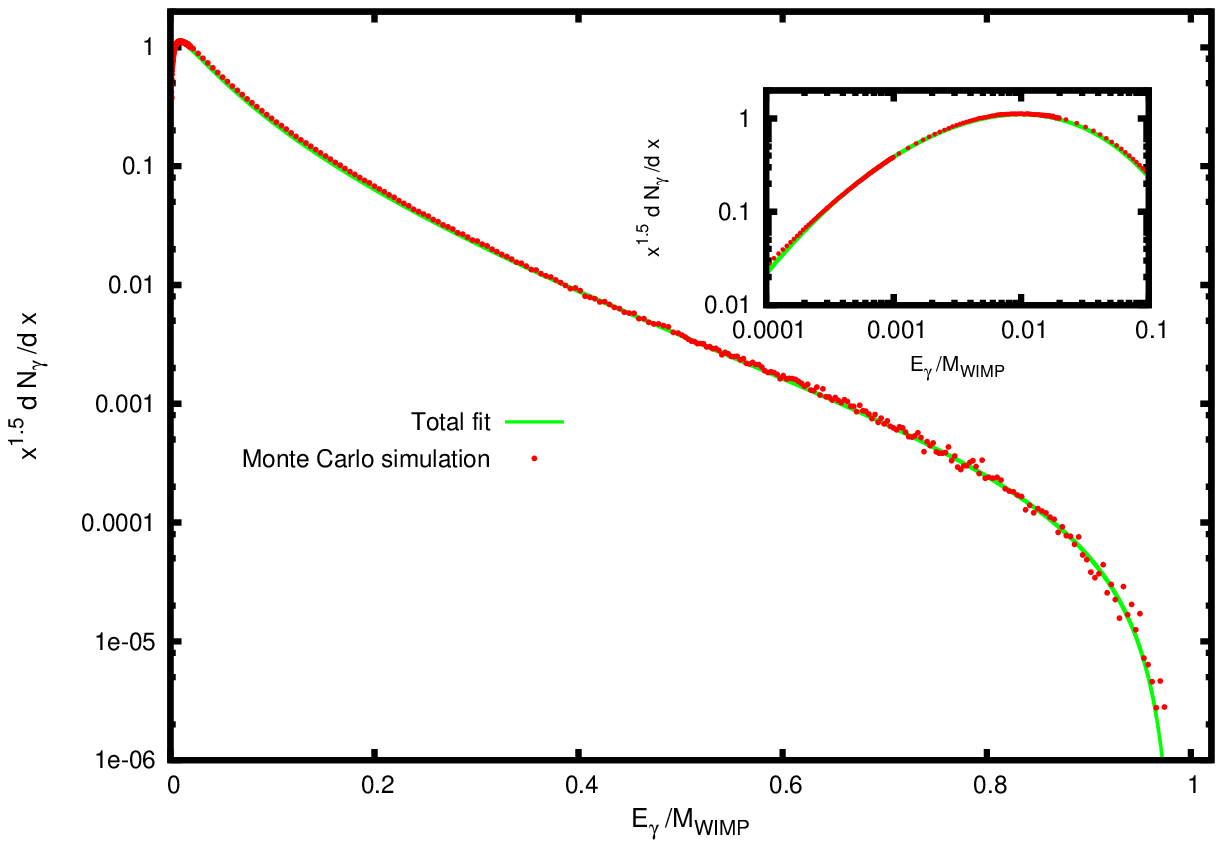}
\end{overpic}
}
\label{top_figures_1}
\caption{Photon spectra for four different WIMP masses (200, 250, 500 and 1000 GeV) in the $t \bar t$ annihilation channel. Red dotted points are PHYTIA simulations and solid lines correspond to the proposed fitting functions.}
\end{figure}
\newpage
\vspace{-0.4cm}
\begin{figure}[!hbp]
\subfigure[ \hspace{1ex} $b_1$ parameter of expression \eqref{general_formula} for $t \bar t$ channel.]{
\begin{overpic}[width=8.520cm]{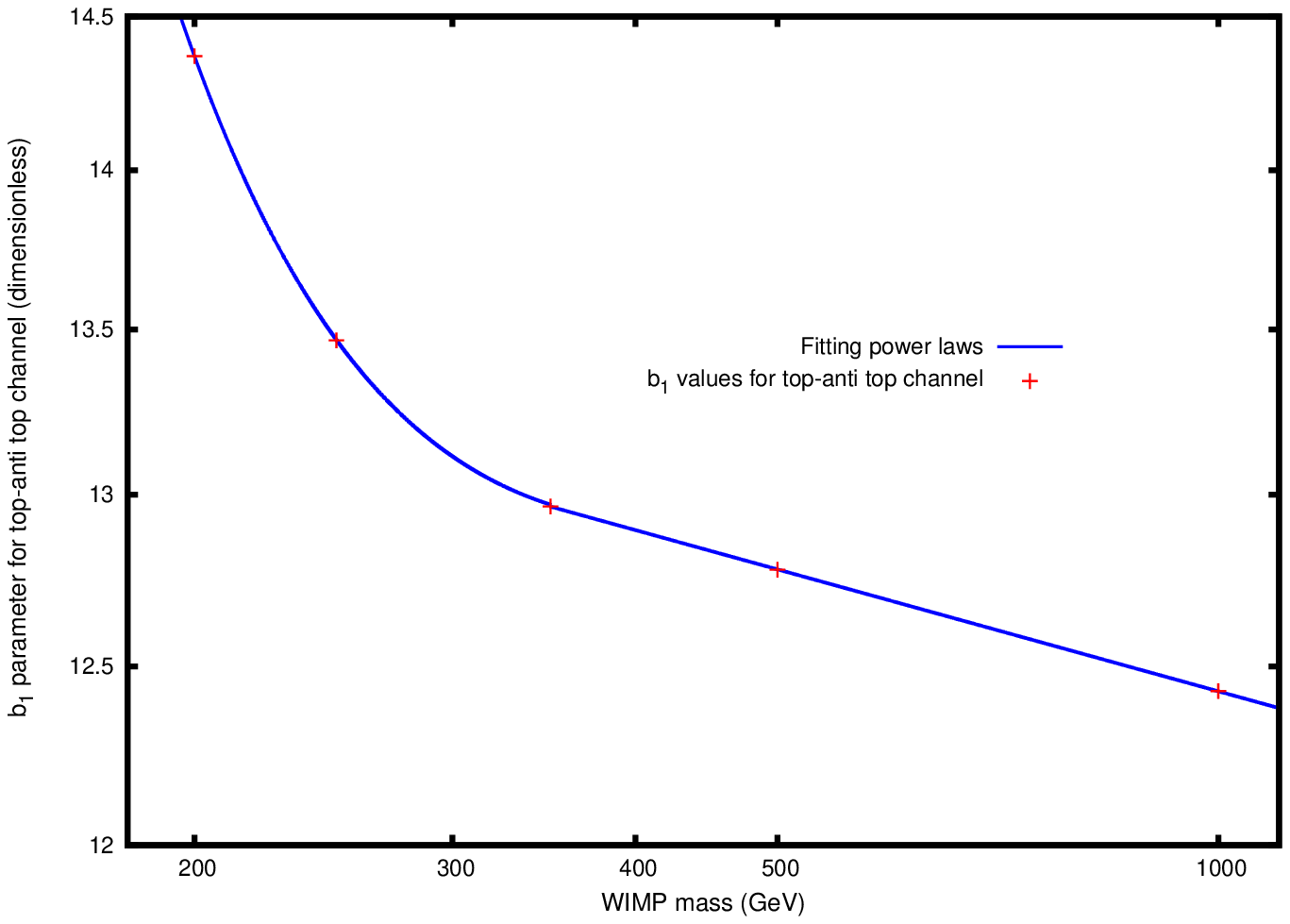}
\end{overpic}
}
\subfigure[ \hspace{1ex} $n_1$ parameter of expression \eqref{general_formula} for $t \bar t$ channel.]{
\begin{overpic}[width=8.520cm]{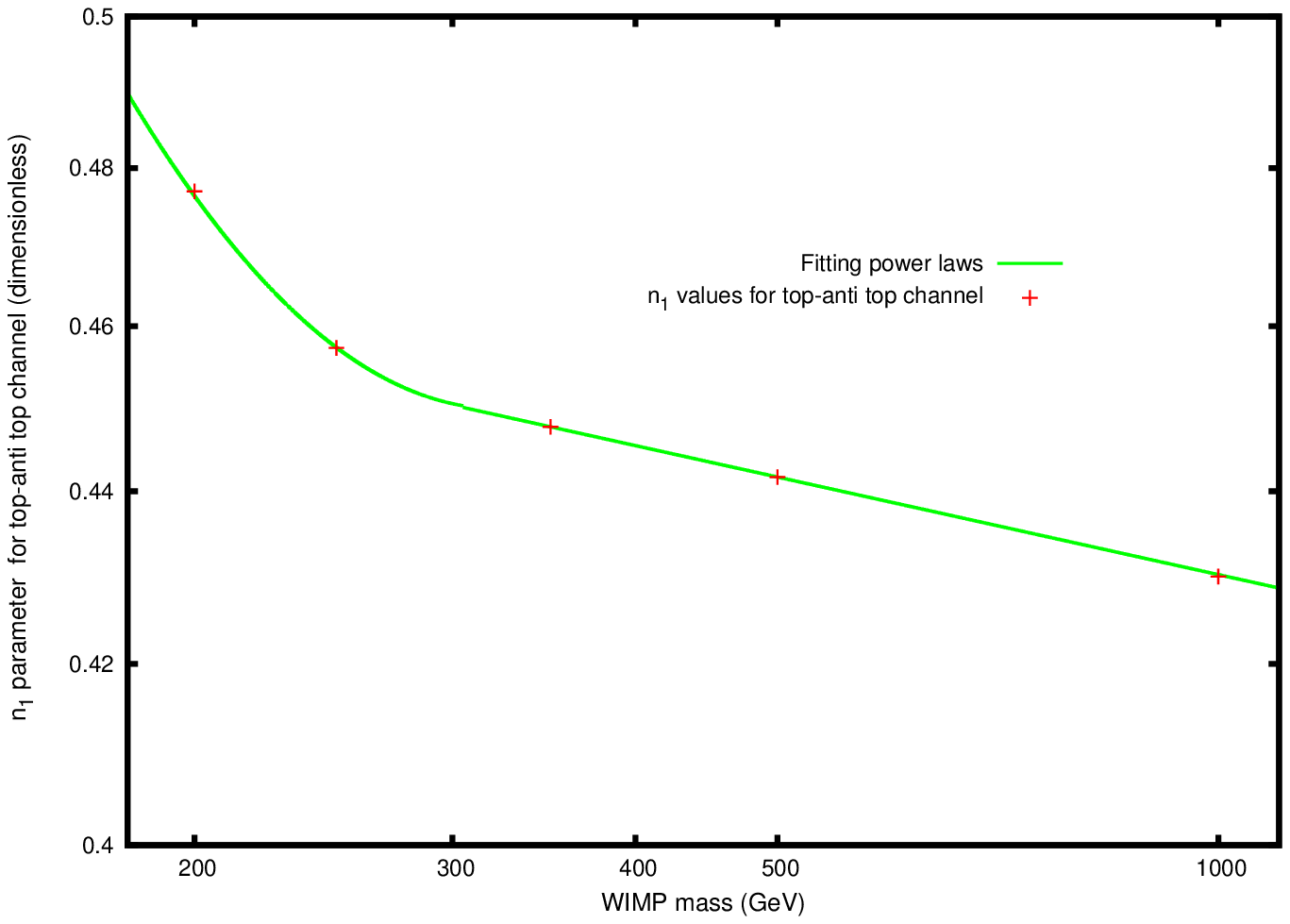}
\end{overpic}
}
\end{figure}
\vspace{-0.4cm}
\begin{figure}[!hbp]
\subfigure[ \hspace{1ex} $c_2$ parameter of expression \eqref{general_formula} for $t \bar t$ channel.]{
\begin{overpic}[width=8.520cm]{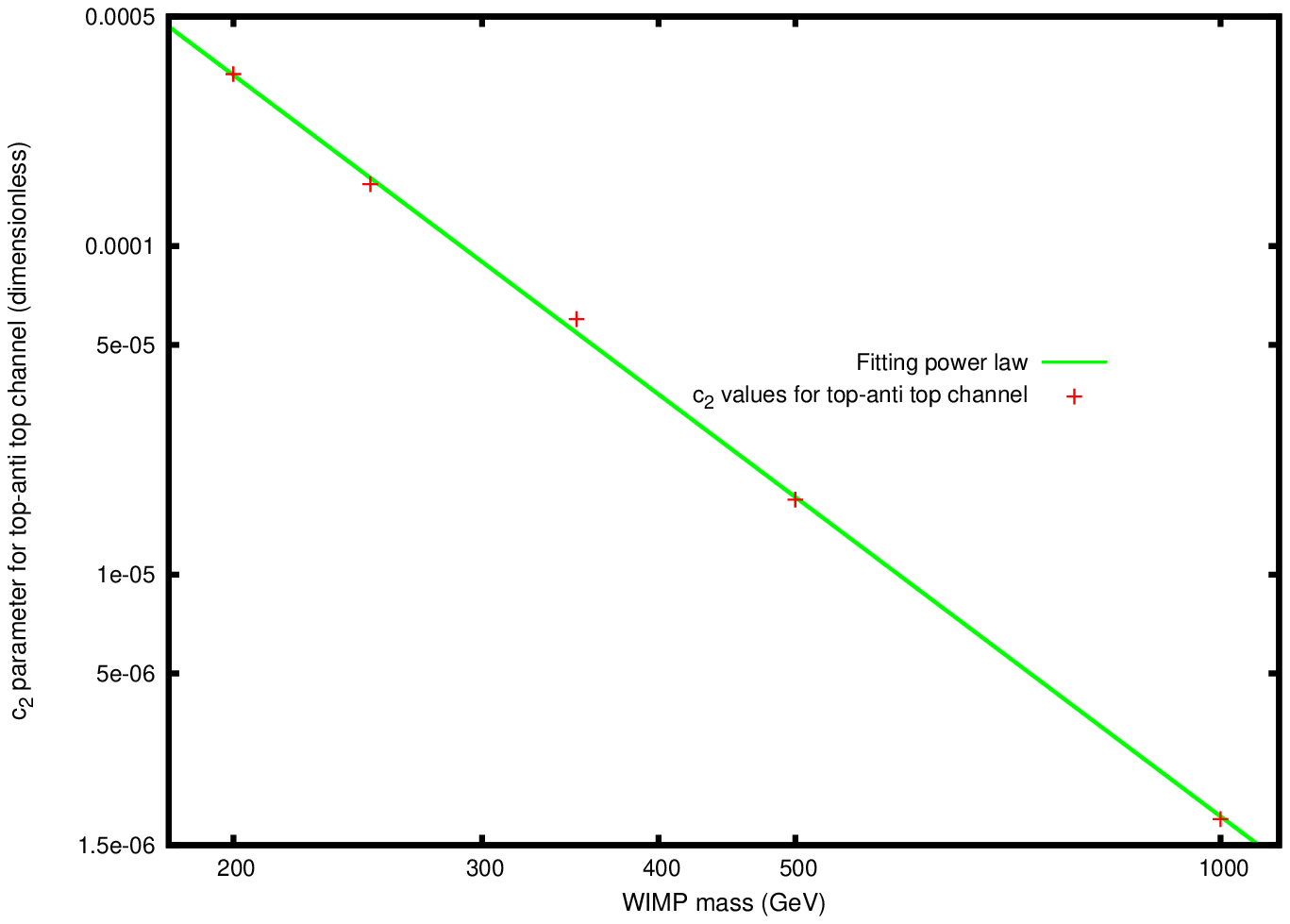}
\end{overpic}
}
\subfigure[ \hspace{1ex} $p$ parameter of expression \eqref{general_formula} for $t \bar t$ channel.]{
\begin{overpic}[width=8.520cm]{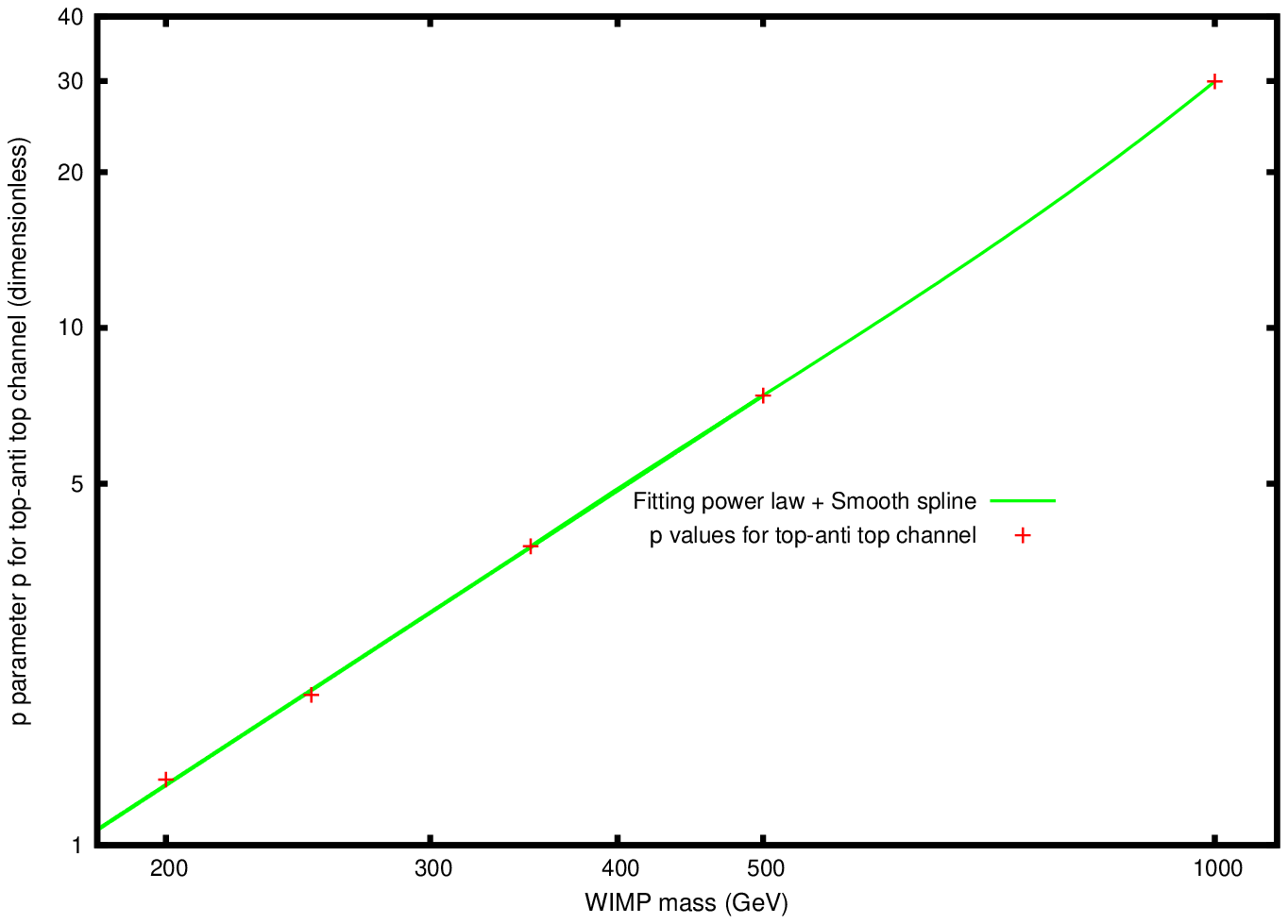}
\end{overpic}
}
\end{figure}
\vspace{-0.4cm}
\begin{figure}[!hbp]
\subfigure[ \hspace{1ex} $q$ parameter of expression \eqref{general_formula} for $t \bar t$ channel.]{
\begin{overpic}[width=8.520cm]{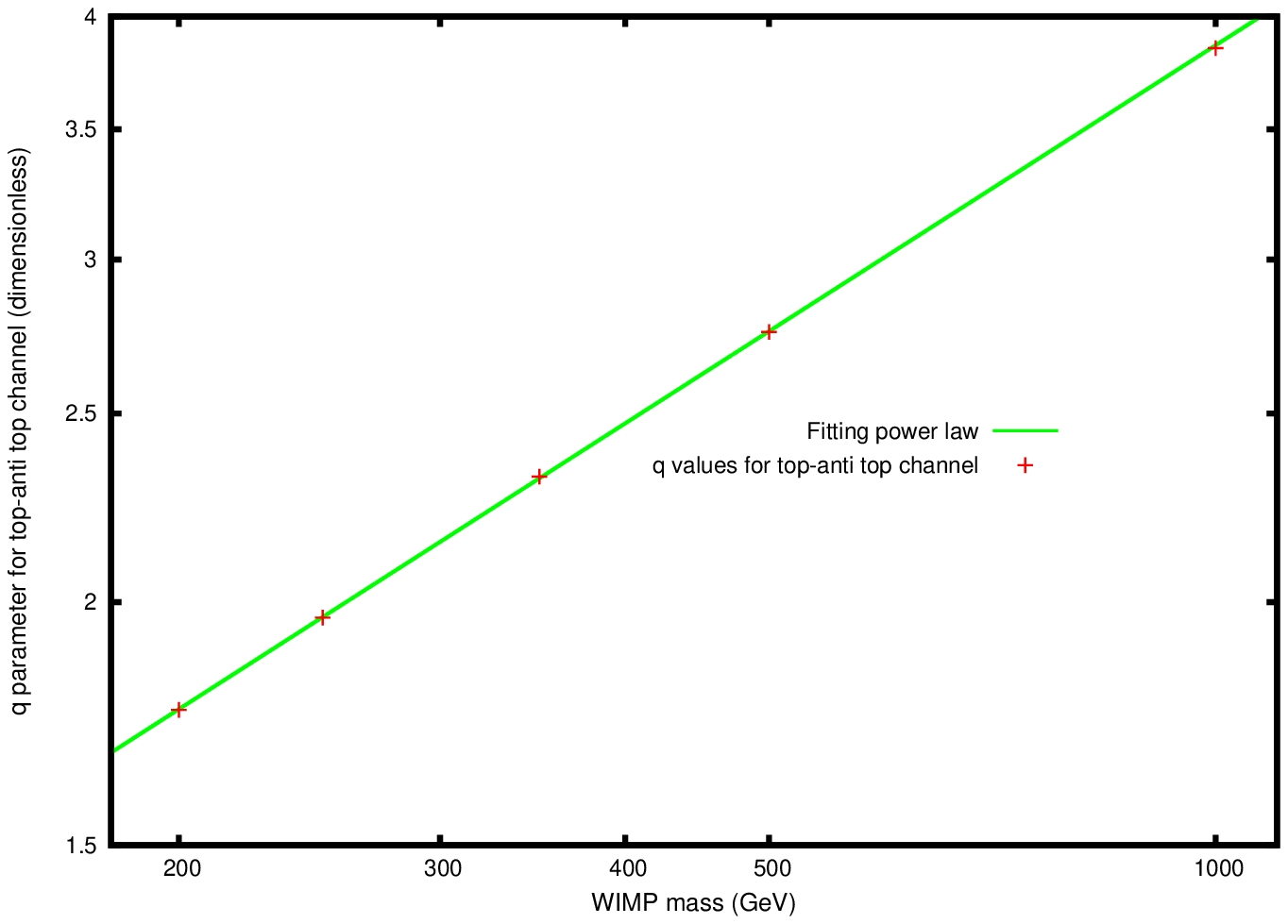}
\end{overpic}
}
\subfigure[ \hspace{1ex} $l$ parameter of expression \eqref{general_formula} for $t \bar t$ channel.]{
\begin{overpic}[width=8.520cm]{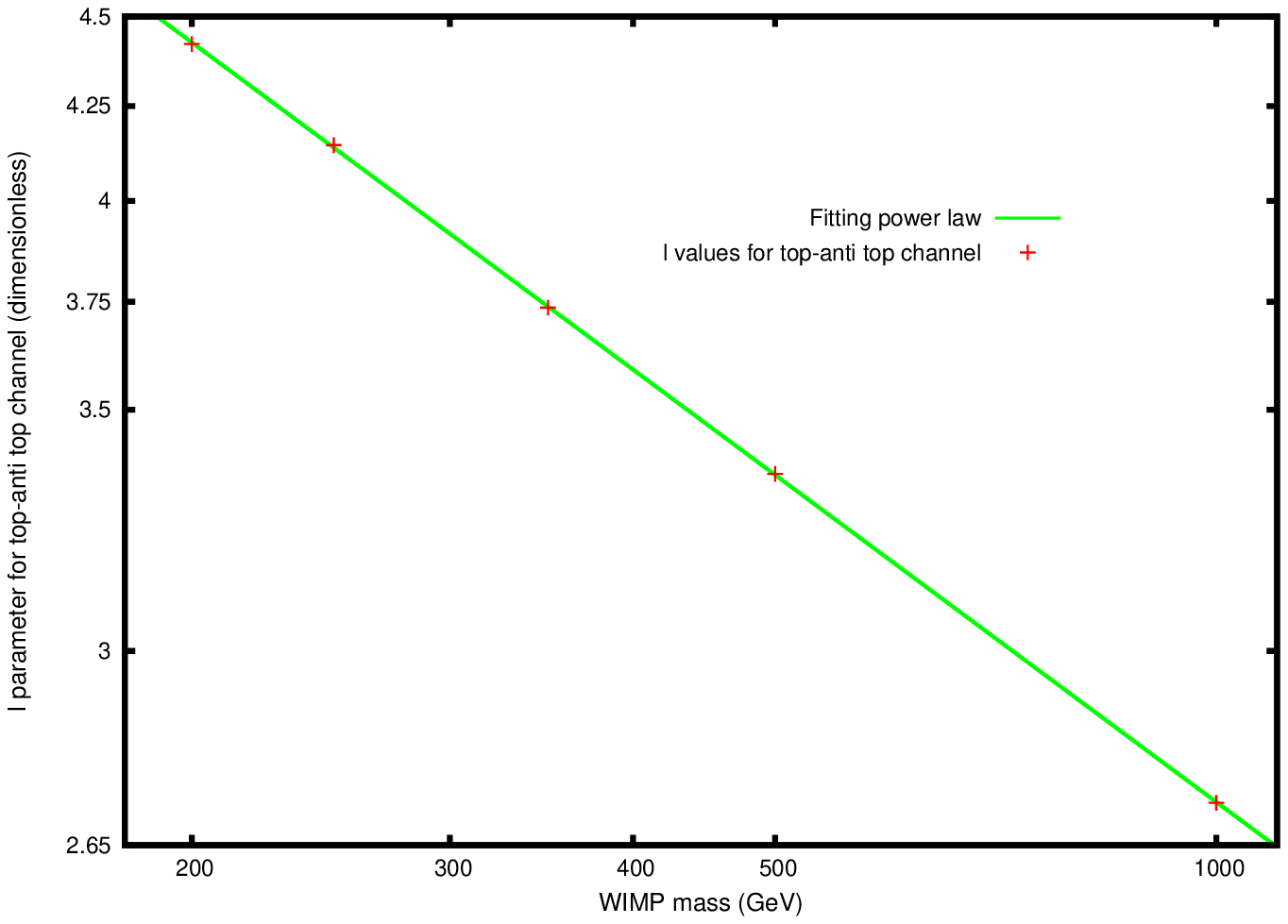}
\end{overpic}
}
\label{top_figures_2}
\caption{Mass dependence of $b_1$, $n_1$, $c_2$, $p$, $q$ and $l$ parameters for $t \bar t$ annihilation channel. Crossed points are parameters values found after the fitting process for each WIMP mass and solid lines correspond to the proposed fitting functions.}
\end{figure}
\newpage

\begin{center}
\bf{C. Plots for $\tau$  lepton}
\end{center}
\vspace{-0.4cm}
\begin{figure}[!hbp]
\subfigure[ \hspace{1ex} Photon spectrum for $M=25\,\text{GeV}$ for $\tau^{+}\tau^{-}$ channel.
]{
\begin{overpic}[width=8.12cm]{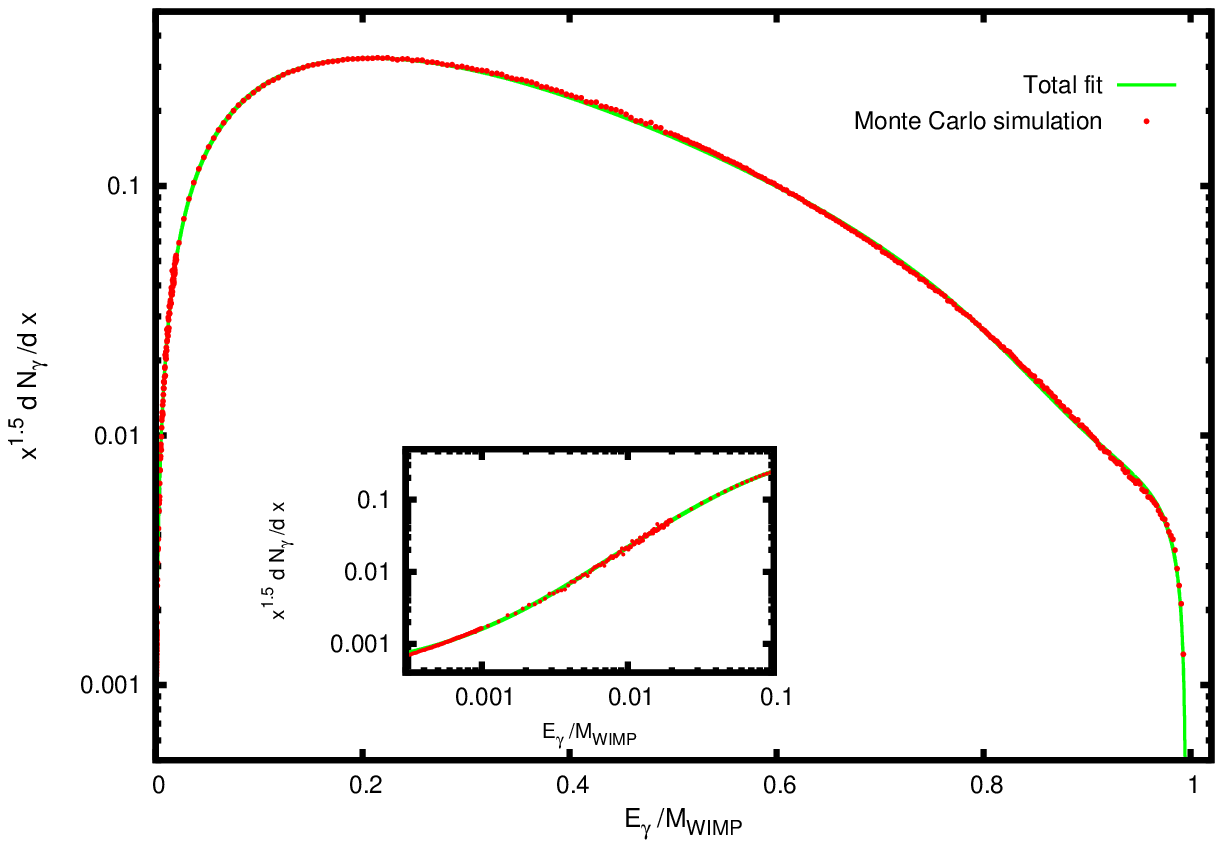}
\end{overpic}
}
\subfigure[ \hspace{1ex} Photon spectrum for $M=100\,\text{GeV}$ for $\tau^{+}\tau^{-}$ channel.]{
\begin{overpic}[width=8.12cm]{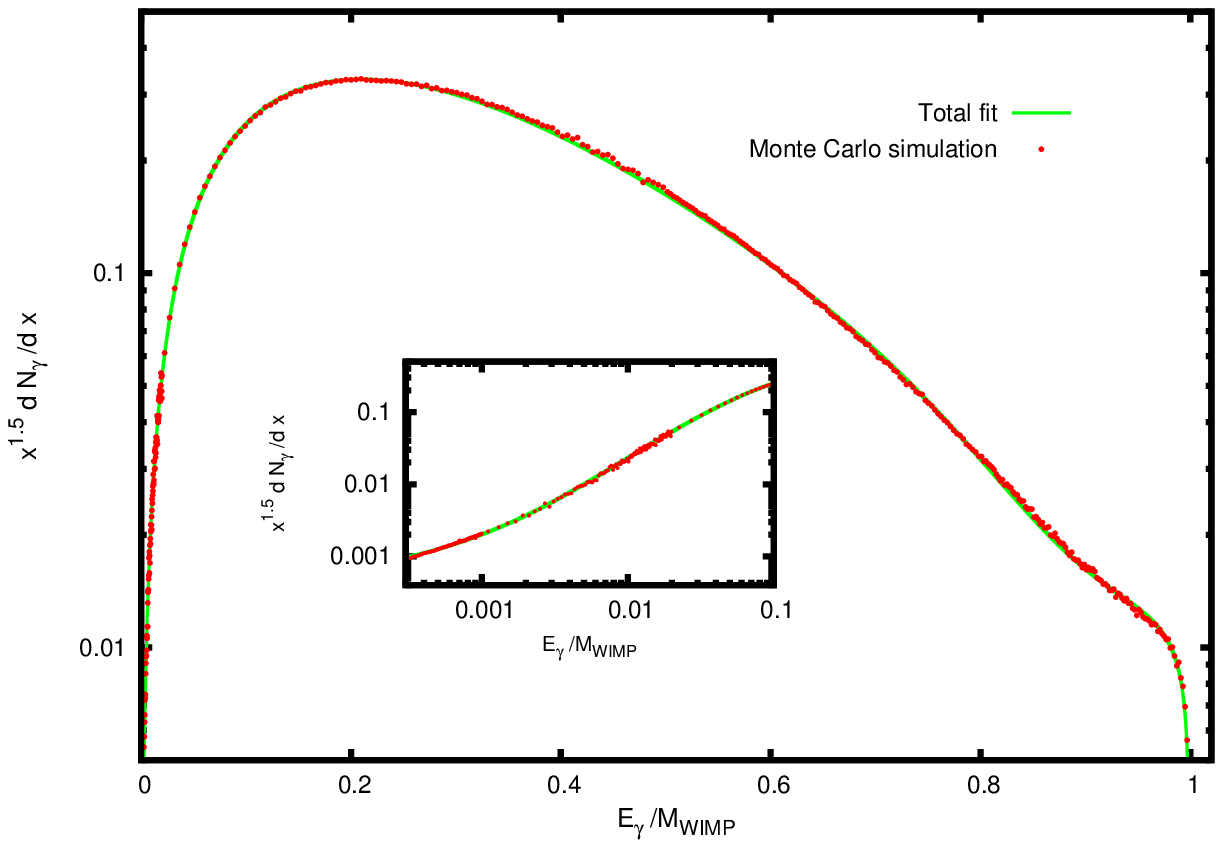}
\end{overpic}
}
\end{figure}
\vspace{-0.4cm}
\begin{figure}[!hbp]
\subfigure[ \hspace{1ex} Photon spectrum for $M=1000\,\text{GeV}$ for $\tau^{+}\tau^{-}$ channel.]{
\begin{overpic}[width=8.12cm]{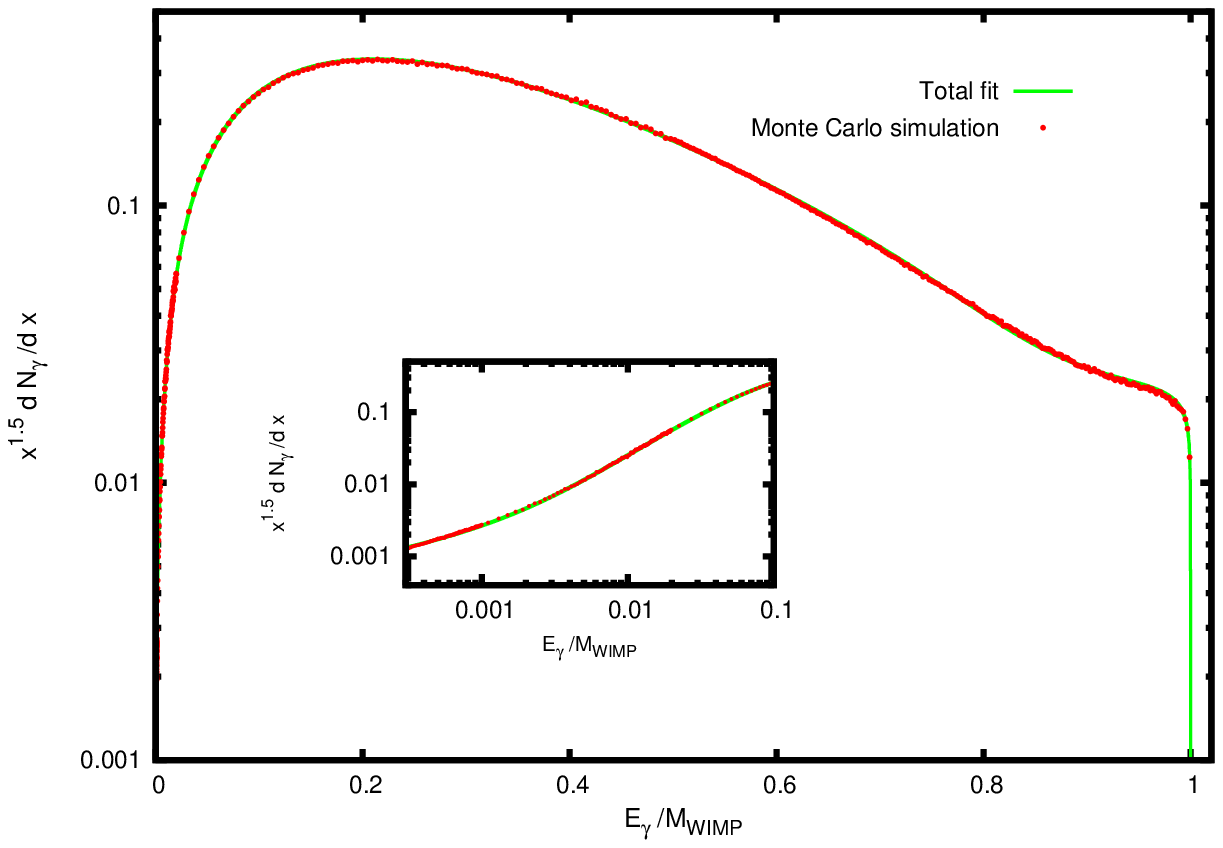}
\end{overpic}
}
\subfigure[ \hspace{1ex} Photon spectrum for $M=5\cdot10^{4}\,\text{GeV}$ for $\tau^{+}\tau^{-}$ channel.]{
\begin{overpic}[width=8.12cm]{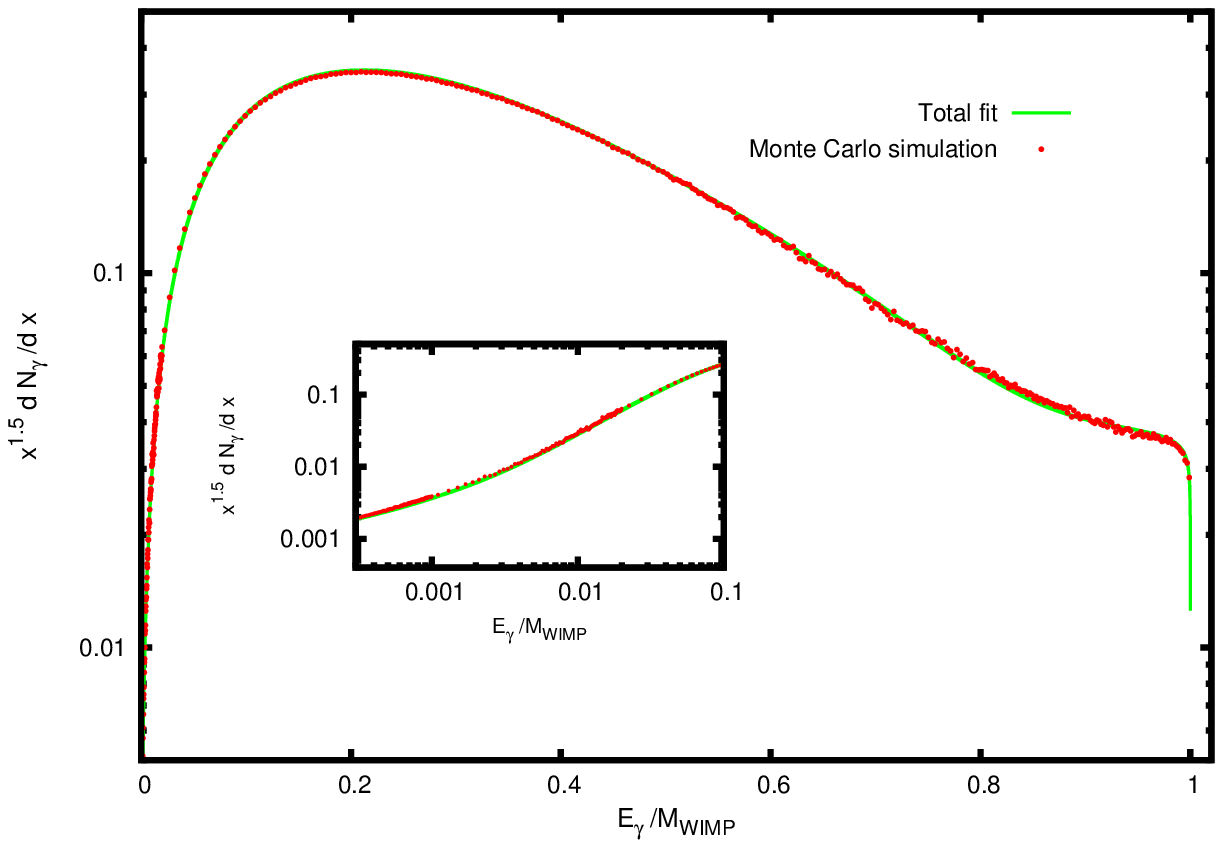}
\end{overpic}
}
\label{tau_figures_1}
\caption{Photon spectra for four different WIMP masses (25, 100, 1000 and $5\cdot10^4$ GeV) in the $\tau^{+}\tau^{-}$ annihilation channel. Red dotted points are PHYTIA simulations and solid lines correspond to the proposed fitting functions.}
\end{figure}
\begin{figure}[!hbp]
\subfigure[ \hspace{1ex} $n_1$ parameter of expression \eqref{general_formula} for $\tau^{+}\tau^{-}$ channel.]{
\begin{overpic}[width=8.12cm]{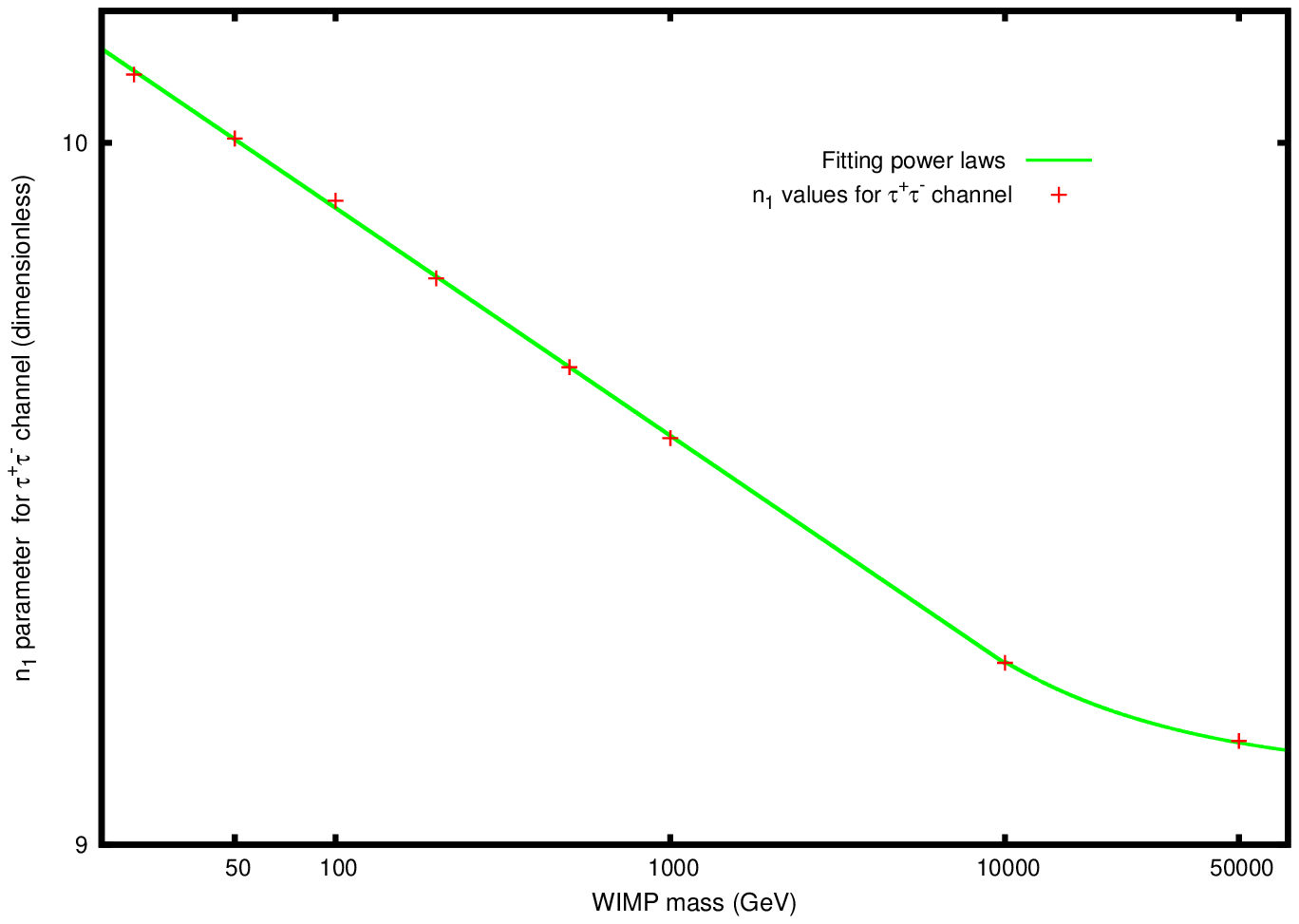}
\end{overpic}
}
\subfigure[ \hspace{1ex} $p$ parameter of expression \eqref{general_formula} for $\tau^{+}\tau^{-}$ channel.]{
\begin{overpic}[width=8.12cm]{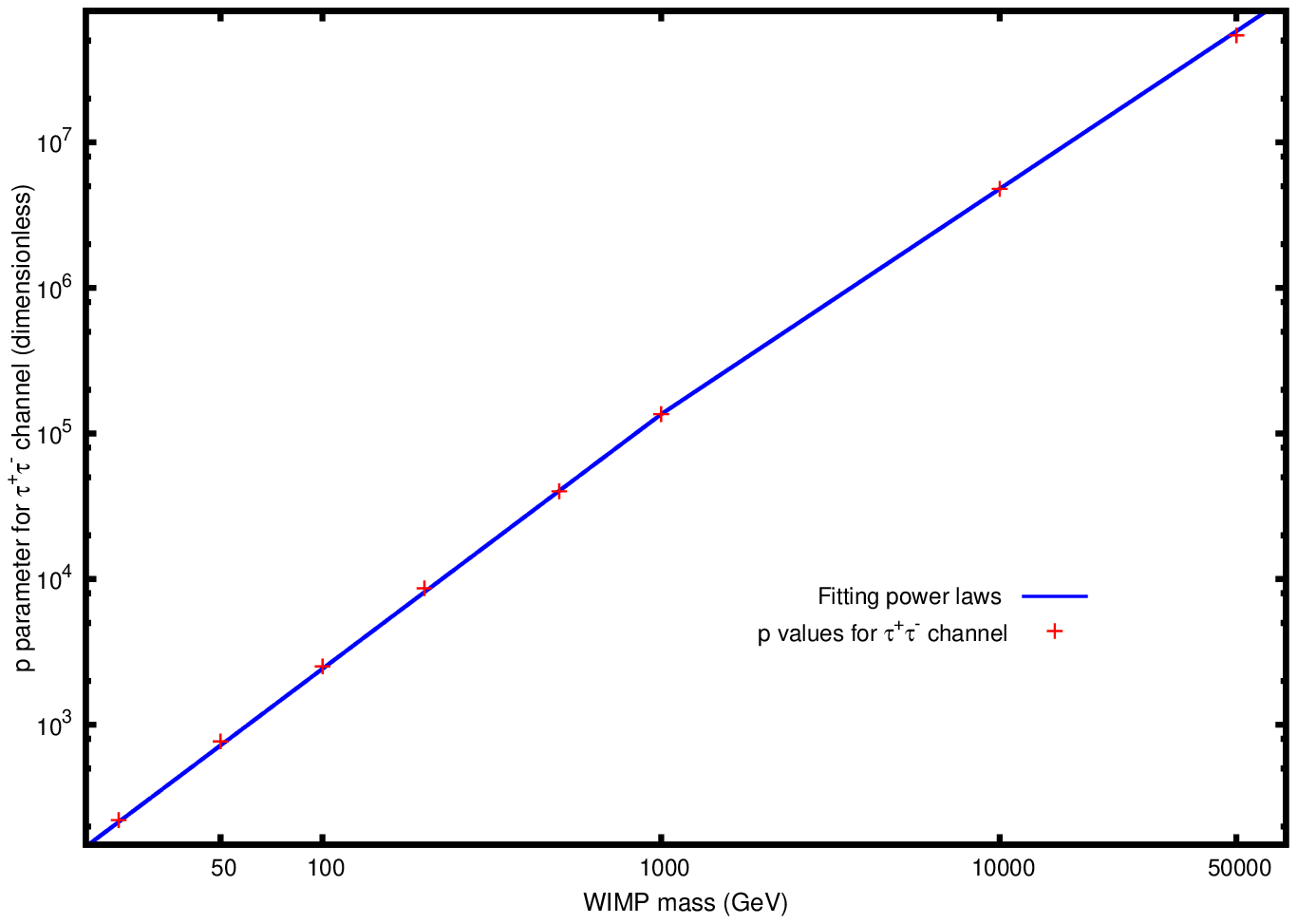}
\end{overpic}
}
\label{tau_figures_2}
\caption{Mass dependence of $n_1$ and $p$ parameters for $\tau^{+}\tau^{-}$ annihilation channel. Crossed points are parameters values found after the fitting process for each WIMP mass and solid lines correspond to the proposed fitting functions.}
\end{figure}
\newpage
\begin{center}
\bf{D. Plots for $b$ quark}
\end{center}

\begin{figure}[!hbp]
\subfigure[ \hspace{1ex} Photon spectrum for $M=50\,\text{GeV}$ for $b\bar{b}$ channel.
]{
\begin{overpic}[width=8.520cm]{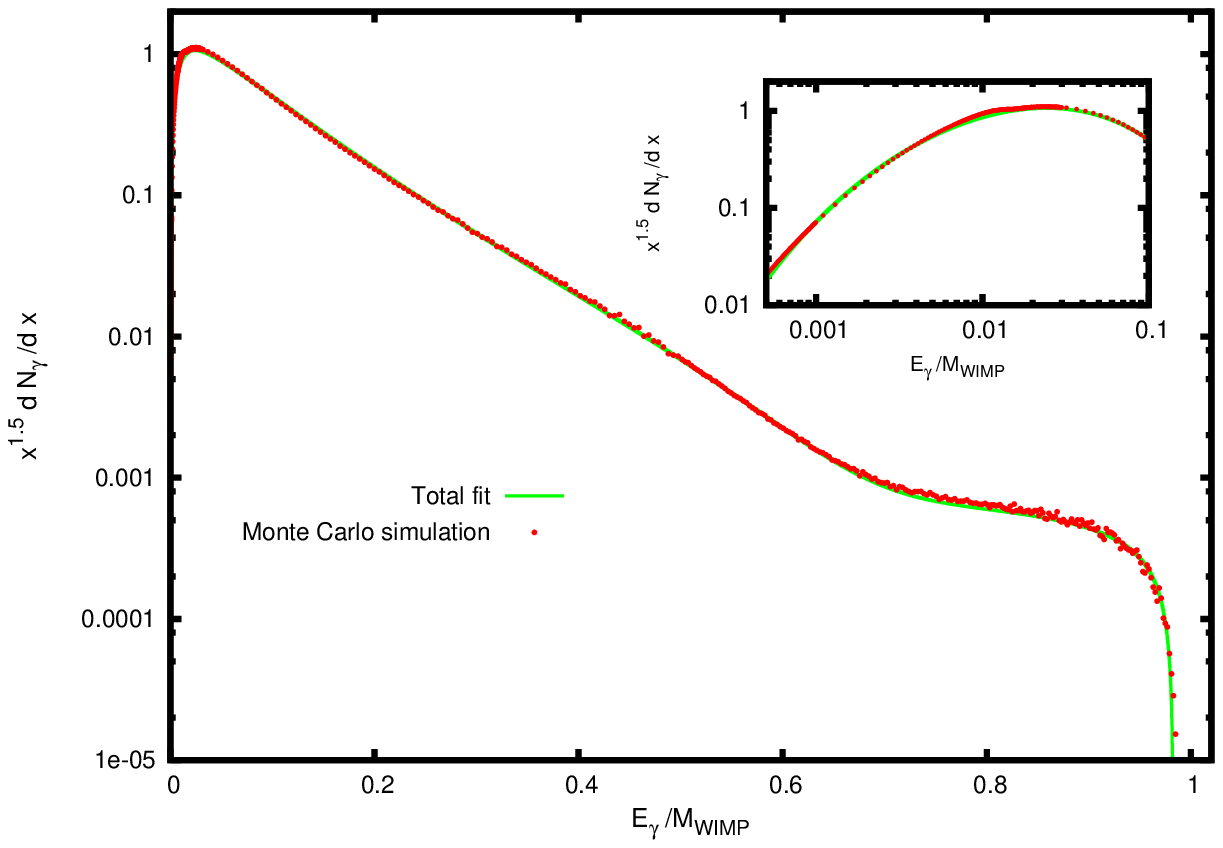}
\end{overpic}
}
\subfigure[ \hspace{1ex} Photon spectrum for $M=200\,\text{GeV}$ for $b\bar{b}$ channel.]{
\begin{overpic}[width=8.520cm]{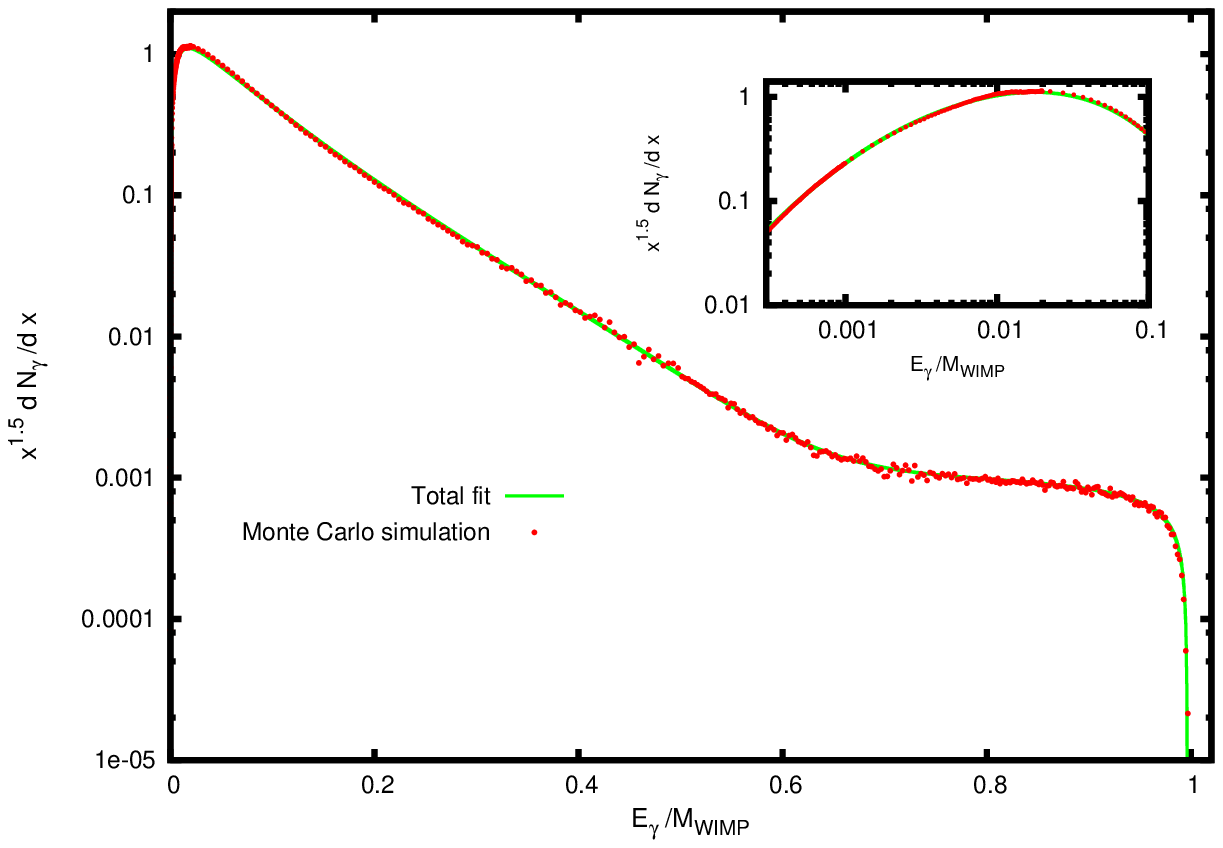}
\end{overpic}
}
\end{figure}
\begin{figure}[!hbp]
\subfigure[ \hspace{1ex} Photon spectrum for $M=1000\,\text{GeV}$ for $b\bar{b}$ channel.]{
\begin{overpic}[width=8.520cm]{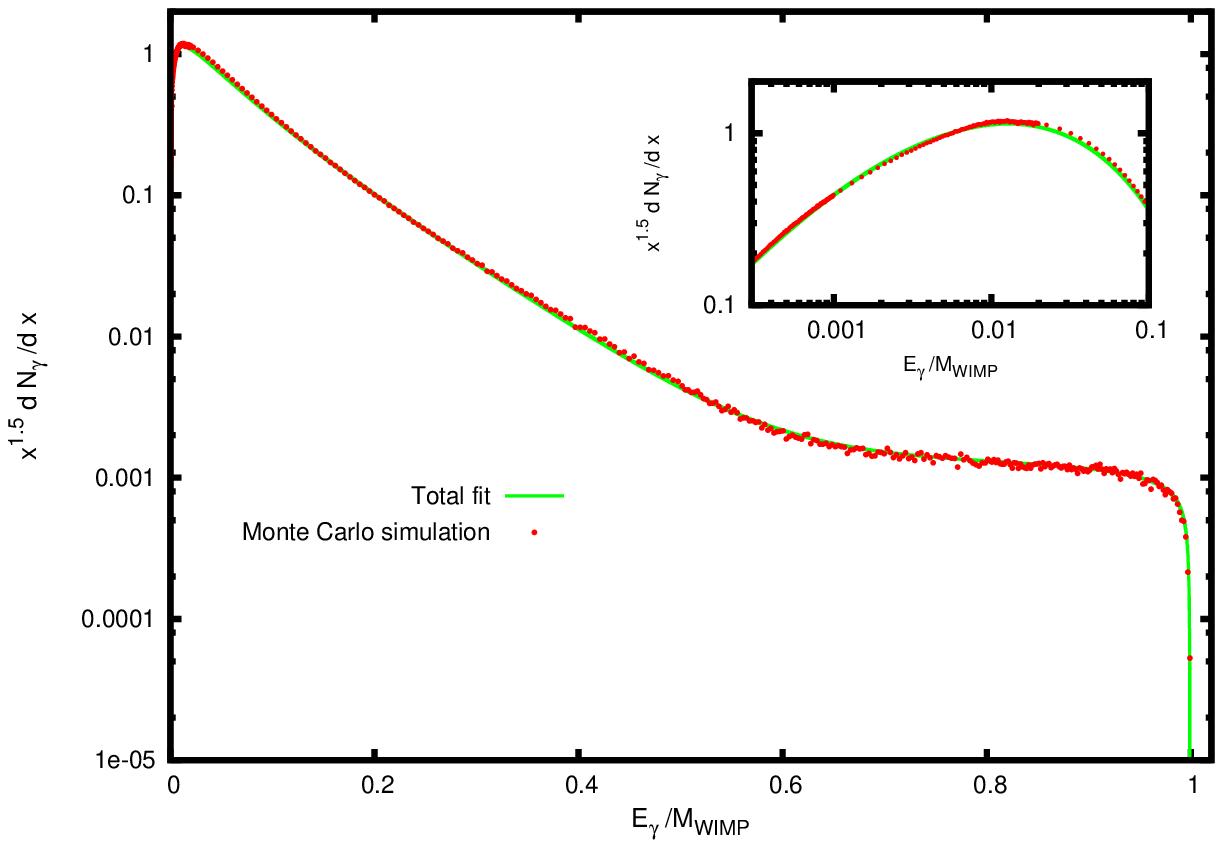}
\end{overpic}
}
\subfigure[ \hspace{1ex} Photon spectrum for $M=5000\,\text{GeV}$ for $b\bar{b}$ channel.]{
\begin{overpic}[width=8.520cm]{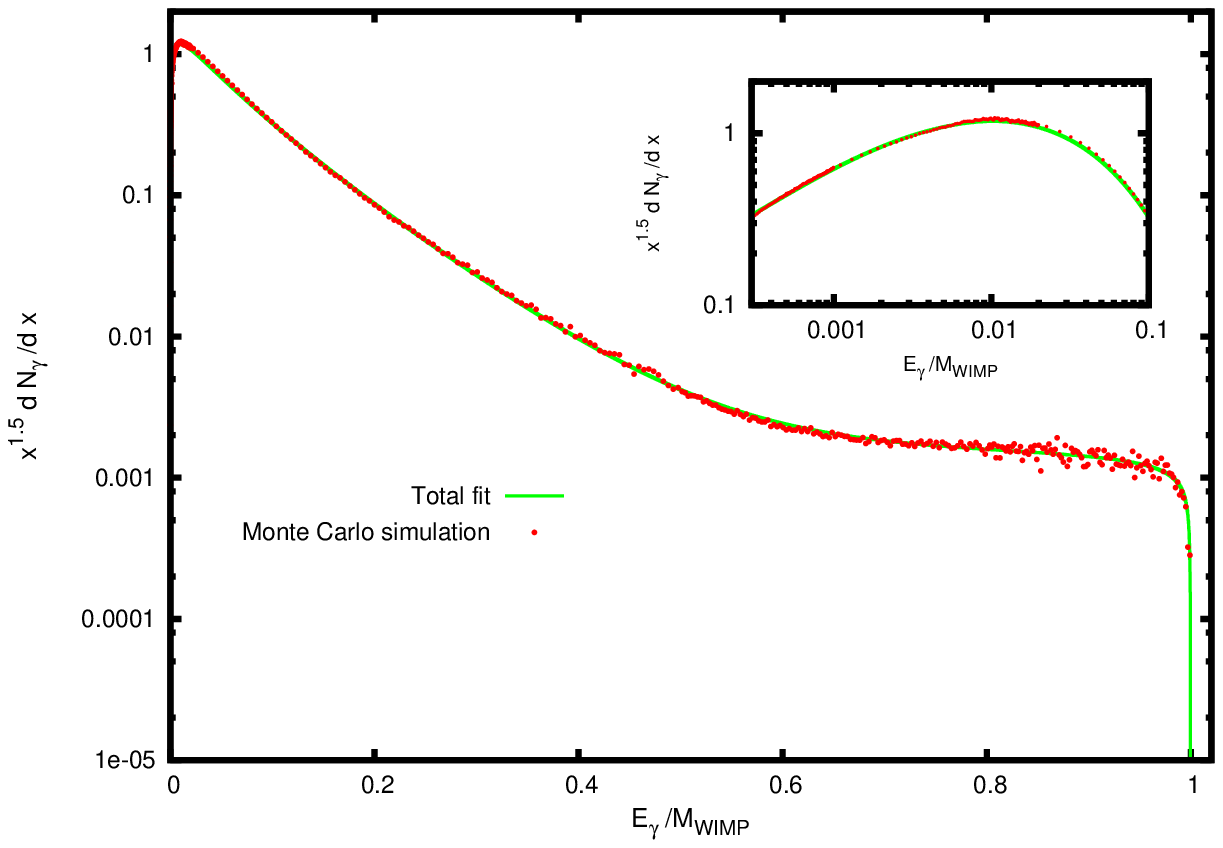}
\end{overpic}
}
\label{b_figures_1}
\caption{Photon spectra for four different WIMP masses (50, 200, 1000 and 5000 GeV) in the $b\bar{b}$ annihilation channel. Red dotted points are PHYTIA simulations and solid lines correspond to the proposed fitting functions.}
\end{figure}
\newpage
\begin{figure}[!hbp]
\subfigure[ \hspace{1ex} $b_1$ parameter of expression \eqref{general_formula} for $b\bar{b}$ channel.]{
\begin{overpic}[width=8.520cm]{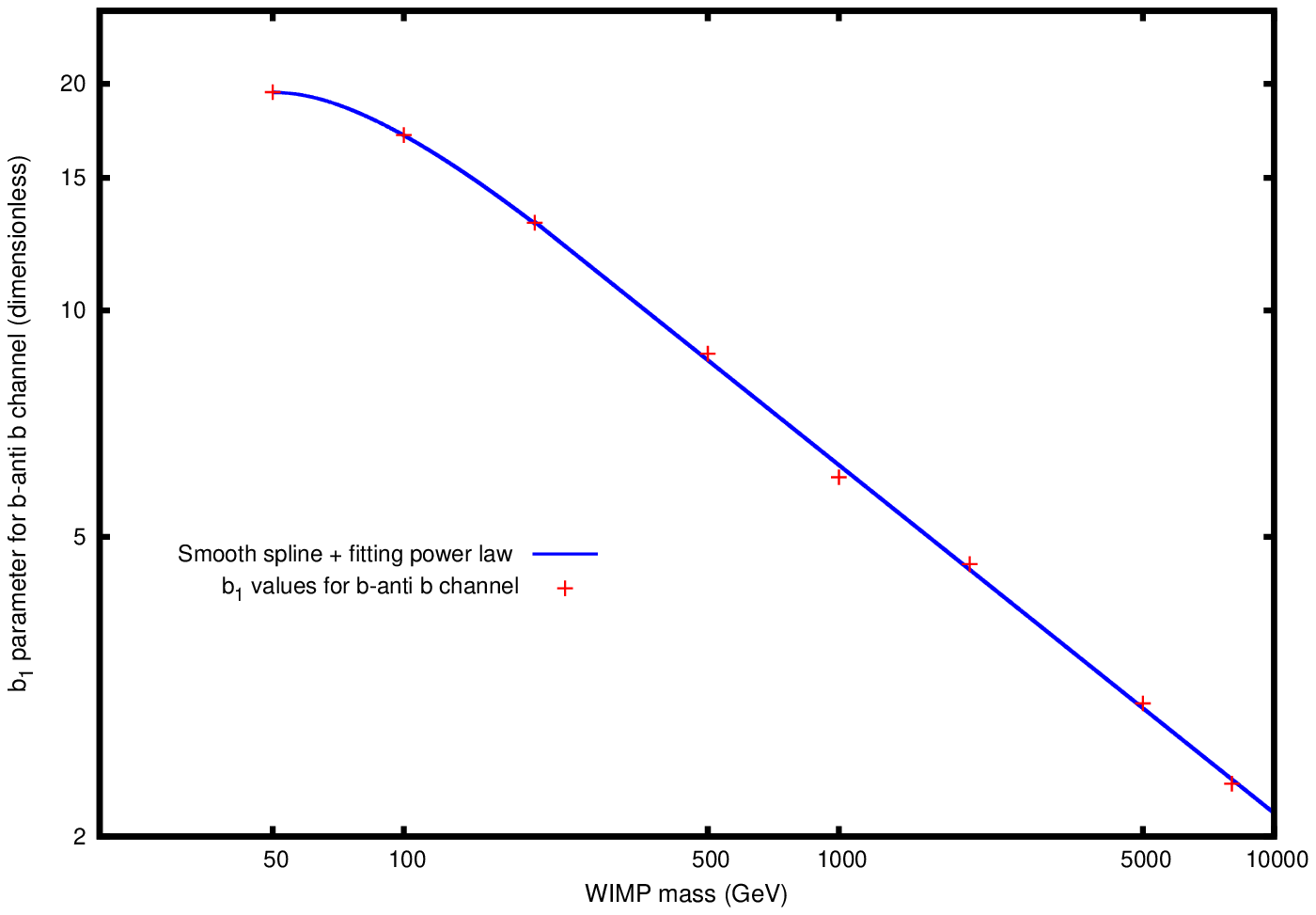}
\end{overpic}
}
\subfigure[ \hspace{1ex} $n_1$ parameter of expression \eqref{general_formula} for $b\bar{b}$ channel.]{
\begin{overpic}[width=8.520cm]{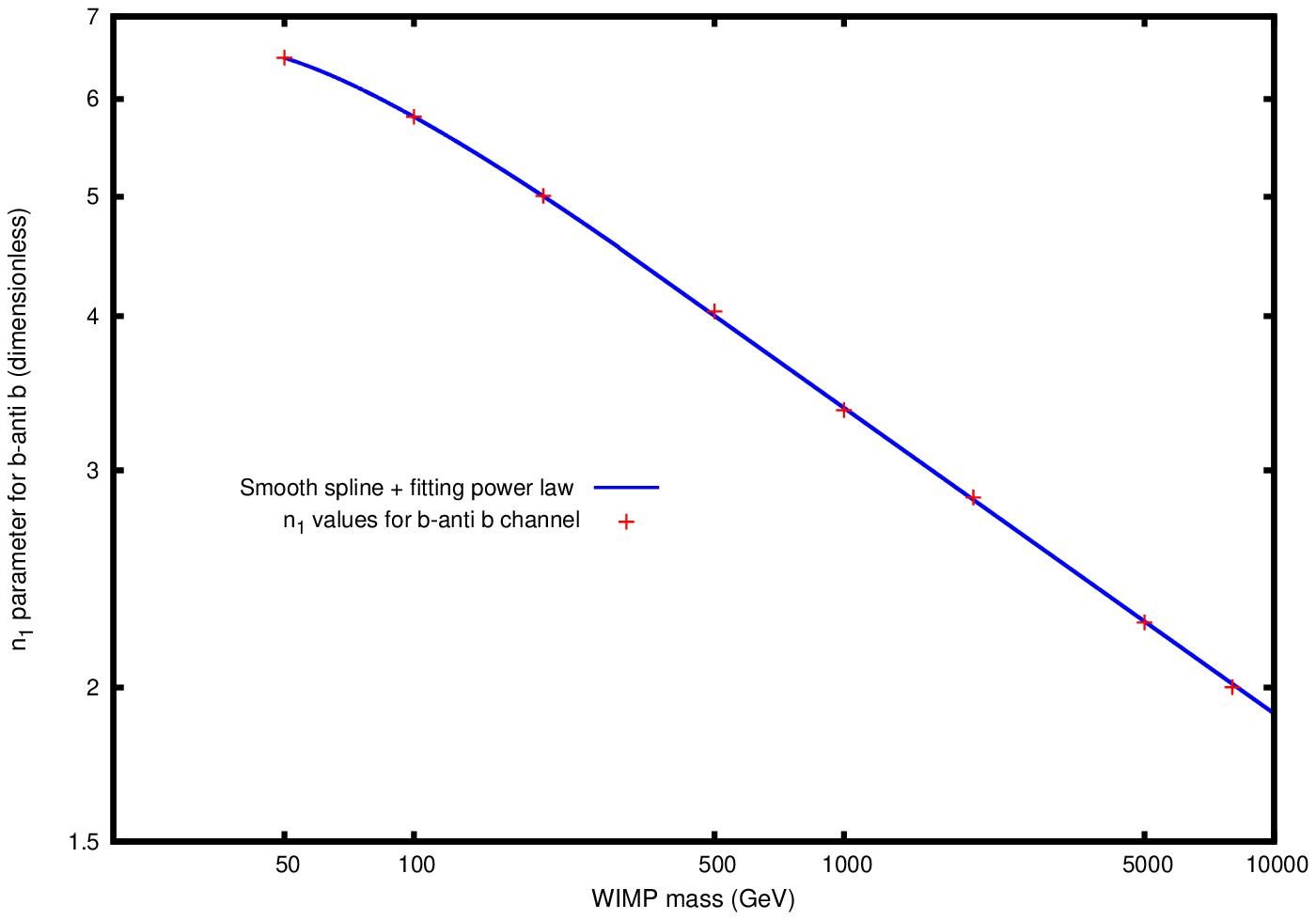}
\end{overpic}
}
\end{figure}
\begin{figure}[!hbp]
\subfigure[ \hspace{1ex} $n_2$ parameter of expression \eqref{general_formula} for $b\bar{b}$ channel.]{
\begin{overpic}[width=8.50cm]{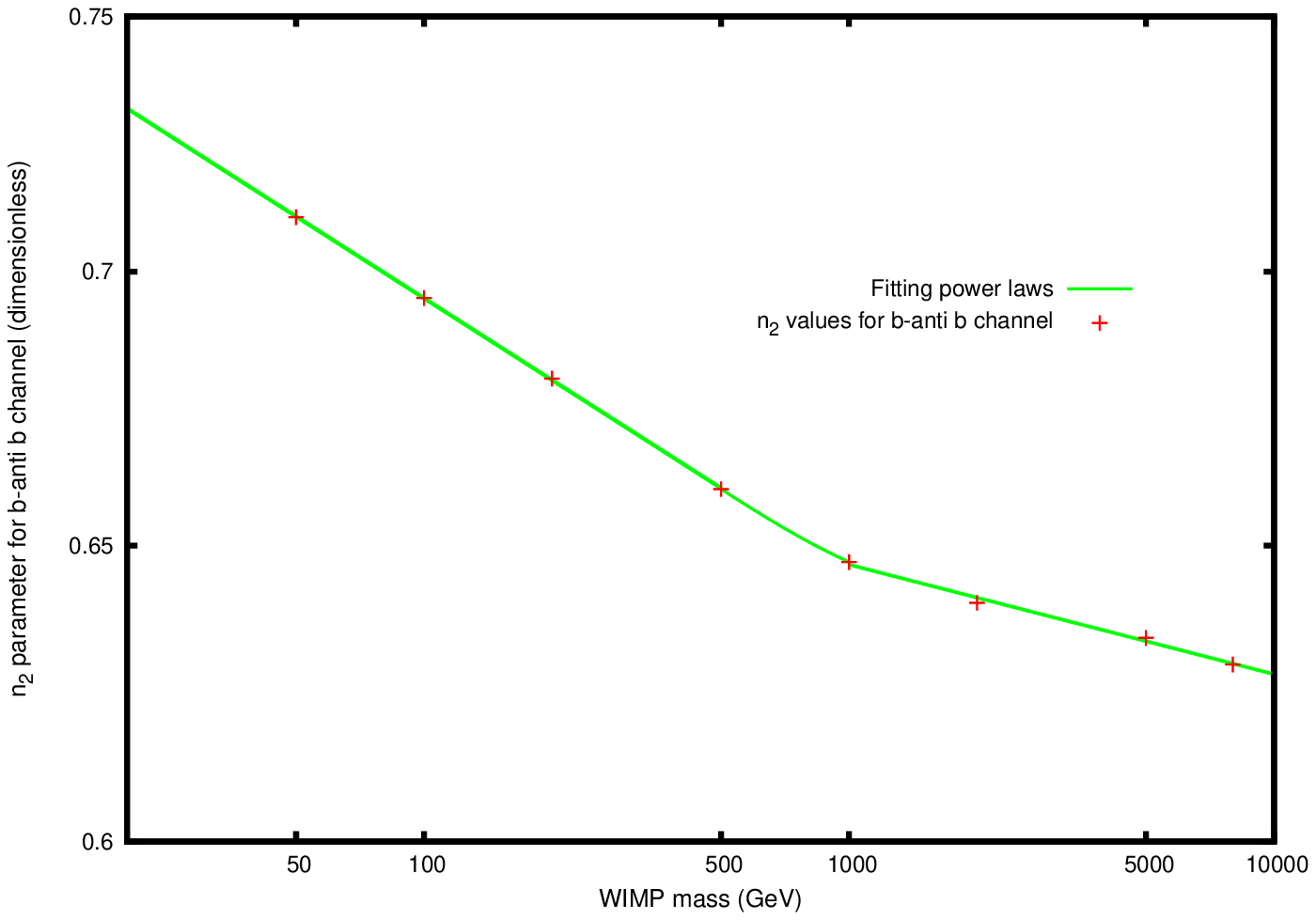}
\end{overpic}
}
\subfigure[ \hspace{1ex} $c_1$ parameter of expression \eqref{general_formula} for $b\bar{b}$ channel.]{
\begin{overpic}[width=8.50cm]{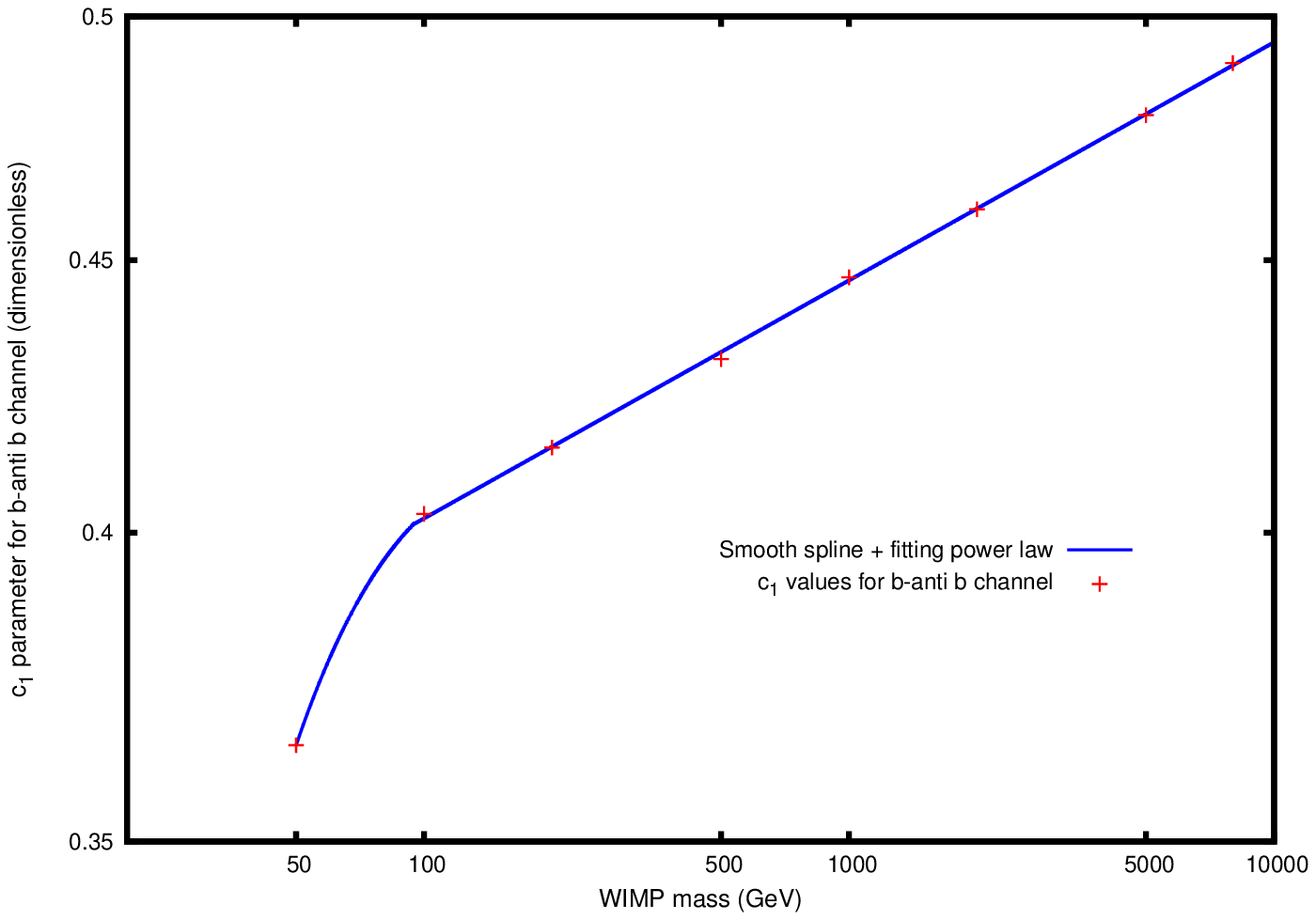}
\end{overpic}
}
\end{figure}
\begin{figure}[!hbp]
\subfigure[ \hspace{1ex} $d_1$ parameter of expression \eqref{general_formula} for $b\bar{b}$ channel.]{
\begin{overpic}[width=8.50cm]{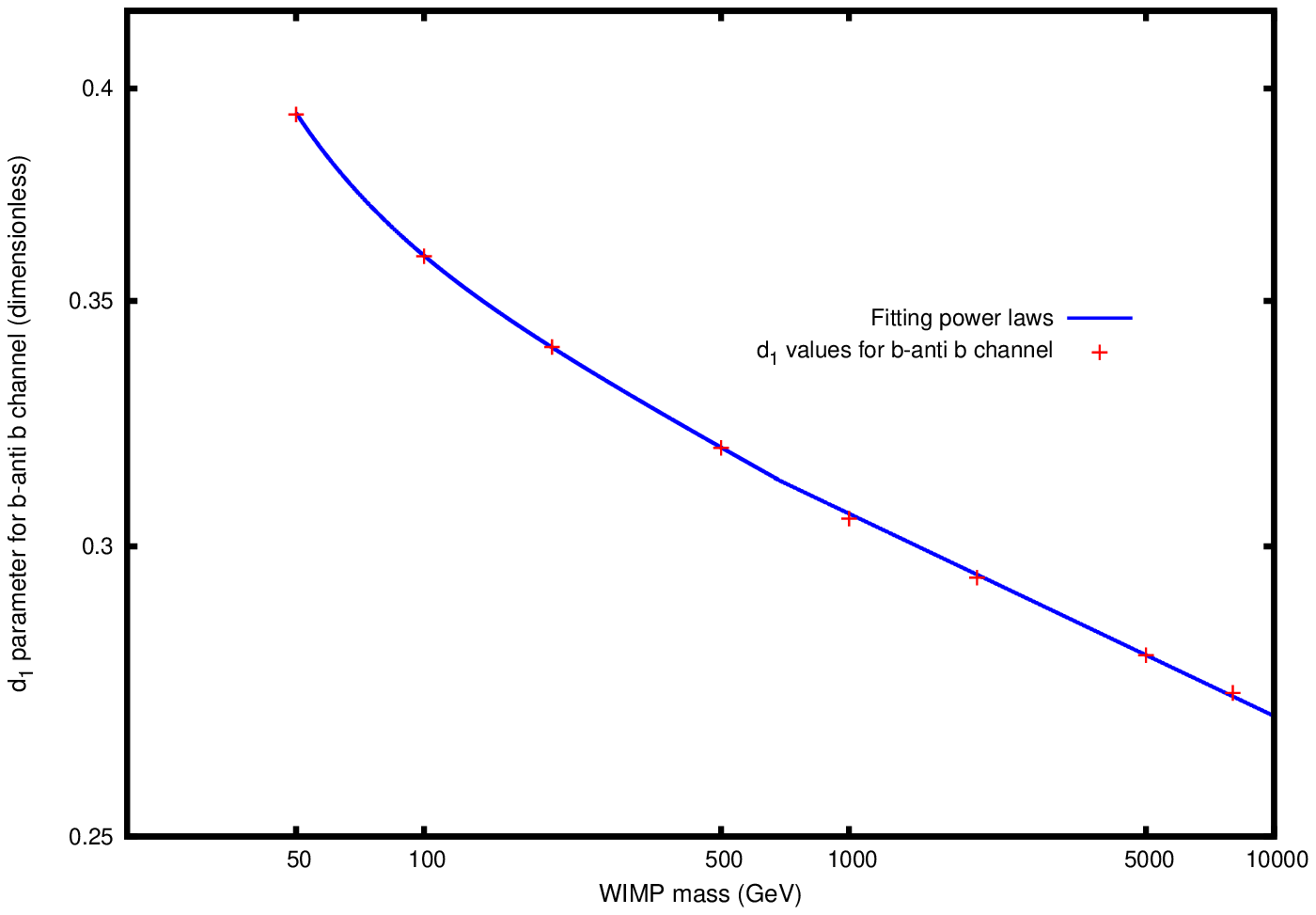}
\end{overpic}
}
\subfigure[ \hspace{1ex} $p$ parameter of expression \eqref{general_formula} for $b\bar{b}$ channel.]{
\begin{overpic}[width=8.50cm]{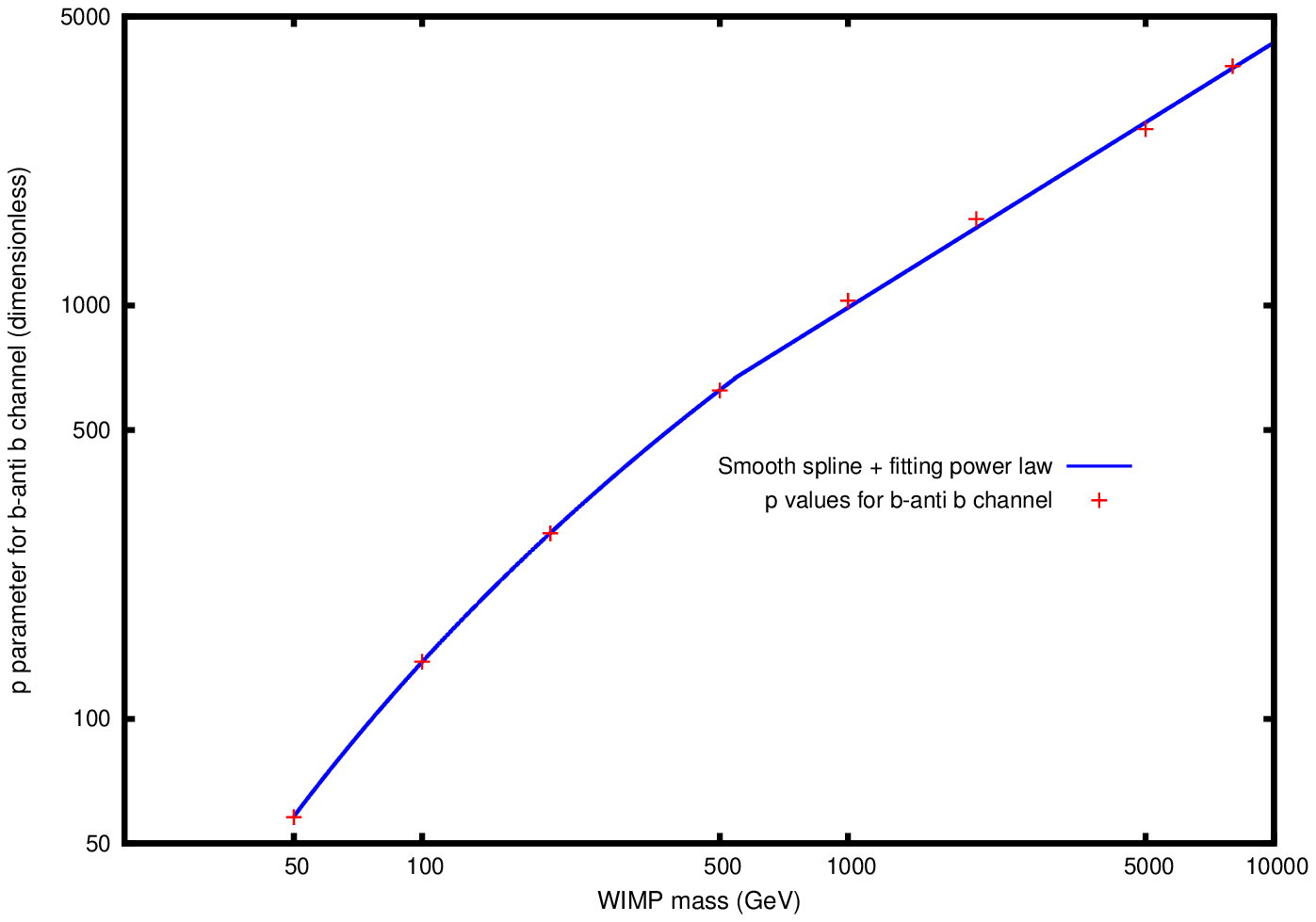}
\end{overpic}
}
\label{b_figures_2}
\caption{Mass dependence of $b_1$, $n_1$, $n_2$, $c_1$, $d_1$ and $p$ parameters for the $b\bar{b}$ annihilation channel. Crossed points are parameters values found after the fitting process for each WIMP mass and solid lines correspond to the proposed fitting functions.}
\end{figure}
\newpage
\begin{center}
\bf{E. Photon number per WIMPs annihilation}
\end{center}

\begin{figure}[!hbp]
\subfigure[ \hspace{1ex} Total number of photons per WIMP annihilation
in lepton-antilepton pairs.]{
\begin{overpic}[width=8.40cm]{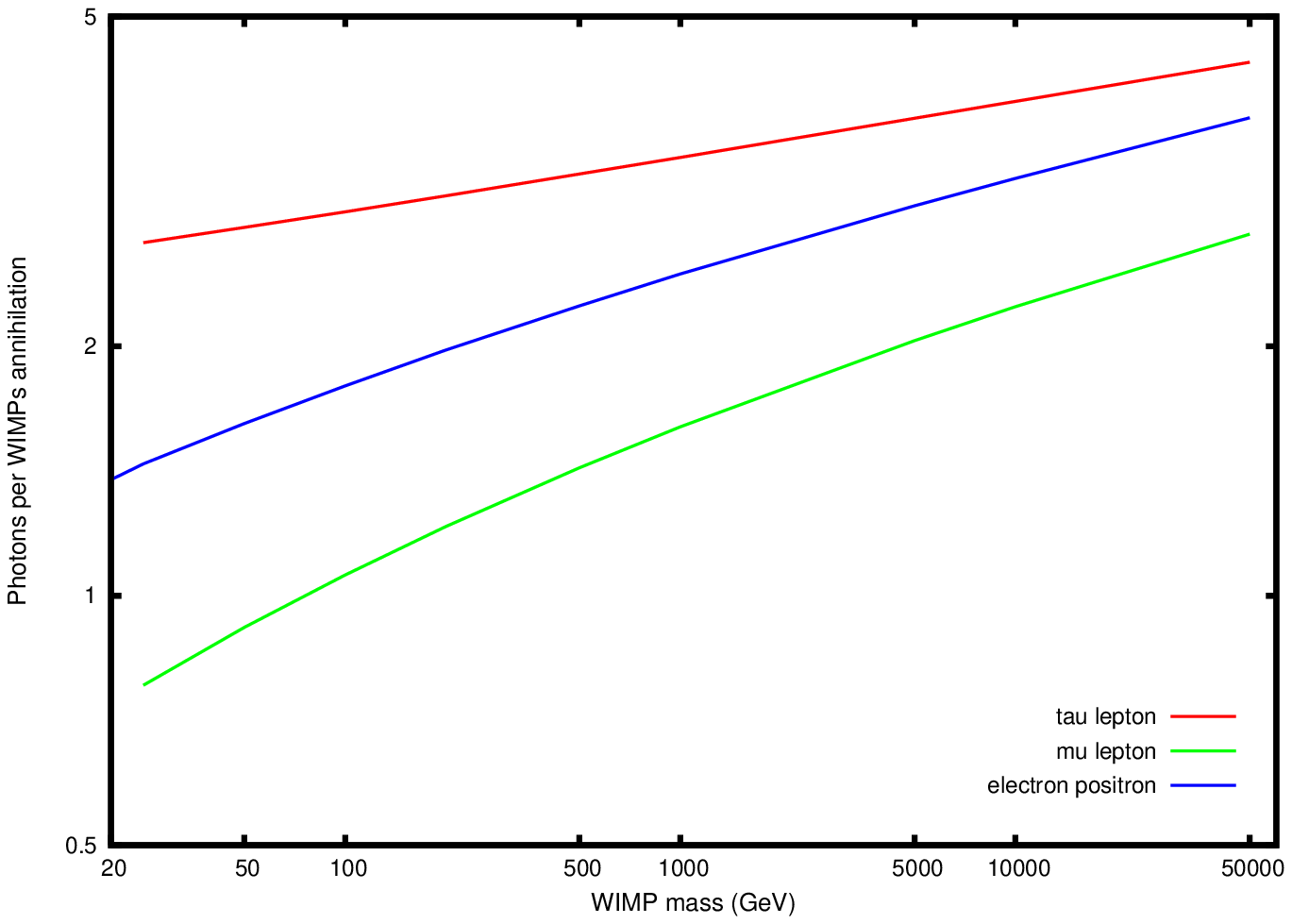}
\end{overpic}
}
\subfigure[ \hspace{1ex} Total number of photons per WIMP annihilation
in gauge bosons and quark-antiquark pairs.]{
\begin{overpic}[width=8.40cm]{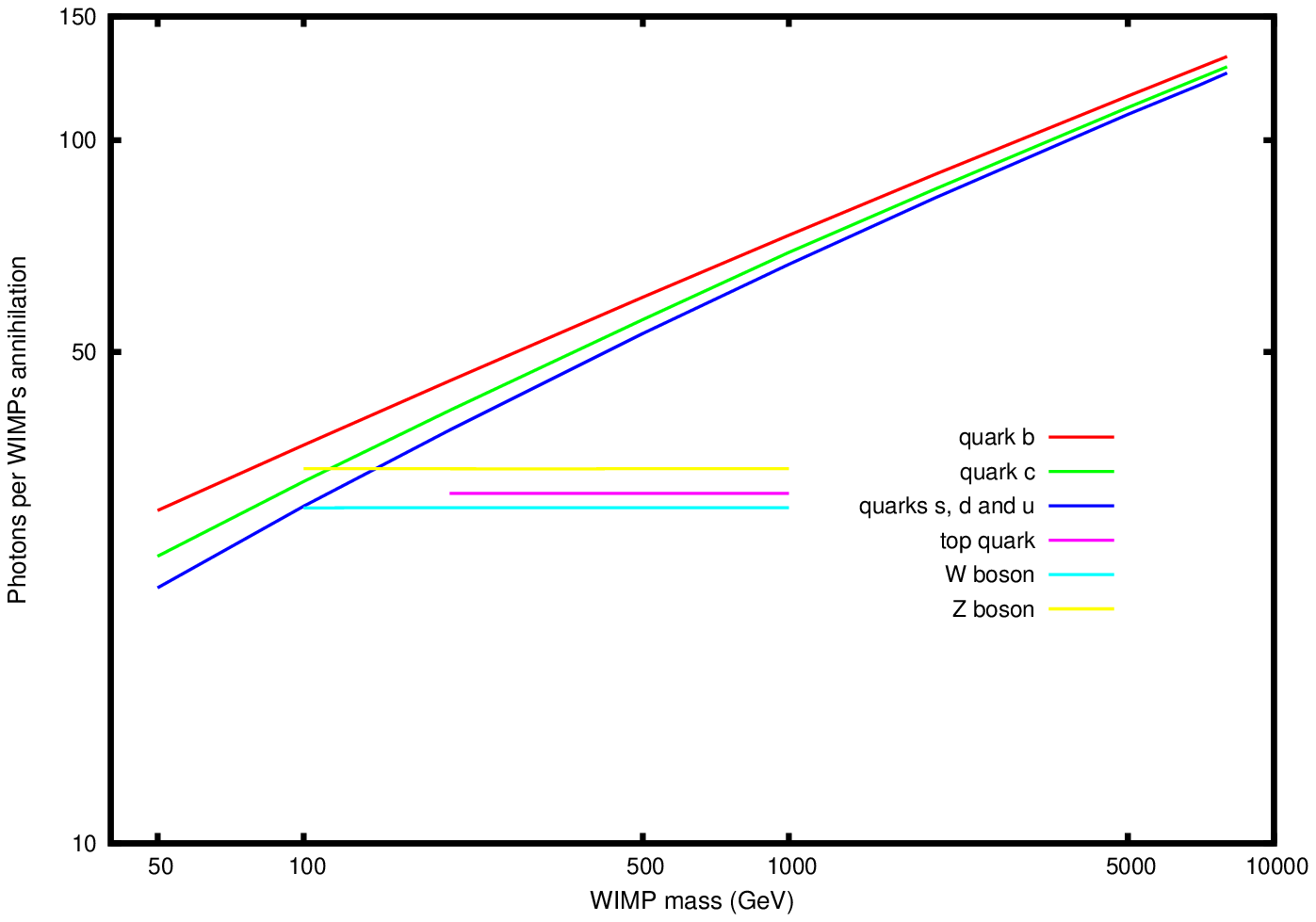}
\end{overpic}
}
\label{photons_number_figures}
\caption{Total photon number per WIMP annihilation:
On the left (a) the leptonic channels are presented
whereas on the right (b) both gauge bosons and quarks results are plotted.}
\end{figure}
\end{document}